\documentclass[aps,twocolumn,showpacs,preprintnumbers,amsmath,amssymb,nofootinbib,showkeys]{revtex4-1}

\RequirePackage[T1]{fontenc}

\usepackage{epsfig,graphicx,amsmath,amssymb}
\RequirePackage{mathptmx}      
\RequirePackage{flushend}
\RequirePackage{color}
\usepackage{changes}
\usepackage{multirow}
\RequirePackage{hyperref}
\hypersetup{
	linktocpage,
    colorlinks,
    citecolor=blue,
    filecolor=black,
    linkcolor=blue,
    urlcolor=blue,
}

\begin{document}

\title{Effective-field theory analysis of the \boldmath $\tau^-\to K^-(\eta^{(\prime)},K^0) \nu_\tau$ decays}


\author{Sergi Gonz\`{a}lez-Sol\'{i}s$^{1,2}$}\email{sgonzal@iu.edu}
\author{Alejandro Miranda$^{3}$} \email{jmiranda@fis.cinvestav.mx}
\author{Javier Rend\'on$^{3}$} \email{jrendon@fis.cinvestav.mx}
\author{Pablo Roig$^{3}$} \email{proig@fis.cinvestav.mx}

\affiliation{$^1$Department of Physics, Indiana University, Bloomington, IN 47405, USA\\
$^2$Center for Exploration of Energy and Matter, Indiana University, Bloomington, IN 47408, USA\\
$^3$Departamento de F\'isica, Centro de Investigaci\'on y de Estudios Avanzados del IPN,\\
Apdo. Postal 14-740,07000 Ciudad de M\'exico, M\'exico}    


\begin{abstract}

We analyze the decays $\tau^-\to K^-(\eta^{(\prime)},K^0) \nu_\tau$ within an effective field theory that includes the most general interactions between Standard Model fields up to dimension six, assuming left-handed neutrinos.  
In particular, we examine different interesting phenomenological observables i.e. decay spectra and branching ratio, Dalitz plot distributions and the forward-backward asymmetry, to explore the sensitivity of the corresponding decays to the effects of non-standard interactions.
A controlled theoretical input on the Standard Model hadronic form factors, based on chiral symmetry, dispersion relations, data and asymptotic QCD properties, has allowed us to set bounds on the New Physics scalar and tensor effective couplings using the measured branching ratios.
These are found to be in line with the findings of our series of previous analyses of two-meson tau decays and less precise than the constraints obtained from semileptonic kaon decays.
In order to set stringent limits on these couplings, we will use all available experimental data of all possible di-meson tau decays.
This is our next step plan, that we hope to be of interest for future experimental analyses of these decays.


\end{abstract}


\maketitle

\section{Introduction}\label{section1}

Hadronic tau decays provide an important source of experimental information about QCD at low and intermediate energies.
These decays have the advantage of containing hadrons in the final state thus avoiding the complications arising from having them in the initial state as well.
At the exclusive level, they can be used to understand specific properties of pions, kaons, $\eta$ and $\eta^{\prime}$ mesons, and the interactions among them.
So far, we have a good knowledge over decays into a pair of pseudoscalar mesons, the Standard Model (SM) input of which is encoded in terms of hadronic form factors.
An ideal roadmap to describe meson form factors would require a model-independent approach demanding a full knowledge of QCD in both its perturbative and non-perturbative regimes, knowledge not yet unraveled.
An alternative to such enterprise would pursuit a synergy between theoretical calculations and experimental data.
In this respect, dispersion relations are a powerful tool to direct oneself towards a model-independent description of meson form factors.
For example, the analyses of the decays $\pi^{-}\pi^{0}$ \cite{Guerrero:1997ku,Pich:2001pj,Dumm:2013zh,Gonzalez-Solis:2019iod} and $K_{S}\pi^{-}$ \cite{Jamin:2006tk,Jamin:2008qg,Boito:2008fq,Boito:2010me,Escribano:2014joa}, carried out by exploiting the synergy between Resonance Chiral Theory \cite{Ecker:1988te} and dispersion theory, are found to be in a nice agreement with the rich data provided by the experiments.
Accord with experimental measurements is also found for the $K^{-}K_{S}$ \cite{Gonzalez-Solis:2019iod} and $K^{-}\eta$ \cite{Escribano:2013bca,Escribano:2014joa} decay modes, although higher-quality data on these processes is required to constrain the corresponding theories or models.

Several recent works \cite{Garces:2017jpz,Miranda:2018cpf,Rendon:2019awg,Cirigliano:2018dyk} have put forward that semileptonic tau decays offer also an interesting scenario to set bounds on non-standard weak charged current interactions complementary to other low-energy semileptonic probes considered before, such nuclear beta decays, purely leptonic lepton, pion and kaon decays or hyperon decays (see e.g.\,Refs.\,\cite{Cirigliano:2009wk,Bhattacharya:2011qm,Cirigliano:2012ab,Cirigliano:2013xha,Chang:2014iba,Courtoy:2015haa,Gonzalez-Alonso:2016etj,Gonzalez-Alonso:2016sip,Gonzalez-Alonso:2017iyc,Alioli:2017ces,Gonzalez-Alonso:2018omy}).
The aim of this work is to extent our previous analyses of the decays $\tau^{-}\to\pi^{-}\pi^{0}\nu_{\tau}$ \cite{Miranda:2018cpf}, $\tau^{-}\to(K\pi)^{-}\nu_{\tau}$ \cite{Rendon:2019awg} and $\tau^{-}\to\pi^{-}\eta^{(\prime)}\nu_{\tau}$ \cite{Garces:2017jpz}, which we studied using the most general effective Lagrangian for weak charge current interactions up to dimension six on a number of phenomenological interesting observables, to the $\tau^{-}\to K^{-}(\eta^{(\prime)},K^{0})\nu_{\tau}$ decays.

On the theory side, a controlled theoretical determination, with a robust error band, of the corresponding form factors within the SM is required in order to increase the accuracy of the search for non-standard interactions
At present, we have such a knowledge for the vector and-to a great extent-the scalar form factors, but there are no experimental data that can help us constructing the tensor form factor and, therefore, it will be built under theoretical considerations only. 

On the experimental side, our study is presently limited by the following facts: $i)$ for the decay $\tau^{-}\to K^{-}K^{0}\nu_{\tau}$, while the PDG reports a branching ratio of $1.486(34)\times10^{-3}$ \cite{PhysRevD.98.030001}, no measurement of the corresponding decay spectrum has been released by the $B$-factories; $ii)$ the associated errors of the brother process $\tau^{-}\to K^{-}K_{S}\nu_{\tau}$ BaBar data \cite{BaBar:2018qry} are still relatively large;
$iii)$ unfolding detector effects has not been performed for the $\tau^-\to K^- \eta\nu_\tau$ Belle data \cite{Inami:2008ar}\footnote{This decay was also measured by BaBar \cite{delAmoSanchez:2010pc}. 
However, the person in charge of the analysis left the field and the data file was lost, unfortunately.}; 
$iv)$ and, finally, the decay $\tau^{-}\to K^{-}\eta^{\prime}\nu_\tau$ has not been detected yet, although an upper limit at the $90\%$ confidence level was placed by BaBar \cite{Lees:2012ks}.
We will not thus attempt to extract new physics bounds from the corresponding experimental data as competitive as those coming from other low-energy probes, like the ones mentioned before,
but rather explore the size of the deviations from the SM predictions that one could expect in these decay channels.
For these reasons, we hope that our paper strengths the case for a (re)analysis, with a larger data sample, of the $K^{-}K^{0}$, $K^{-}K_{S}$ and $K^{-}\eta$ decay spectra and encourage experimental groups to measure the $K^{-}\eta^{\prime}$ decay mode.
All this should be well within the reach of Belle-II \cite{Kou:2018nap}, and of other future $Z$, tau-charm and $B$-factories where new measurements should be possible.

Our paper is organized as follows.
The theoretical framework is given in section \ref{section2} where we briefly present the effective Lagrangian and discuss the different effective weak currents contributing to the decays.
The hadronic matrix element and the participant form factors are also defined in this section.
The latter are the matter subject of section \ref{section3}, where we pay special attention to the tensor form factor.
In section \ref{section4}, we discuss the different interesting phenomenological observables i.e. decay spectra and branching ratio, Dalitz plot distributions and the forward-backward asymmetry, that can help us setting bounds on non-SM interactions.
We derive these bounds in section \ref{Limits}. 
Finally, our conclusions are presented in section \ref{conclusions}.

\section{Effective field theory analysis and decay amplitude of $\tau^{-}\to\nu_\tau\bar{u}D$ $(D=d,s)$}\label{section2}

We start out writing the effective Lagrangian including dimension six operators that describes semileptonic $\tau^{-}\to \nu_\tau \bar{u} D$ strangeness-conserving ($D=d$) or changing ($D=s$) charged current transitions with left-handed neutrinos.
Such Lagrangian reads \cite{Garces:2017jpz,Miranda:2018cpf,Rendon:2019awg}:
\begin{eqnarray}
\mathcal{L}_{CC}&=&-\frac{G_F}{\sqrt{2}}V_{uD}(1+\epsilon_L+\epsilon_R)\big[\bar{\tau}\gamma_\mu(1 -\gamma^5)\nu_{\tau}\,\nonumber\\[1ex]
&&\cdot\bar{u}\big[\gamma^\mu-(1-2\hat{\epsilon}_R)\gamma^\mu\gamma^5\big]D\nonumber\\[1ex]
&&+\bar{\tau}(1-\gamma^5) \nu_{\tau}\,\bar{u}(\hat{\epsilon}_S -\hat{\epsilon}_P\gamma^5)D\nonumber\\[1ex]
&&+2\hat{\epsilon}_T\bar{\tau}\sigma_{\mu\nu}(1-\gamma^5) \nu_{\tau}\,\bar{u}\sigma^{\mu\nu}D\big]+h.c.\,,
\label{Lcc}
\end{eqnarray}
where $G_{F}$ is the tree-level definition of the Fermi constant.
In the previous Lagrangian, we have defined $\hat{\epsilon}_{i}=\epsilon_{i}/(1+\epsilon_{L}+\epsilon_{R})$ for $i=R,S,P,T$, with $\epsilon_{L,R}$ and $\epsilon_{i}$ being effective couplings characterizing NP that can be taken real since we are only interested in $CP$ conserving quantities.
Of course, if we set them to zero i.e. $\epsilon_{L,R}=\hat{\epsilon}_{R,S,P,T}=0$, we recover the SM Lagrangian. 
This factorized form of Eq.\,(\ref{Lcc}) is useful as long as conveniently normalized rates allow to cancel the overall factor $(1+\epsilon_L+\epsilon_R)$.
Note that since $\epsilon_{i}=\hat{\epsilon}_{i}$ at linear order in $\hat{\epsilon}_{i}'s$, we may use $\epsilon_{i}$ instead of $\hat{\epsilon}_{i}$ when comparing to works which use the former instead of the latter \cite{Bhattacharya:2011qm}.
A more detailed derivation of the Lagrangian of Eq.\,(\ref{Lcc}) can be found in our previous publications \cite{Garces:2017jpz,Miranda:2018cpf,Rendon:2019awg} and we therefore have decided not repeat it here once again.

The decay amplitude for $\tau^-\left(P\right)\to K^-\left(p_K\right) \eta^{(\prime)}(p_{\eta^{(\prime)}})\nu_\tau\left(P'\right)$ that arises from the Lagrangian in Eq.\,(\ref{Lcc}) contains a vector $(V)$, an scalar $(S)$ and a tensor $(T)$ contribution.
The resulting amplitude can be expressed as\footnote{The short-distance electroweak radiative corrections encoded in $S_{EW}$ \cite{Erler:2002mv}, do not affect the scalar and tensor contributions.
However, the error made by taking $\sqrt{S_{EW}}$ as an overall factor in Eq.\,(\ref{DecayAmplitude}) is negligible.}
\begin{eqnarray}
\mathcal{M}&=&\mathcal{M}_V + \mathcal{M}_S + \mathcal{M}_T\nonumber\\[1ex]
&=&\frac{G_F V_{us} \sqrt{S_{EW}}}{\sqrt{2}}(1+\epsilon_L+\epsilon_R)\nonumber\\[1ex]
&&\times\bigl[L_\mu H^\mu+\hat{\epsilon}_S LH+2\hat{\epsilon}_T L_{\mu\nu}H^{\mu\nu}\bigr]\,,
\label{DecayAmplitude}
\end{eqnarray}
where the leptonic currents are defined by:
\begin{eqnarray}
L_\mu&=&\bar{u} (P')\gamma_\mu (1-\gamma^5)u(P)\,,\\[1ex]
L&=&\bar{u} (P')(1+\gamma^5)u(P)\,,\\[1ex]
L_{\mu\nu}&=&\bar{u} (P')\sigma_{\mu\nu} (1+\gamma^5)u(P)\,.
\end{eqnarray}
The scalar $H$, vector $(H^{\mu})$ and tensor $(H^{\mu\nu})$ hadronic matrix elements in Eq.\,(\ref{DecayAmplitude}) can be decomposed in terms of allowed Lorentz structures and a number of form factors encoding the hadronization procedure as
\begin{eqnarray}\label{ad:2}
H&=&\langle  K^-\eta^{(\prime)} \vert \bar{s} u \vert 0\rangle\equiv F_S^{ K^-\eta^{(\prime)}}(s)\,,\label{ScalarCurrent}\\[1ex]
H^\mu&=&\langle K^-\eta^{(\prime)} \vert \bar{s}\gamma^\mu u \vert 0\rangle=C^V_{ K^-\eta^{(\prime)}} Q^\mu F_+^{ K^-\eta^{(\prime)}}(s)\nonumber\\[1ex]
&+&C^S_{ K^-\eta^{(\prime)}} \left(\frac{\Delta_{K\pi}}{s}\right)q^\mu F_0^{ K^-\eta^{(\prime)}}(s)\,,\label{VectorCurrent}\\[1ex]
H^{\mu\nu}&=&\langle  K^-\eta^{(\prime)}\vert \bar{s}\sigma^{\mu\nu} u \vert 0\rangle=iF_T^{ K^-\eta^{(\prime)}}(s)(p^\mu_{\eta^{(\prime)}}p^\nu_{K}-p^\mu_{K}p^\nu_{\eta^{(\prime)}})\,,\nonumber\\
\end{eqnarray}
where $q^\mu=(p_{K}+p_{\eta^{(\prime)}})^\mu$, $Q^\mu=(p_{\eta^{(\prime)}}-p_{K})^\mu+(\Delta_{K\eta^{(\prime)}}/s)q^\mu$, $s=q^2$ and $\Delta_{ij}=m_i^2-m_j^2$, and with the Clebsch-Gordan coefficients: $C^V_{K\eta^{(\prime)}}=-\sqrt{\frac{3}{2}}$, $C^S_{K\eta}=-\frac{1}{\sqrt{6}}$ and $C^S_{K\eta^{'}}=\frac{2}{\sqrt{3}}$.    
The divergence of the vector current Eq.\,(\ref{VectorCurrent}) relates the form factors $F_S(s)$ and $F_0(s)$ via
\begin{equation}
\label{FS_F0_relation}
F_S(s)=\frac{C^S_{K\eta^{(\prime)}}\Delta_{K\pi}}{m_s-m_u}F_0^{K\eta^{(\prime)}}(s)\,.
\end{equation}

As in \cite{Garces:2017jpz,Miranda:2018cpf,Rendon:2019awg}, the scalar and vector contributions in Eqs.\,(\ref{ScalarCurrent}) and Eq.\,(\ref{VectorCurrent}), respectively, can be treated jointly by doing the following replacement 
\begin{equation}
 C^S_{K\eta^{(\prime)}}\frac{\Delta_{K\pi}}{s}\to C^S_{K\eta^{(\prime)}}\frac{\Delta_{K\pi}}{s}\left(1+\frac{s\, \hat{\epsilon}_S }{m_\tau(m_s-m_u)}\right)\,,
\end{equation}
in Eq.\,(\ref{VectorCurrent}).
For the decay $\tau^{-}\rightarrow K^{-}K^{0}\nu_{\tau}$, the associated amplitude is that of Eq.\,(\ref{DecayAmplitude}) but replacing $p_{\eta^{(')}}\rightarrow p_{K^{0}}$, $\Delta_{K^{-}\eta^{(')}}\rightarrow \Delta_{K^{-}K^{0}}$, and $m_{s}\rightarrow m_{d}$ along the lines of the previous equations, and with the Clebsch-Gordan coefficients $C^{V}_{KK}=C^{S}_{KK}=-1$.

The parametrization of the three independent form factors i.e. $F_{0}(s),F_{+}(s)$ and $F_{T}(s)$, is the subject of the next section.

\section{Hadronization of the scalar, vector and tensor currents}\label{section3}

In this section, we provide a brief overview of the description of the scalar, vector and tensor form factors that we need for our analysis.
It is fundamental to have good control over them since they are used as SM inputs for binding the non-standard interactions.
The setup approach to describe the $K^{-}\eta^{(\prime)}$ vector form factor is the following.
They are calculated within the context of Resonance Chiral Theory taking into account the effects of the $K^{*}(892)$ and the $K^{*}(1410)$ vector resonances, and are connected to the $K\pi$ vector form factor through $F_{+}^{K\eta^{(\prime)}}(s)=\cos\theta_{P}(\sin\theta_{P})F_{+}^{K\pi}(s)$ \cite{Escribano:2013bca}, where $\theta_{P}$ is the $\eta$-$\eta^{\prime}$ mixing angle in the octet-singlet basis.
We will thus discuss the illustrative case of the $K\pi$ vector form factor and take $\theta_{P}=(-13.3\pm0.5)^{\circ}$ \cite{Ambrosino:2006gk}.
For our analysis, we follow the representation outlined in Ref.\,\cite{Boito:2008fq}, and briefly summarized below for the convenience of the reader, and write a thrice subtracted dispersion relation
\begin{eqnarray}
F_+^{K\pi}(s)&=&F_+^{K\pi}(0)\exp\Bigg[\alpha_1\frac{s}{m_{\pi}^2}+\frac{1}{2}\alpha_2\frac{s^2}{m_{\pi}^4}\nonumber\\[1ex]
&+&\frac{s^3}{\pi}\int_{s_{K\pi}}^{s_{\rm{cut}}} ds^\prime \frac{\delta_{+}^{K\pi}(s^\prime)}{(s^\prime)^3(s^\prime-s-i0)}\Bigg]\,,
\label{DR3sub}
\end{eqnarray}
where $s_{K\pi}=(m_K+m_\pi)^2$ is the threshold of the $K\pi$ system, while the value of $F_+^{K\pi}(0)$ is extracted from $|V_{us}F_{+}^{K^{-}\pi^{0}}(0)|=0.2165(2)$ \cite{PhysRevD.98.030001}, and $\alpha_1$ and $\alpha_2$ are two subtraction constants that are related to the low-energy expansion of the form factor. 
The use of a three-times subtracted dispersion relation reduces the high-energy contribution of the integral where the phase is less well-known.
In Eq.\,(\ref{DR3sub}), $s_{\rm{cut}}$ is a cut-off whose value is fixed from the requirement that the fitted parameters are compatible within errors with the case $s_{\rm{cut}}\to\infty$.
In Refs.\,\cite{Boito:2008fq,Escribano:2014joa}, the value of $s_{\rm{cut}}=4$ GeV$^{2}$ was found to satisfy this criterion, and variations of $s_{\rm{cut}}$ were used to estimate the associated systematic error.
For the input phase $\delta_{+}^{K\pi}(s)$ we use
\begin{equation}
 \delta_{+}^{K\pi}(s)\,=\,\mathrm{tan}^{-1}\left[\frac{\mathrm{Im}\widetilde{f}^{K\pi}_+(s)}{\mathrm{Re}\widetilde{f}^{K\pi}_+(s)}\right]\,,
\label{KpiPhase} 
\end{equation}
where $\widetilde{f}^{K\pi}_+(s)$ is taken to be of the form \cite{Boito:2008fq}
\begin{equation}\label{Tilded VFF BEJ}
 \widetilde{f}^{K\pi}_+(s)\,=\,\frac{m_{K^\star}^2-\kappa_{K^\star}\widetilde{H}_{K\pi}(0)+\gamma s}{D(m_{K^\star},\,\gamma_{K^\star})}-\frac{\gamma s}{D(m_{K^{\star\prime}},\,
\gamma_{K^{\star^\prime}})}\,,
\end{equation}
where we have included two resonances, the $K^{*}=K^{*}(892)$ and the $K^{*\prime}=K^{*}(1410)$.
The denominators in Eq.\,(\ref{Tilded VFF BEJ}) are
\begin{equation}
 D(m_n,\gamma_n)\equiv m_n^2-s-\kappa_n{\rm{Re}}\left[H_{K\pi}(s)\right]-im_n\gamma_n(s)\,,
\end{equation}
where
\begin{equation}
 \kappa_n\,=\,\frac{192\pi F_KF_\pi}{\sigma_{K\pi}^3(m_n^2)}\frac{\gamma_n}{m_n}\,,\quad \gamma_n(s)\,=\,\gamma_n\frac{s}{m_n^2}\frac{\sigma^3_{K\pi}(s)}{\sigma^3_{K\pi}(m_n^2)}\,,
\end{equation}
and with the two-body phase-space factor given by $\sigma_{K\pi}(s)=2q_{K\pi}(s)/\sqrt{s}$ where
\begin{eqnarray}
q_{K\pi}(s)&=&\frac{1}{2\sqrt{s}}\sqrt{(s-(m_{K}+m_{\pi})^{2})(s-(m_{K}-m_{\pi})^{2})}\nonumber\\[1ex]
&&\theta(s-(m_{K}+m_{\pi})^{2})\,.
\end{eqnarray} 
The scalar one-loop integral functions $H_{K\pi}(s)$ is defined below Eq.\,(3) of Ref.\,\cite{Jamin:2006tk}, however removing the factor $1/F_{\pi}^{2}$ which cancels if $\kappa_{n}$ is expressed in terms of the unphysical width $\gamma_{n}$.
For our analysis, we use the results of the reference fit of Ref.\,\cite{Escribano:2014joa} together with the systematic uncertainty obtained as explained along the lines of the same reference.
One disadvantage of Eq.\,(\ref{DR3sub}) is that the $1/s$ asymptotic fall-off of the form factor \cite{Lepage:1979zb} it is not guaranteed because the subtraction constants are fixed from a fit to experimental data. 
However, we have checked that our form factor parametrization is indeed a decreasing function of $s$ (apart from the $K^{*}(892)$ and $K^{*}(1410)$ peak structures) within the entire range where we apply it.

Regarding the $K\eta^{(\prime)}$ scalar form factors, we employ the well-established results of Ref.\,\cite{Jamin:2001zq} derived from a dispersive analysis with three channels $(K\pi,K\eta,K\eta^{\prime})$ 
\footnote{We are very grateful to Matthias Jamin and Jose Antonio Oller for providing us their solutions in tables.}.

\begin{figure*}
\includegraphics[scale=0.34]{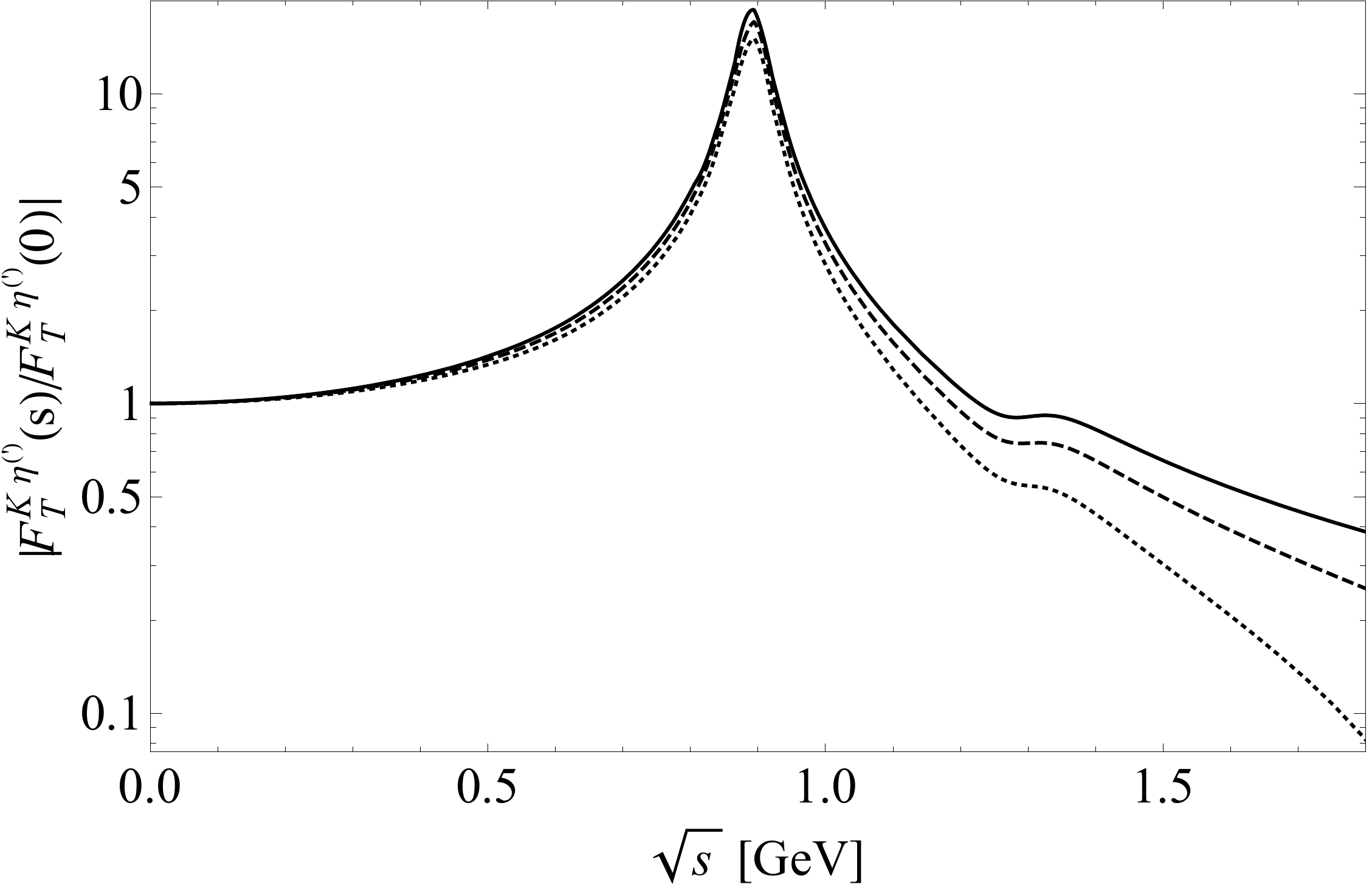}
\includegraphics[scale=0.35]{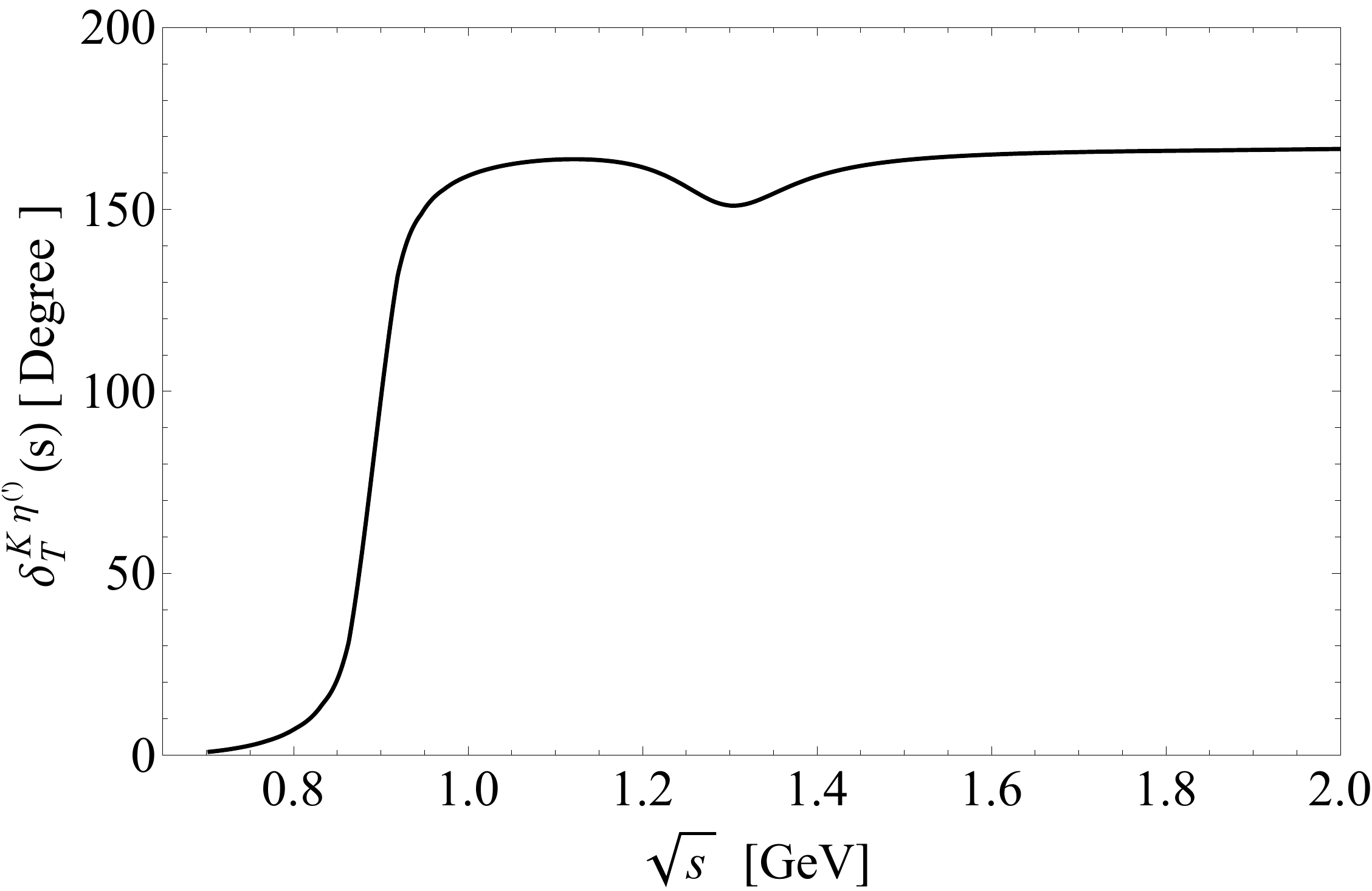}
\centering
\caption{Normalized absolute value of the tensor form factor $F_{T}^{K\eta^{(\prime)}}(s)$ given in Eq.\,(\ref{KetaTensorFF}) (left), for $s_{\rm{cut}}=4$ GeV$^{2}$ (dotted line), $9$ GeV$^{2}$ (dashed line) and $s_{\rm{cut}}\to\infty$ (solid line), and tensor form factor phase $\delta_{T}^{K\eta^{(\prime)}}(s)$ (right).}
\label{PlotsFTs}
\end{figure*}

The tensor form factor is one of the most difficult inputs to be reliable estimated since there are no experimental data that can help constructing $F_{T}^{K\eta^{(\prime)}}(s)$.
Therefore, we shall rely on theoretical considerations only.
The key observation is that the tensor form admits an Omn\`{e}s dispersive representation \cite{Cirigliano:2017tqn,Miranda:2018cpf,Rendon:2019awg}
\begin{equation}\label{KetaTensorFF}
F_T^{K\eta^{(\prime)}}(s)=F_T^{K\eta^{(\prime)}}(0)\exp\left[\frac{s}{\pi}\int_{s_{K\pi}}^{s_{\rm{cut}}}ds^\prime \frac{\delta_T^{K\eta^{(\prime)}}\left(s^\prime\right)}{s^\prime(s^\prime-s-i0)}\right]\,,
\end{equation}
where in the elastic region, the phase of the tensor form factor equals the $P$-wave phase of the $K\pi$ vector form factor i.e. $\delta_T^{K\eta^{(\prime)}}(s)=\delta_+^{K\pi}(s)$, with $\delta_+^{K\pi}(s)$ extracted from Eq.\,(\ref{KpiPhase}).
We will assume the previous relations also hold above the onset of inelasticities until $m_{\tau}^{2}$ where we guide smoothly the tensor phase to $\pi$ as in Ref.\,\cite{Gonzalez-Solis:2019iod} to ensure the asymptotic $1/s$ behavior dictated by perturbative QCD \cite{Lepage:1979zb}.
Lacking of precise low-energy information, we do not increase the number of subtractions in Eq.\,(\ref{KetaTensorFF}), which, in turn, would reduce the importance of the higher-energy part of the integral, but rather cut the integral at different values of $s_{\rm{cut}}$ and take the differing results as an estimate of our theoretical systematic uncertainty for the results presented in section \ref{Limits}.
In Fig.\,\ref{PlotsFTs}, we show the tensor form factor phase $\delta_{T}^{K\eta^{(\prime)}}(s)$ (right panel) together with the (normalized) absolute value of the tensor form factor (left panel) for the cases $s_{\rm{cut}}=4,9$ GeV$^{2}$ and $s_{\rm{cut}}\to\infty$, which is taken as the baseline hypothesis.
The value of the normalization $F_T^{K\eta^{(\prime)}}(0)$ required in Eq.\,(\ref{KetaTensorFF}) can be estimated within ChPT as explained in the following.
The lowest-order ChPT Lagrangian with tensor sources is of $\mathcal{O}(p^4)$ in the chiral counting and reads \cite{Cata:2007ns}
\begin{equation}\label{keta:eq3.4}
\mathcal{L}=\Lambda_1\left\langle t_+^{\mu\nu} f_{+\mu\nu}\right\rangle-i\Lambda_2 \left\langle t_+^{\mu\nu} u_\mu u_\nu\right\rangle+\Lambda_3 \left\langle t_+^{\mu\nu} t^+_{\mu\nu}\right\rangle+\Lambda_4 \left\langle f_+^{\mu\nu}\right\rangle^2\,,
\end{equation}
where $t_+^{\mu\nu}=u^\dagger t^{\mu\nu} u^\dagger+u t^{\mu\nu\dagger} u$ includes the tensor source and its adjoint, and $\left\langle\cdots\right\rangle$ stands for a flavor space trace. 
Only terms proportional to $\Lambda_2$ contribute to the decays we are considering.
The chiral tensors entering Eq.\,(\ref{keta:eq3.4}) are given by: $u_\mu=i\left[u^\dagger (\partial_\mu-ir_\mu)u-u(\partial_\mu-i l_\mu)u^\dagger\right]$, where $l_\mu$ and $r_\mu$ are the left- and right-handed sources, and $f_+^{\mu\nu}=u F_L^{\mu\nu} u^\dagger +u^\dagger F_R^{\mu\nu} u$, that includes the left- and right-handed field-strength tensors for $l_\mu$ and $r_\mu$, $F_{L,R}^{\mu\nu}$.
The non-linear representation of the pseudo-Goldstone bosons is given by $u=\exp\left[\frac{i}{\sqrt{2}F}\phi\right]$ \cite{Coleman:1969sm,Callan:1969sn}, where
\begin{equation}
\phi=\left(\begin{array}{ccc}
\frac{\pi^3+\eta_q}{\sqrt{2}} & \pi^+ & K^+\\
\pi^- & \frac{-\pi^3+\eta_q}{\sqrt{2}} & K^0 \\
K^- & \bar{K}^{0} & \eta_s\\
\end{array}\right)\,,
\end{equation}
where $\eta_q=C_q\eta +C_{q^\prime}\eta^\prime$ and $\eta_s=-C_s\eta +C_{s^\prime}\eta^\prime$ are the light and strange quark components of the $\eta$ and $\eta^\prime$ mesons, respectively. 
$\pi^3$ coincides with the $\pi^0$ when the isospin-breaking terms are neglected (as done along this article). 
The constants describing the mixing between $\eta_q$ and $\eta_s$ states are given by \cite{Leutwyler:1997yr, Kaiser:2000gs}
\begin{eqnarray}
C_q&\equiv&\frac{F_{\pi}}{\sqrt{3}\cos(\theta_8-\theta_0)}\left(\frac{\cos \theta_0}{f_8}-\frac{\sqrt{2}\sin \theta_8}{f_0}\right)\,,\\[1ex]
C_{q^\prime}&\equiv&\frac{F_{\pi}}{\sqrt{3}\cos(\theta_8-\theta_0)}\left(\frac{\sqrt{2}\cos \theta_8}{f_0}+\frac{\sin \theta_0}{f_8}\right)\,,\\[1ex]
C_s&\equiv&\frac{F_{\pi}}{\sqrt{3}\cos(\theta_8-\theta_0)}\left(\frac{\sqrt{2}\cos \theta_0}{f_8}+\frac{\sin \theta_8}{f_0}\right)\,\\[1ex]
C_{s^\prime}&\equiv&\frac{F_{\pi}}{\sqrt{3}\cos(\theta_8-\theta_0)}\left(\frac{\cos \theta_8}{f_0}-\frac{\sqrt{2}\sin \theta_0}{f_8}\right),
\end{eqnarray}
and for the corresponding mixing parameters we use \cite{Escribano:2015yup,Guevara:2018rhj}
\begin{eqnarray}
\theta_8&=&(-21.2\pm1.9)^\circ,\quad \theta_0=(-6.9\pm2.4)^\circ\,,\\[1ex] 
f_8&=&(1.27\pm0.02)F_{\pi}, \quad f_0=(1.14\pm 0.05)F_{\pi}\,,
\end{eqnarray}
with $F_{\pi}=92.2\,\mathrm{MeV}$ being the pion decay constant.

The tensor source ($\overline{t}^{\mu\nu}$) is related to its chiral projections ($t^{\mu\nu}$ and $t^{\mu\nu\dagger}$) by \cite{Cata:2007ns}
\begin{equation}
t^{\mu\nu}=P_L^{\mu\nu\lambda\rho}\,\overline{t}_{\lambda\rho}, \quad 4P_L^{\mu\nu\lambda\rho}=\left(g^{\mu\lambda}g^{\nu\rho}-g^{\mu\rho}g^{\nu\lambda}+i\epsilon^{\mu\nu\lambda\rho}\right),
\end{equation}
where $\overline{\Psi}\sigma_{\mu\nu}\overline{t}^{\mu\nu}\Psi$ is the tensor quark current.
Taking the functional derivative of eq. (\ref{keta:eq3.4}) with respect to the tensor source $\bar{t}_{\mu\nu}$, we get
\begin{eqnarray}
\Big\langle K^-\eta\Big| \frac{\delta \mathcal{L}_{\chi PT}^4}{\delta \bar{t}_{\mu\nu}}\Big| 0\Big\rangle&=&i\left(\frac{C_q}{\sqrt{2}}+C_s\right)\frac{\Lambda_2}{F_{\pi}^2}\left(p_\eta^\mu p_K^\nu-p_K^\mu p_\eta^\nu\right)\,,\nonumber\\[1ex]
\Big\langle K^-\eta'\Big|\frac{\delta \mathcal{L}_{\chi PT}^4}{\delta \bar{t}_{\mu\nu}}\Big| 0\Big\rangle&=&i\left(\frac{C_{q'}}{\sqrt{2}}-C_{s'}\right)\frac{\Lambda_2}{F_{\pi}^2}\left(p^{\mu}_{\eta^{\prime}}p_K^\nu-p_K^\mu p^{\nu}_{\eta^{\prime}}\right)\,.\nonumber\\
\label{FunctDer}
\end{eqnarray}
Ref.\,\cite{Baum:2011rm} evaluated $F_{T}^{K\pi}(0)=2m_{\pi}F_{T}(0)$ on the lattice.
Their result $F_{T}^{K\pi}(0)=0.417\pm0.015$, together with the fact that 
\begin{eqnarray}
F_{T}^{K^{-}\eta}(0)&=&\left(\frac{C_q}{\sqrt{2}}+C_s\right)\frac{\Lambda_2}{F_{\pi}^2}\,,\\[1ex]
F_{T}^{K^{-}\eta^{\prime}}(0)&=&\left(\frac{C_q'}{\sqrt{2}}-C_s'\right)\frac{\Lambda_2}{F_{\pi}^2}\,,
\label{FT0}
\end{eqnarray}
 yields $\Lambda_2=(11.1\pm0.4)\,\mathrm{MeV}$, that we will use for our analysis.
This value is consistent within one sigma with the one employed for the $\pi\pi$ channel \cite{Miranda:2018cpf}.

We turn next to describe the form factors required for $\tau^{-}\to K^{-}K^{0}\nu_{\tau}$.
We will not discuss them at length here but rather provide a compilation of the main formulae to make this work self-contained.
For the kaon vector form factor, we follow Ref.\,\cite{Gonzalez-Solis:2019iod}, where a three-times dispersion relation was formulated, and write 
\begin{equation}
F_{+}^{KK}(s)=\exp\left[\tilde{\alpha}_{1}s+\frac{\tilde{\alpha}_{2}}{2}s^{2}+\frac{s^{3}}{\pi}\int_{4m_{\pi}^{2}}^{s_{\rm{cut}}}ds^{\prime}\frac{\delta_{+}^{KK}(s)}{(s^{\prime})^{3}(s^{\prime}-s-i0)} \right]\,,
\label{KaonVFF}
\end{equation}
where $\tilde{\alpha}_{1}$ and $\tilde{\alpha}_{2}$, are two subtraction constants related to the slope and curvature appearing in the low-energy expansion of the form factor of the kaon.
To get a model for the form factor phase, $\delta_{+}^{KK}(s)$ in Eq.\,(\ref{KaonVFF}), we adopt the so-called exponential Omn\`{e}s representation of the form factor \cite{Gonzalez-Solis:2019iod}:
\begin{widetext}
\begin{eqnarray}
f_{+}^{KK}(s)&=&\frac{M_{\rho}^{2}+s\left(\tilde{\gamma} e^{i\tilde{\phi}_{1}}+\tilde{\delta} e^{i\tilde{\phi}_{2}}\right)}{M_{\rho}^{2}-s-iM_{\rho}\Gamma_{\rho}(s)}\exp\Bigg\lbrace {\rm{Re}}\Bigg[-\frac{s}{96\pi^{2}F_{\pi}^{2}}\left(A_{\pi}(s)+\frac{1}{2}A_{K}(s)\right)\Bigg]\Bigg\rbrace\nonumber\\[2mm]
&&-\frac{\tilde{\gamma}\,s\,e^{i\tilde{\phi}_{1}}}{M_{\rho^{\prime}}^{2}-s-iM_{\rho^{\prime}}\Gamma_{\rho^{\prime}}(s)}\exp\Bigg\lbrace-\frac{s\Gamma_{\rho^{\prime}}(M_{\rho^{\prime}}^{2})}{\pi M_{\rho^{\prime}}^{3}\sigma_{\pi}^{3}(M_{\rho^{\prime}}^{2})}{\rm{Re}}A_{\pi}(s)\Bigg\rbrace-\frac{\tilde{\delta}\,s\,e^{i\tilde{\phi}_{2}}}{M_{\rho^{\prime\prime}}^{2}-s-iM_{\rho^{\prime\prime}}\Gamma_{\rho^{\prime\prime}}(s)}\exp\Bigg\lbrace-\frac{s\Gamma_{\rho^{\prime\prime}}(M_{\rho^{\prime\prime}}^{2})}{\pi M_{\rho^{\prime\prime}}^{3}\sigma_{\pi}^{3}(M_{\rho^{\prime\prime}}^{2})}{\rm{Re}}A_{\pi}(s)\Bigg\rbrace\,.\nonumber\\
\label{FFExpThreeResKaon}
\end{eqnarray}
\end{widetext}
In Eq.\,(\ref{FFExpThreeResKaon}), the mixing between resonances is taken with respect to the $\rho$ with relative strengths $1,\tilde{\gamma},\tilde{\delta}$
These parameters are in general complex thus carrying a phase that it is denoted by $\tilde{\phi}_{1}$ and $\tilde{\phi}_{2}$, respectively.
Taking $\tilde{\gamma}$ and $\tilde{\delta}$ real would demand a perfect knowledge of the amplitudes of the $\rho^{\prime}$ and $\rho^{\prime\prime}$ contributions and, as this is not the case, we consider a more flexible scenario and add a phase that can absorb part of the associated shortcomings. 
The $\rho$-meson resonance width is accounted for through \cite{GomezDumm:2000fz}
\begin{eqnarray}
\Gamma_{\rho}(s)&=&-\frac{M_{\rho}s}{96\pi^{2}F_{\pi}^{2}}{\rm{Im}}\left[A_{\pi}(s)+\frac{1}{2}A_{K}(s)\right]\nonumber\\[1ex]
&=&\frac{M_{\rho}s}{96\pi F_{\pi}^{2}}\left[\sigma_{\pi}^{3}(s)\theta(s-4m_{\pi}^{2})+\frac{1}{2}\sigma_{K}^{3}(s)\theta(s-4m_{K}^{2})\right]\,,\nonumber\\
\label{RhoWidth}
\end{eqnarray}
while for the energy-dependent width of the $\rho^{\prime}$ and $\rho^{\prime\prime}$ we do not take intermediate states other than $\pi\pi$
\begin{eqnarray}
\Gamma_{\rho^{\prime},\rho^{\prime\prime}}(s)&=&\Gamma_{\rho^{\prime},\rho^{\prime\prime}}\frac{s}{M_{\rho^{\prime},\rho^{\prime\prime}}^{2}}\frac{\sigma^{3}_{\pi}(s)}{\sigma^{3}_{\pi}(M_{\rho^{\prime},\rho^{\prime\prime}}^{2})}\theta(s-4m_{\pi}^{2})\,.
\end{eqnarray}
From Eq.\,(\ref{FFExpThreeResKaon}) we extract its phase through
\begin{equation}
\tan\delta_{+}^{KK}(s)=\frac{{\rm{Im}}f^{KK}_{+}(s)}{{\rm{Re}}f^{KK}_{+}(s)}\,.
\label{KKphase}
\end{equation}
In fact, we only use the phase thus extracted to describe the energy region that goes from 1 GeV$^{2}$ to $m_{\tau}^{2}$.
From $4m_{\pi}^{2}$ to 1 GeV$^{2}$ we employ the $P$-wave phase shift of the pion-pion scattering solution of the Roy equations \cite{GarciaMartin:2011cn} that we match to the phase in Eq.\,(\ref{KKphase}) at 1 GeV$^{2}$, while for the region $m_{\tau}^{2}\leq s$ we guide smoothly the phase to $\pi$ such that the correct $1/s$ high-energy behavior of the form factor is ensured (see Ref.\,\cite{Gonzalez-Solis:2019iod} for more details).
For our analysis, we employ the numerical values given under the label of Fit $i)$ of Table 7 of Ref.\,\cite{Gonzalez-Solis:2019iod} for the corresponding parameters.

For the $K^{-}K^{0}$ scalar form factor, we use the results of Ref.\,\cite{Guo:2011pa, Guo:2012yt, Guo:2016zep}\footnote{We thank very much Zhi-Hui Guo for providing us tables with the unitarized $\pi\eta$, $\pi\eta^{\prime}$ and $K^{0}\bar{K}^{0}$ scalar form factors. 
We translate the result of $K^{0}\bar{K}^{0}$ to the $K^{-}K^{0}$ concerning us through the relation $F^{K^{-}K^{0}}_{0}(s)=-F^{K^{0}\bar{K}^{0}}_{0}(s)/\sqrt{2}$.}.
These were obtained after the unitarization, based on the method of $N/D$, of the complete one-loop calculation of the strangeness conserving scalar form factors within $U(3)$ ChPT.

\begin{figure*}
\includegraphics[scale=0.35]{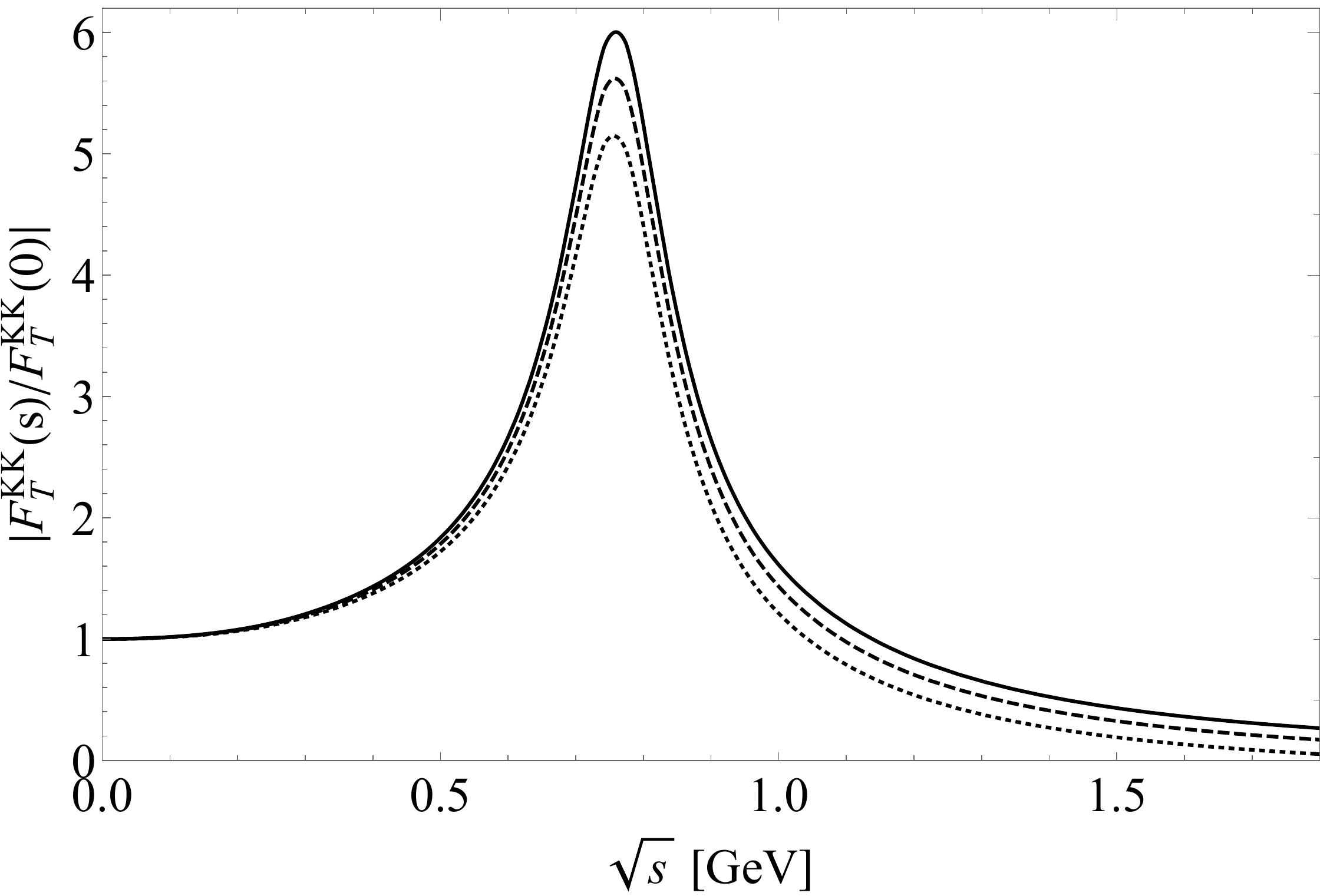}
\includegraphics[scale=0.365]{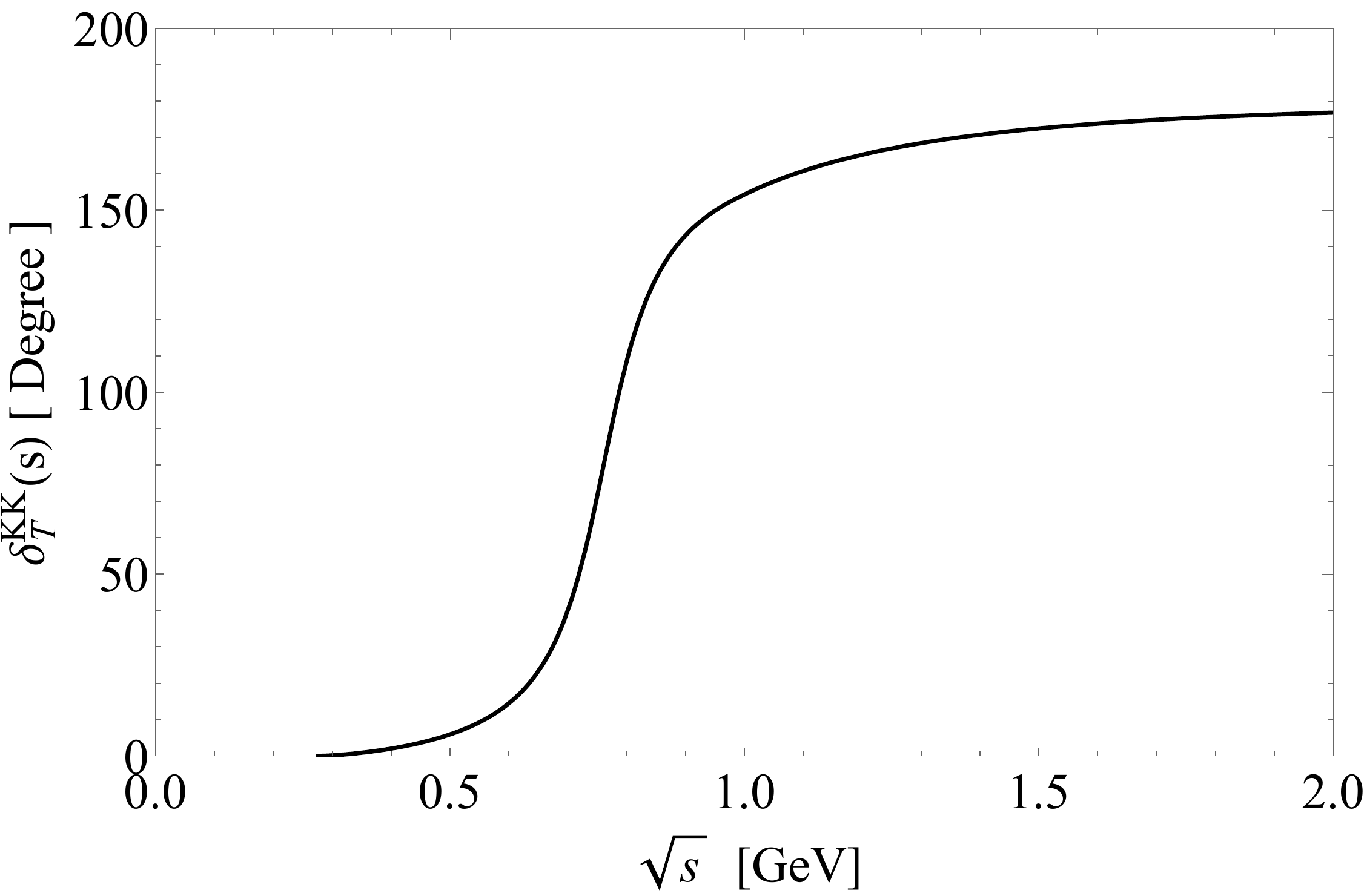}
\caption{Normalized absolute value of the tensor form factor $F_{T}^{KK}(s)$ given in Eq.\,(\ref{KKTensorFF}) (left), for $s_{\rm{cut}}=4$ GeV$^{2}$ (dotted line), $9$ GeV$^{2}$ (dashed line) and $s_{\rm{cut}}\to\infty$ GeV$^{2}$ (solid line), and tensor form factor phase $\delta_{T}^{KK}(s)$ (right).}
\label{PlotsFTs2}
\end{figure*}

Finally, for the tensor form factor $F_{T}^{K^{-}K^{0}}(s)$ we proceed in a similar fashion as for the $\tau^{-}\rightarrow K^{-}\eta^{(')}\nu_{\tau}$ and write
\begin{equation}
F^{K^{-}K^{0}}_{T}(s)=F^{K^{-}K^{0}}_{T}(0)\exp\left[\frac{s}{\pi}\int_{4m_{\pi}^{2}}^{s_{\rm{cut}}}ds^{'}\frac{\delta_{T}^{KK}(s^{'})}{s^{'}(s^{'}-s-i\epsilon)}\right]\,,
\label{KKTensorFF}    
\end{equation}
where we take $\delta_{T}^{KK}(s)=\delta_{+}^{\pi\pi}(s)$ in the elastic region i.e. until 1 GeV$^{2}$, with $\delta_{+}^{\pi\pi}(s)$ being the $P$-wave $\pi\pi$ scattering phase (see text below Eq.\,(\ref{KKphase})). 
As for the $K\eta^{(\prime)}$ case, we will assume this relation also holds above the onset of inelasticities and guide smoothly the tensor phase to $\pi$ at 1 GeV$^{2}$ to fulfill the expected $1/s$ asymptotic behavior.
Similar to Eq.\,(\ref{FunctDer}), the functional derivate 
\begin{equation}\label{tensor_form_at_zero}
     i\Big\langle K^{-}K^{0}\Big|\frac{\delta\mathcal{L}^{4}_{\chi PT}}{\delta\bar{t}_{\alpha\beta}}\Big|0\Big\rangle=\frac{\Lambda_{2}}{F_{\pi}^2}\left(p^{\alpha}_{K^{0}}p^{\beta}_{K^{-}}-p^{\alpha}_{K^{-}}p^{\beta}_{K^{0}}\right)\,,
\end{equation}
yields $F^{K^{-}K^{0}}_{T}(0)=\frac{\Lambda_{2}}{F_{\pi}^{2}}$, with $\Lambda_{2}$ given under Eq.\,(\ref{FT0}).
In Fig.\,\ref{PlotsFTs2}, we show the tensor phase $\delta_{T}^{KK}(s)$ (right panel) and the (normalized) absolute value $|F_{T}^{KK}(s)|$ for the cases $s_{\rm{cut}}=4,9$ GeV$^2$ and $s_{\rm{cut}}\to\infty$, which is taken as the baseline hypothesis.
As before, the variations due to $s_{\rm{cut}}$ will be taken into account as a source of systematic uncertainty in section \ref{Limits}.

\section{Decay observables}\label{section4}

In this section, we focus in the possible NP effects, characterized by the effective weak couplings described in section \ref{section2}, in the following $\tau^{-}\to K^{-}(\eta^{(\prime)},K^{0})\nu_{\tau}$ decay observables: Dalitz plots, angular and decay distributions, and the forward-backward asymmetry. 
The doubly differential decay width for $\tau^{-}\to K^{-}\eta^{(\prime)}\nu_{\tau}$, in the rest frame of the tau lepton, reads
\begin{equation}\label{DoublyDifferentialDecayWidth}
\frac{d^2\Gamma}{ds\,dt}=\frac{1}{32\left(2\pi\right)^3m_\tau^3}\overline{\left\vert \mathcal{M}\right\vert^2}\,,
\end{equation}
where $\overline{\left\vert \mathcal{M}\right\vert^2}$ is the unpolarized spin-averaged squared matrix element, $s$ is the invariant mass of the $K^- \eta^{(\prime)}$ system, limited in the interval $(m_{\eta^{(\prime)}}+m_K)^2 \leq s \leq m_\tau^2$, and $t=(P'+p_{\eta^{(\prime)}})^2=(P-p_{K})^2$ with kinematic boundaries given by $t^-(s)\leq t \leq t^+(s)$, with
\begin{eqnarray}
t^\pm (s)&=&\frac{1}{2s}\Big[2s\,m_{\eta^{(\prime)}}^2+(m_\tau^{2}-s)(s+m_{\eta^{(\prime)}}^2-m_K^{2})\nonumber\\[1ex]
&\pm&(m_\tau^2-s)\sqrt{\lambda(s,m_{\eta^{(\prime)}}^2,m_{K^{2}})}\Big]\,,
\end{eqnarray}
and where $\lambda(x,y,z)=x^2+y^2+z^2-2xy-2xz-2yz$ is the usual Kallen function.
The kinematic limits in $s$ and $t$ for the decay $\tau^{-}\to K^{-}K^{0}\nu_{\tau}$ are obtained by replacing $m_{\eta^{(')}}\rightarrow m_{K^{0}}$ above.

\subsection{Dalitz plot}

The unpolarized spin-averaged squared amplitude yields
\begin{eqnarray}\label{SquaredAmplitude}
\overline{\left\vert \mathcal{M}\right\vert^2}&=&\frac{G_F^2 \vert V_{us}\vert^2 S_{EW}}{s^2}\left(1+\epsilon_L+\epsilon_R\right)^2\nonumber\\[1ex]
&\times&\left[ M_{00}+M_{++}+M_{0+}+M_{T+}+M_{T0}+M_{TT}\right],
\end{eqnarray}
where $M_{00}$, $M_{++}$ and $M_{TT}$ are, respectively, the scalar, vector and tensor amplitudes, whereas $M_{0+}$, $M_{T+}$ and $M_{T0}$ are their corresponding interferences.
Their expressions are given by:
\begin{widetext}
\begin{equation}
\begin{array}{ccc}
\begin{array}{lccl}
M_{0+}=-2C^S_{K\eta^{(\prime)}}C^V_{K\eta^{(\prime)}}m_\tau^2 \mathrm{Re}[F_+^{K\eta^{(\prime)}}(s)(F_0^{K\eta^{(\prime)}}(s))^{*}] \Delta_{K\pi}\left(1+\frac{s\,\hat{\epsilon}_S}{m_\tau\left(m_s-m_u\right)}\right)\left(s(m_\tau^2-s-2t+\Sigma_{K\eta^{(\prime)}})-m_\tau^2\Delta_{K\eta^{(\prime)}}\right)\,,\\[3ex]
M_{T+}=-4C^V_{K\eta^{(\prime)}}\hat{\epsilon}_T m_\tau^3 s \mathrm{Re}[F_T^{K\eta^{(\prime)}}(s)(F_+^{K\eta^{(\prime)}}(s))^{*}] \left(1-\frac{s}{m_\tau^2}\right)\lambda(s,m_{\eta^{(\prime)}}^2,m_K^2)\,,\\[3ex]
M_{T0}=4C^S_{K\eta^{(\prime)}}\hat{\epsilon}_T\Delta_{K\pi}m_\tau s  \mathrm{Re}[F_T^{K\eta^{(\prime)}}(s)(F_0^{K\eta^{(\prime)}}(s))^{*}] \left(1+\frac{s\,\hat{\epsilon}_S}{m_\tau\left(m_s-m_u\right)}\right)\left(s(m_\tau^2-s-2t+\Sigma_{K\eta^{(\prime)}})-m_\tau^2\Delta_{K\eta^{(\prime)}}\right)\,,\\[3ex]
M_{00}=(C^{S}_{K\eta^{(\prime)}})^2 \Delta_{K\pi}^2 m_\tau^4 \left(1-\frac{s}{m_\tau^2}\right)|F_0^{K\eta^{(\prime)}}(s)|^2  \left(1+\frac{s\,\hat{\epsilon}_S}{m_\tau\left(m_s-m_u\right)}\right)^2\,,\\[3ex]
M_{++}=(C^{V}_{K\eta^{(\prime)}})^2|F_+^{K\eta^{(\prime)}}(s)|^2\left\lbrace m_\tau^4(s-\Delta_{K\eta^{(\prime)}})^2+4m_k^2 s^2(m_{\eta^{(\prime)}}^2-t)+4s^2 t(s+t-m_{\eta^{(\prime)}}^2)\right.\\[3ex]
\left.\qquad-m_\tau^2 s\left(s\left(s+4t\right)-2\Delta_{K\eta^{(\prime)}}(s+2t-2m_{\eta^{(\prime)}}^2) +\Delta_{K\eta^{(\prime)}}^2\right)\right\rbrace\,,\\[3ex]
M_{TT}=4\hat{\epsilon}_T^2|F_T^{K\eta^{(\prime)}}(s)|^2 s^2 \left\lbrace m_K^4(m_\tau^2-s)-m_{\eta^{(\prime)}}^4(3m_\tau^2+s)-s\left((s+2t)^2-m_\tau^2(s+4t)\right)\right.\\[3ex]
\left.\qquad +2m_{\eta^{(\prime)}}^2\left((s+2t)(s+m_\tau^2)-2m_\tau^4\right)-2m_K^2(m_\tau^2-s)(s+2t-m_{\eta^{(\prime)}}^2)\right\rbrace\,,
\label{Amplitudes}
\end{array}
\end{array}
\end{equation}
\end{widetext}
where we have defined $\Delta_{PQ}=m_P^2-m_Q^2$ and $\Sigma_{PQ}=m_P^2+m_Q^2$. 
The corresponding expressions for $\tau^{-}\to K^{-}K^{0}\nu_{\tau}$ are obtained by replacing $V_{us}\to V_{ud}$ in Eq.\,(\ref{SquaredAmplitude}), and $m_{\eta^{(')}}\rightarrow m_{K^{0}}$, $m_{s}\rightarrow m_{d}$ and $C_{K\eta^{(\prime)}}^{S,V}\to C_{KK}^{S,V}$ in Eq.\,(\ref{Amplitudes}).
For this latter case, we would like to notice that those contributions involving the scalar form factor i.e. $M_{0+},M_{T0}$ and $M_{00}$, are always suppressed since they are proportional to $\Delta_{K^{-}K^{0}}$, an isospin-violating factor which is tiny.
This makes its effect negligible even for $|\hat{\epsilon}_{S}|\sim1$ (low-energy processes limit $|\hat{\epsilon}_{S}|\leq3.4\cdot10^{-3}$ \cite{Gonzalez-Alonso:2018omy} under the reasonable assumption of lepton flavor universality).

In order to study possible NP signatures in Dalitz plots distributions, we introduce the following observable \cite{Miranda:2018cpf}
\begin{equation}
\widetilde\Delta\left(\hat{\epsilon}_S,\hat{\epsilon}_T\right)=\frac{\left\vert \overline{\left\vert \mathcal{M}\left(\hat{\epsilon}_S,\hat{\epsilon}_T\right) \right\vert^2}-\overline{\left\vert \mathcal{M}\left(0,0\right)\right\vert^2}\right\vert}{\overline{\left\vert \mathcal{M}\left(0,0\right)\right\vert^2}}\,,
\label{DeltaObservable}
\end{equation}
which measures deviations between non-SM (either $\hat{\epsilon}_{S}\neq0$ or $\hat{\epsilon}_{T}\neq0$, or both $\hat{\epsilon}_{S,T}\neq0$) and SM ($\hat{\epsilon}_{S,T}=0$) interactions.

\begin{figure*}
\includegraphics[width=7cm]{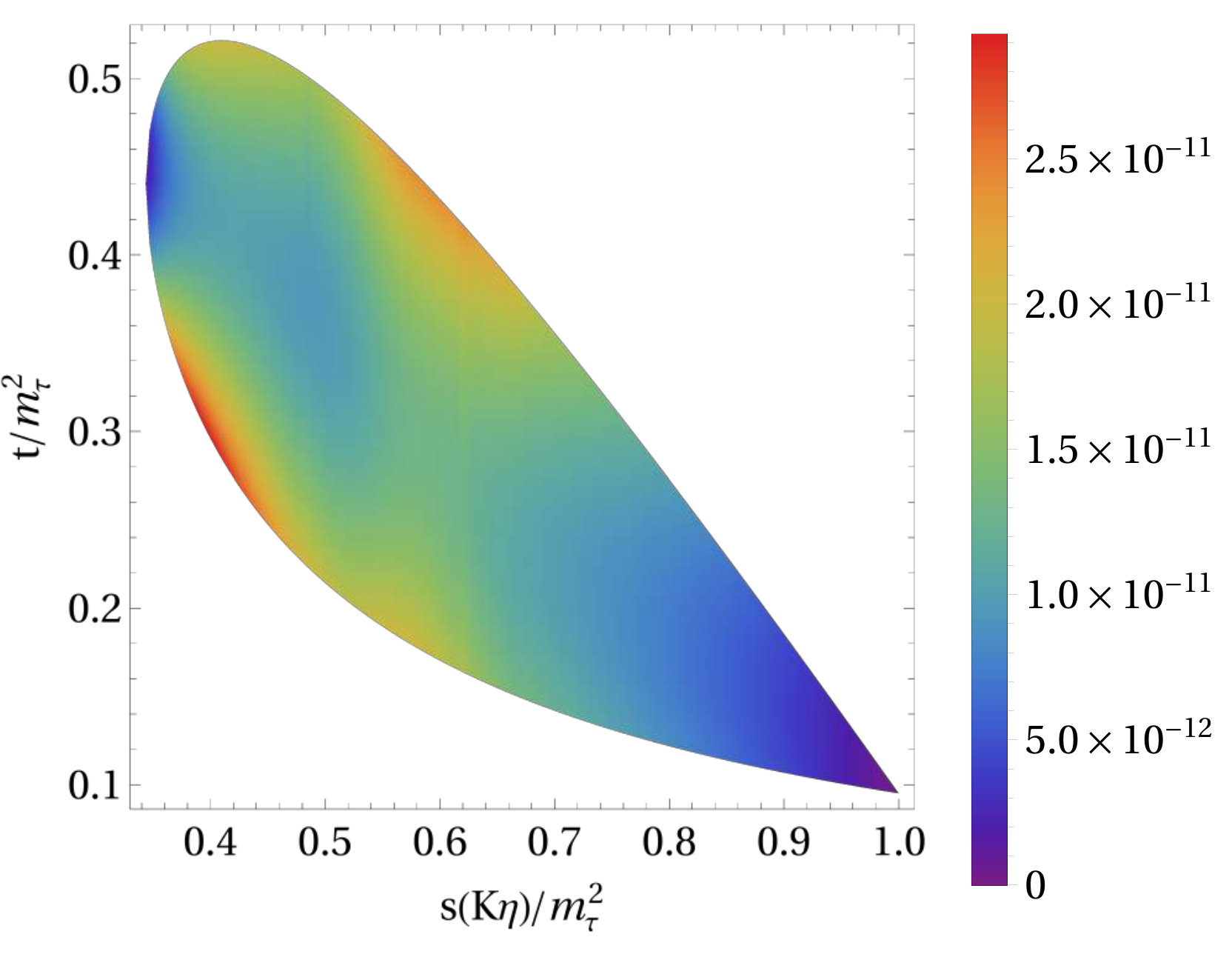}
\includegraphics[width=7cm]{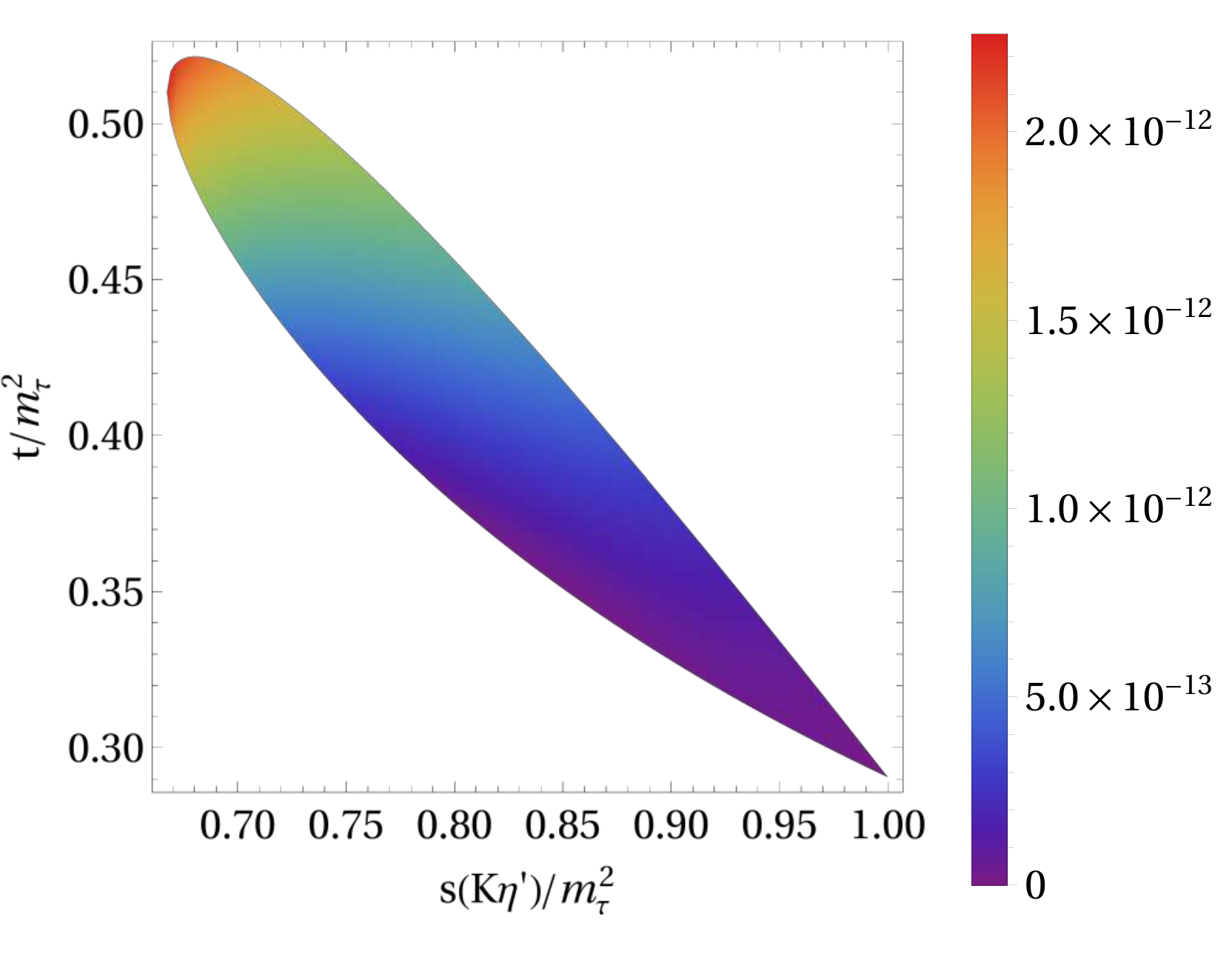}	
\includegraphics[width=7cm]{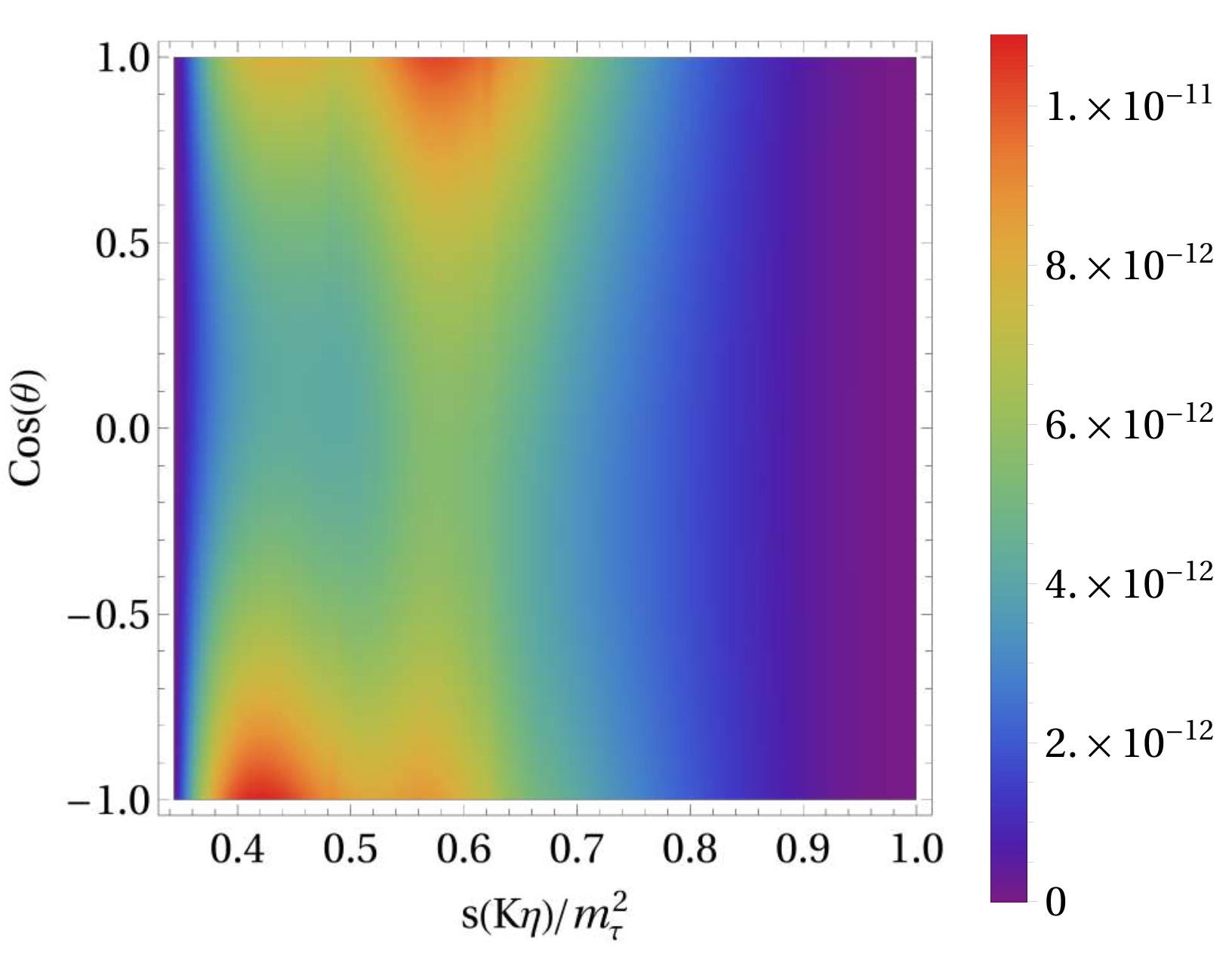}	
\includegraphics[width=7cm]{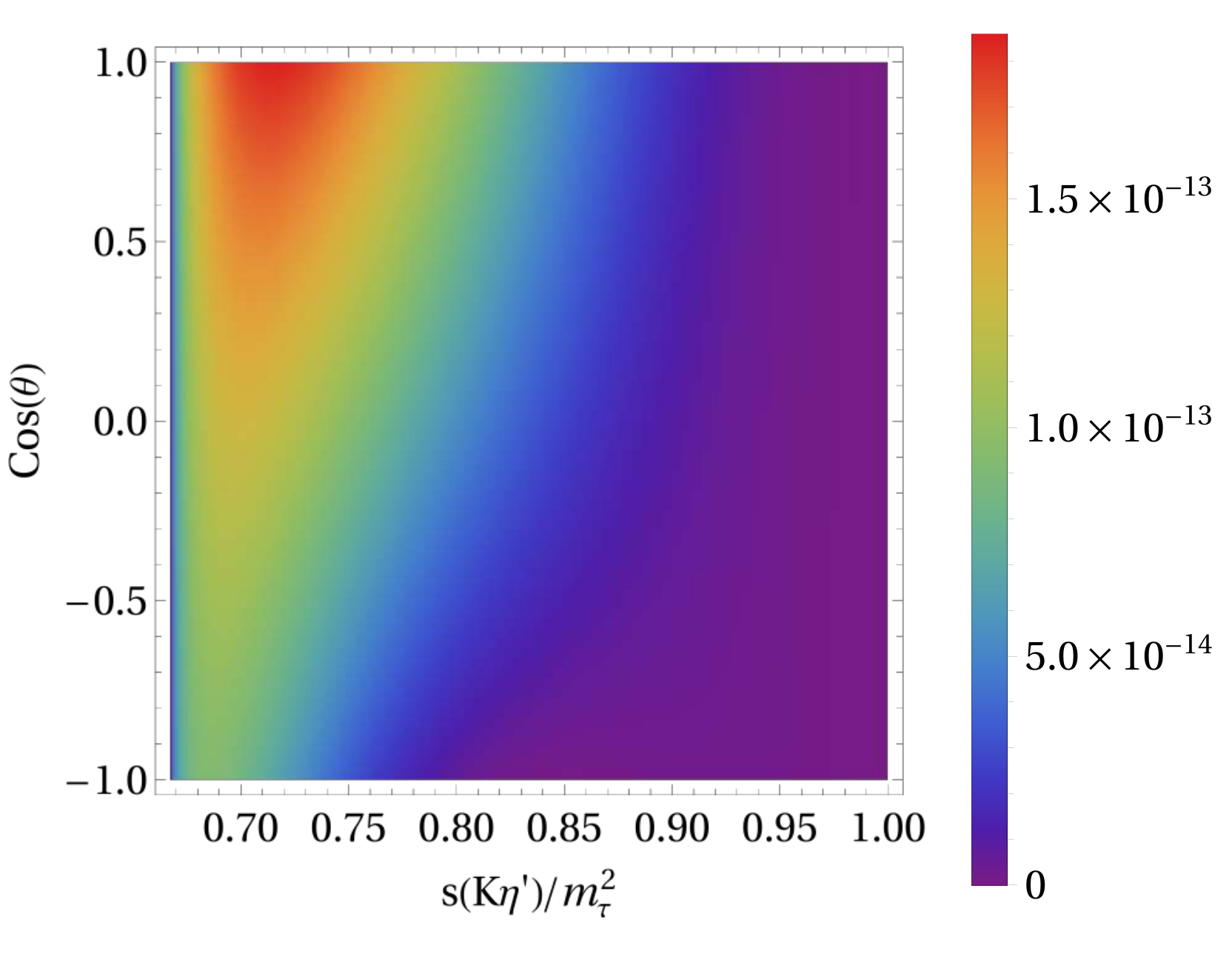}	
\centering			
\caption{Dalitz plot distribution in the SM, $\overline{\left\vert \mathcal{M}\left(0,0\right)\right\vert^2}$ in Eq.\,(\ref{SquaredAmplitude}), for $\tau^-\to K^-\eta\nu_\tau$ (left) and $\tau^-\to K^-\eta^\prime\nu_\tau$ (right) in the $(s,t)$ variables. 
The figures of the lower row show the differential decay distribution in the $(s,\cos \theta)$ variables, Eq.\,(\ref{AngularDistribution}).
The $s$ and $t$ variables are normalized to $m_\tau^2$.}
\label{SMKeta}
\end{figure*}

\begin{figure*}
\includegraphics[width=7cm]{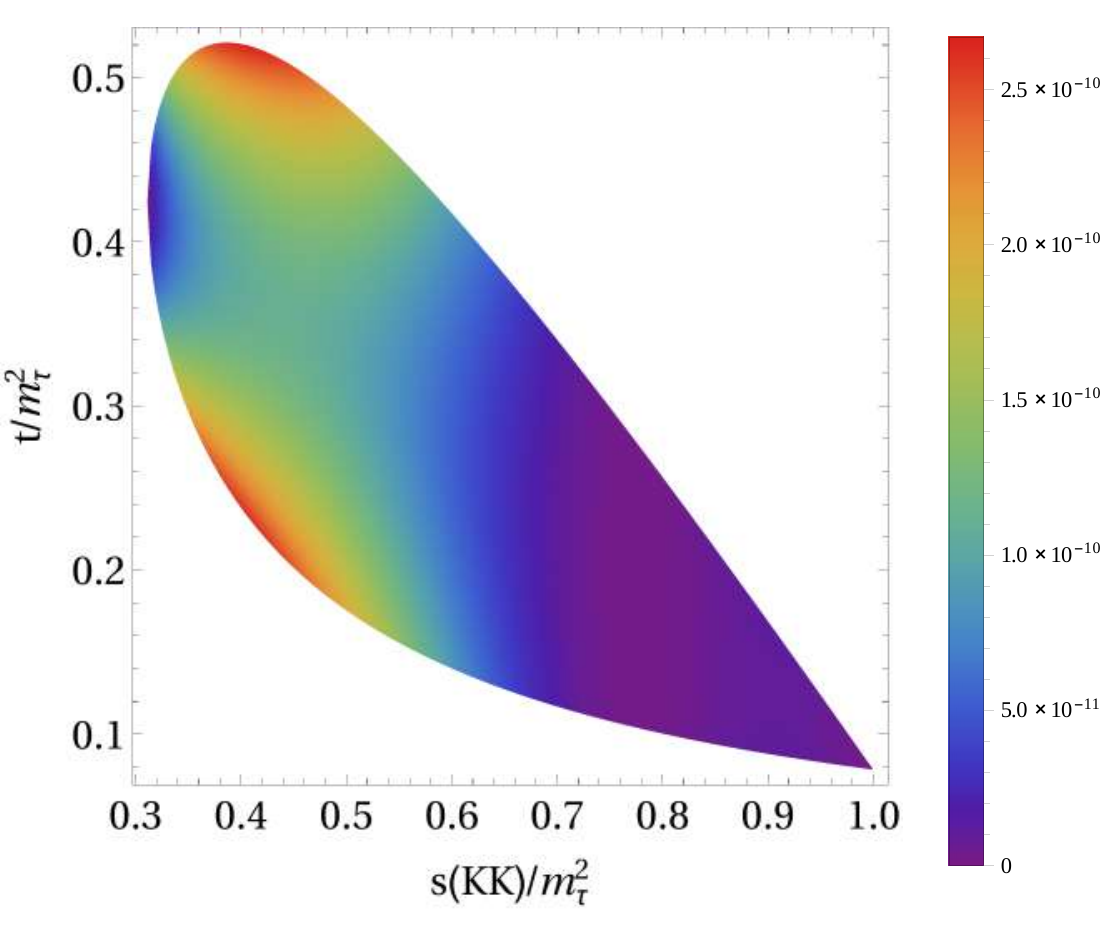}	
\includegraphics[width=7cm]{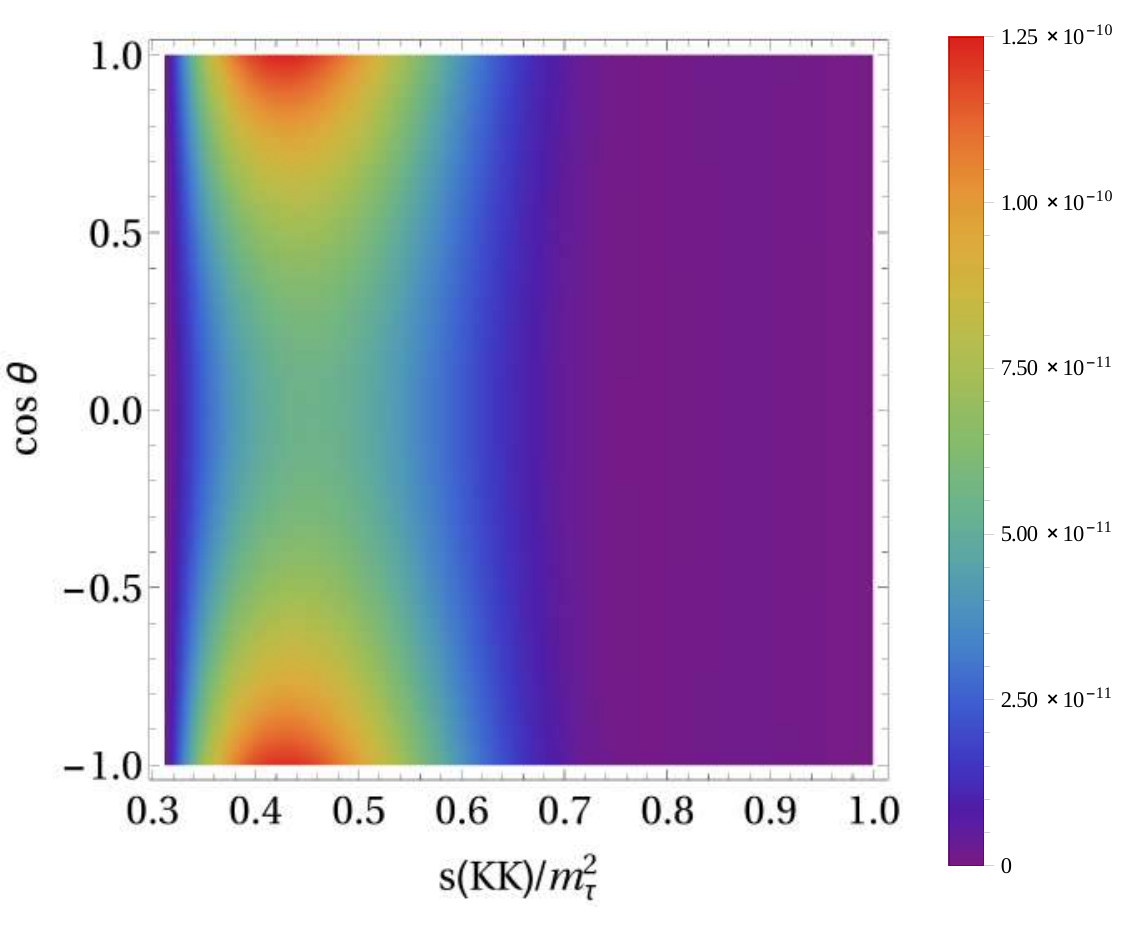}
\centering			
\caption{Dalitz plot distribution in the SM, $\overline{\left\vert \mathcal{M}\left(0,0\right)\right\vert^2}$ in Eq.\,(\ref{SquaredAmplitude}), for $\tau^-\to K^{-}K^{0}\nu_\tau$ in the $(s,t)$ variables (left). 
The figure shown on the right corresponds to the differential decay distribution in the $(s,\cos \theta)$ variables, Eq.\,(\ref{AngularDistribution}).
The $s$ and $t$ variables are normalized to $m_\tau^2$.}
\label{SMKK}
\end{figure*}

Firstly, in Fig.\ref{SMKeta} we provide a graphical account of the Dalitz plot distributions in the SM in the $(s,t)$ variables for the decays $\tau^{-}\to K^{-}\eta\nu_{\tau}$ (upper-left plot) and $\tau^{-}\to K^{-}\eta^{\prime}\nu_{\tau}$ (upper-right plot).
As it can be seen from these plots, there is no evidence for a meson resonance production and only the $K^{*}(892)$-and to lesser extent- the $K^{*}(1410)$, and $K^{0}(1430)$ tails can be appreciated for the $K\eta$ and $K\eta^{\prime}$ decay channels, respectively.
Similarly, the SM plot for the decay $\tau^{-}\to K^{-}K^{0}\nu_{\tau}$ is displayed in Fig.\,\ref{SMKK}, where the tail of the $\rho(1450)$ meson can be seen.   

\begin{figure*}
\includegraphics[width=7cm]{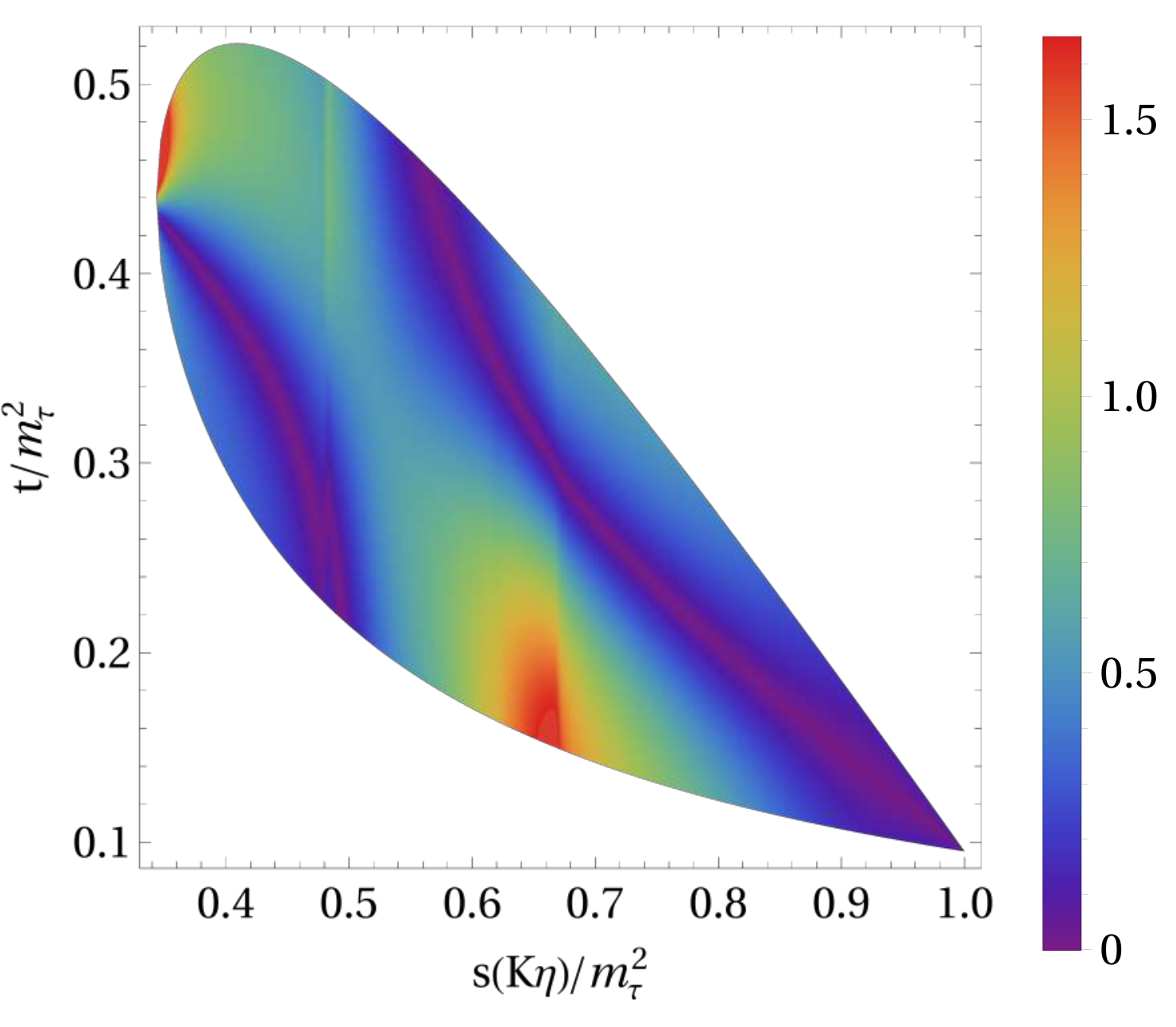}
\includegraphics[width=7cm]{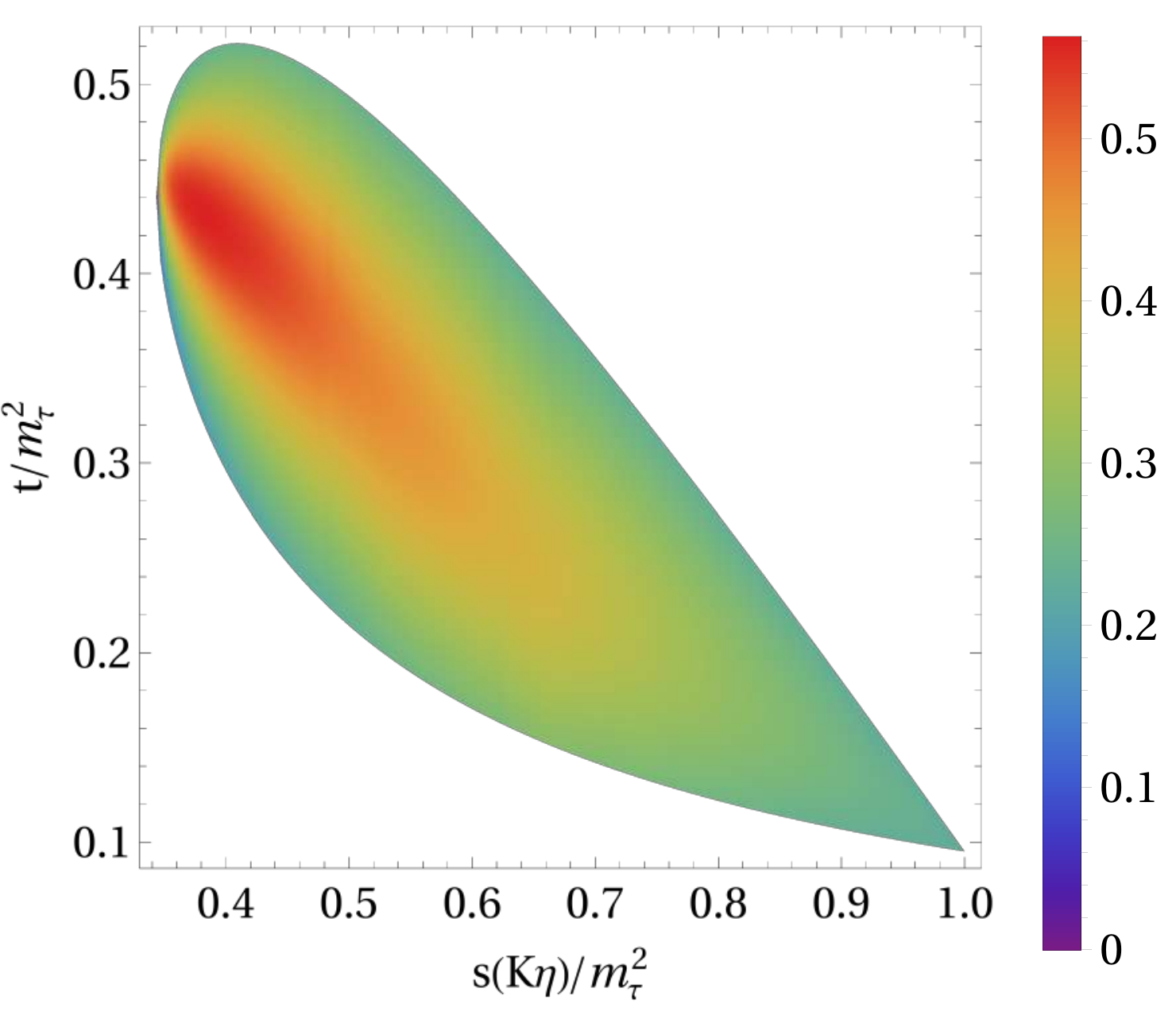}		
\includegraphics[width=7cm]{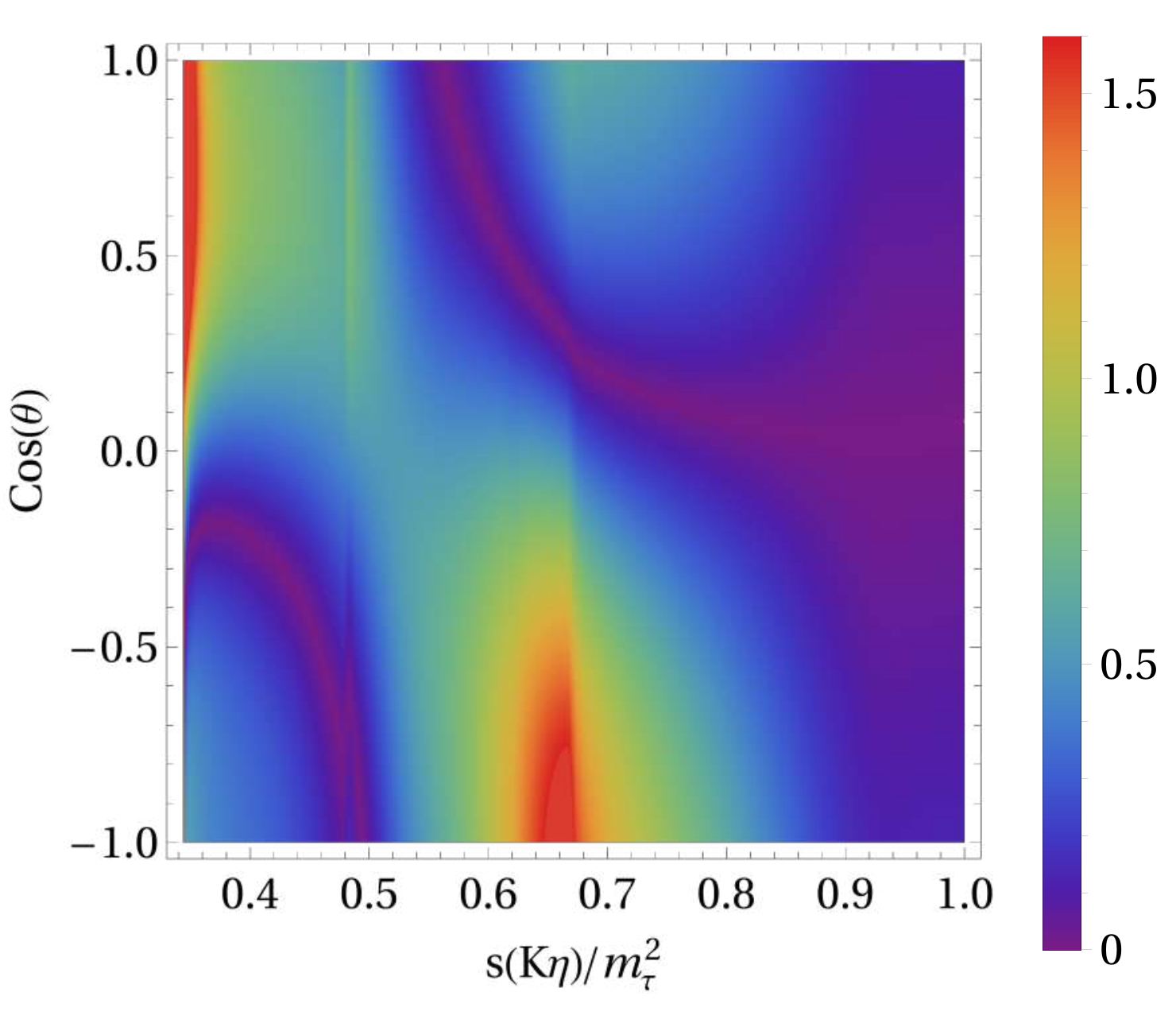}
\includegraphics[width=7cm]{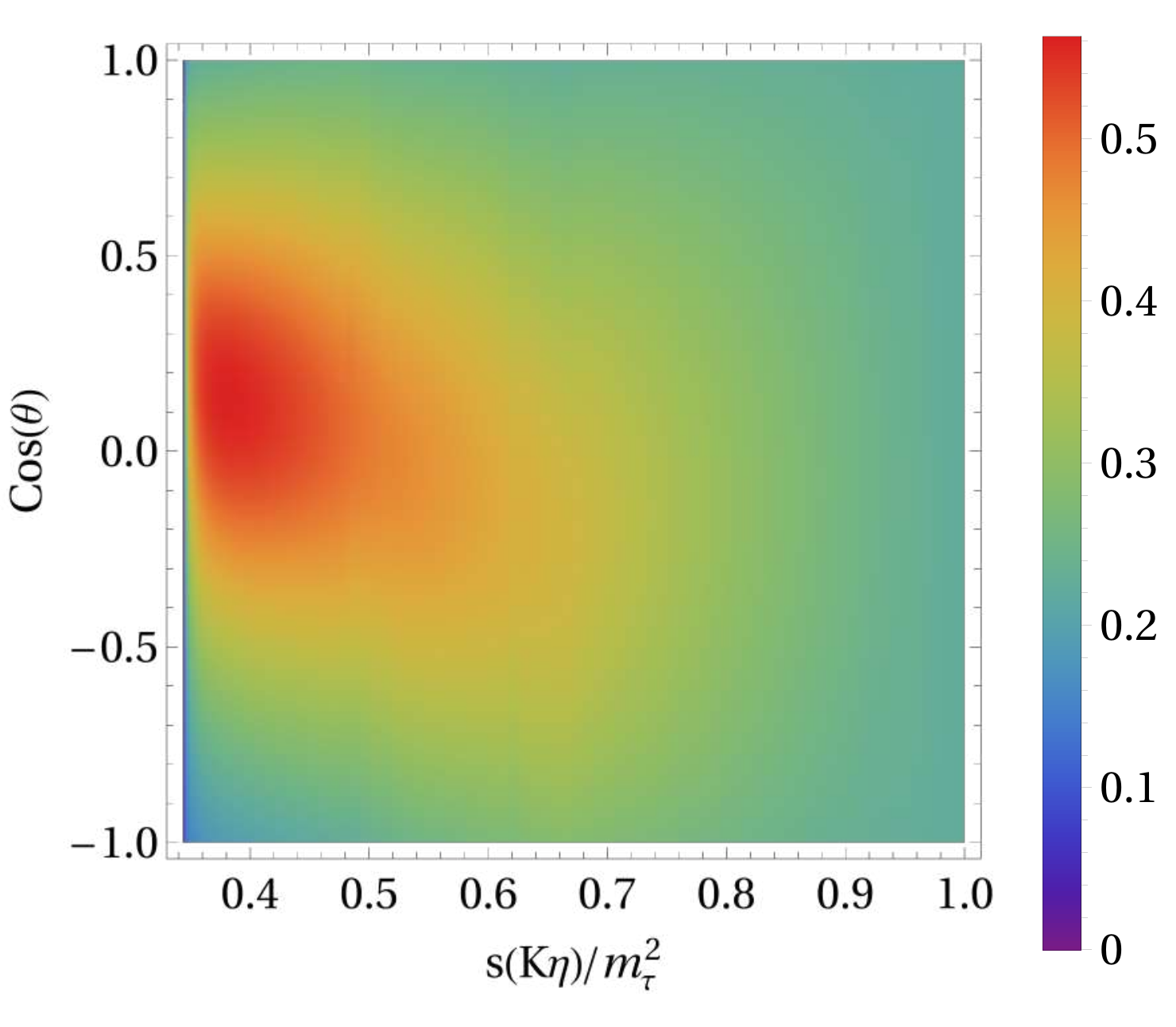}		
\centering			
\caption{Dalitz plot distribution of $\widetilde\Delta\left(\hat{\epsilon}_S,\hat{\epsilon}_T\right)$ in Eq.\,(\ref{DeltaObservable}) for $\tau^-\to K^-\eta\nu_\tau$ with $(\hat{\epsilon}_S=-0.38,\hat{\epsilon}_T=0)$ (left panels) and $(\hat{\epsilon}_S=0,\hat{\epsilon}_T=0.085)$ (right panels). 
The lower row shows the differential decay distribution in the $(s,\cos\theta)$ variables, Eq.\,(\ref{AngularDistribution}). 
The $s$ and $t$ variables are normalized to $m_\tau^2$.}
\label{NPKeta}
\end{figure*}

Secondly, we turn to analyze possible NP signatures by allowing non-zero values of either $\hat{\epsilon}_{S}$ or $\hat{\epsilon}_{T}$.
In Fig.\,\ref{NPKeta}, first row, we show the observable $\widetilde\Delta\left(\hat{\epsilon}_S,\hat{\epsilon}_T\right)$ in Eq.\,(\ref{DeltaObservable}) for the decay $\tau^-\to K^-\eta\nu_\tau$ for two representative values of the set of effective couplings $(\hat{\epsilon}_{s},\hat{\epsilon}_{T})$, that we anticipate from our results in section \ref{Limits}, that are consistent with the measured branching ratio.
For the left plots of the figure we use $(\hat{\epsilon}_S=-0.38,\hat{\epsilon}_T=0)$ and thus the variations with respect to the SM occur due to $M_{0+}$ and $M_{00}$ in Eq.\,(\ref{Amplitudes}), while for the right ones we employ $(\hat{\epsilon}_S=0,\hat{\epsilon}_T=0.085)$ with NP effects entering through $M_{T+},M_{T0}$ and $M_{TT}$ in Eq.\,(\ref{Amplitudes}).
As one can observe, the variations of scalar nature are in general small and occur close to the $s$ minimum i.e. near the $K\eta$ threshold and $t/m_{\tau}^{2}\sim0.47$, and for $s/m_{\tau}^{2}\sim0.66$, while the tensor contributions yield a sizable signal starting near the $K\eta$ threshold and populate the diagonal of the Dalitz plot decreasingly.
However, these contributions arise in zones with very suppressed probability in the SM (see upper-left plot in Fig.\,\ref{SMKeta}) and will thus be very challenging to identify.

\begin{figure*}
\includegraphics[width=7cm]{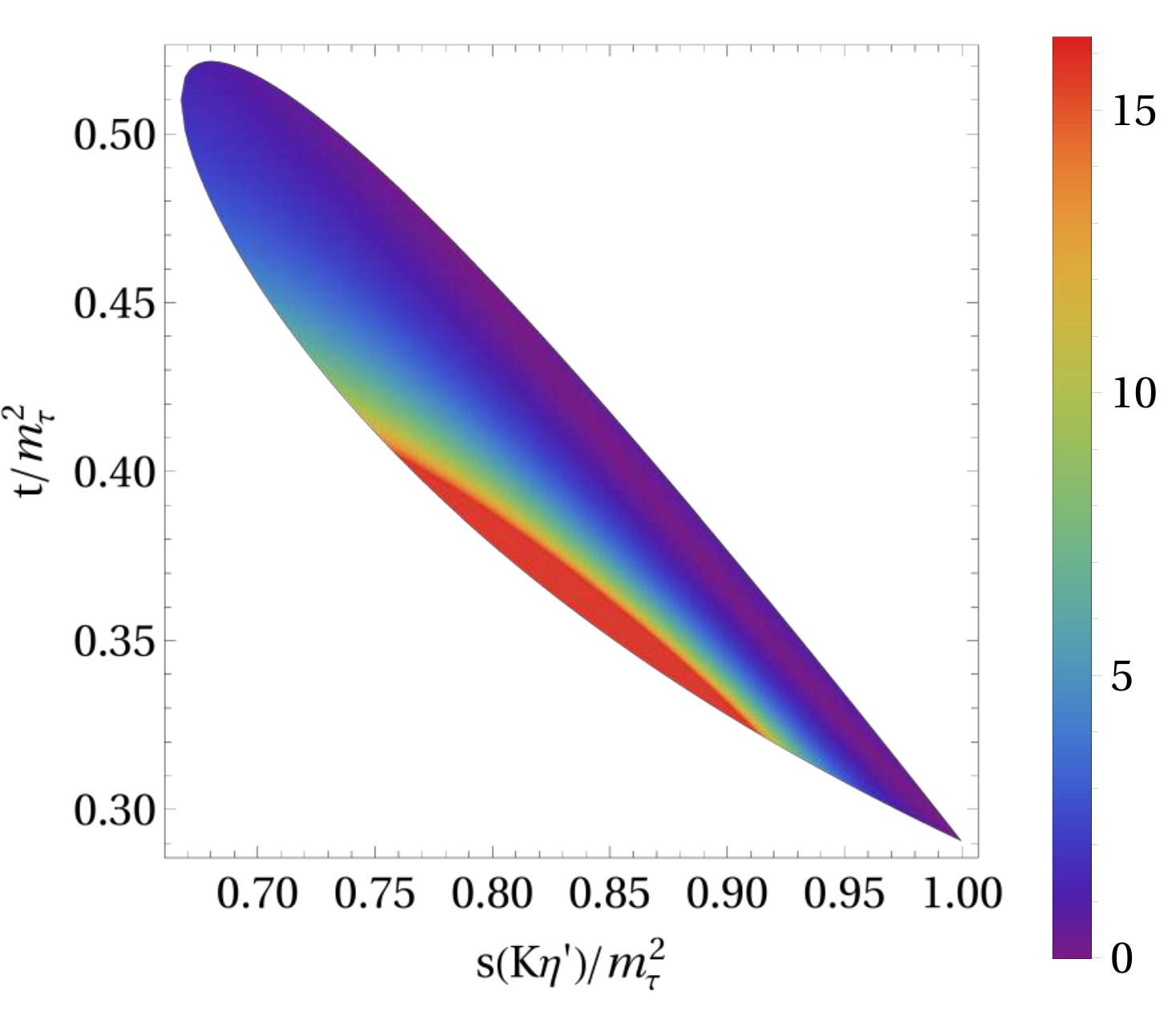}	
\includegraphics[width=7cm]{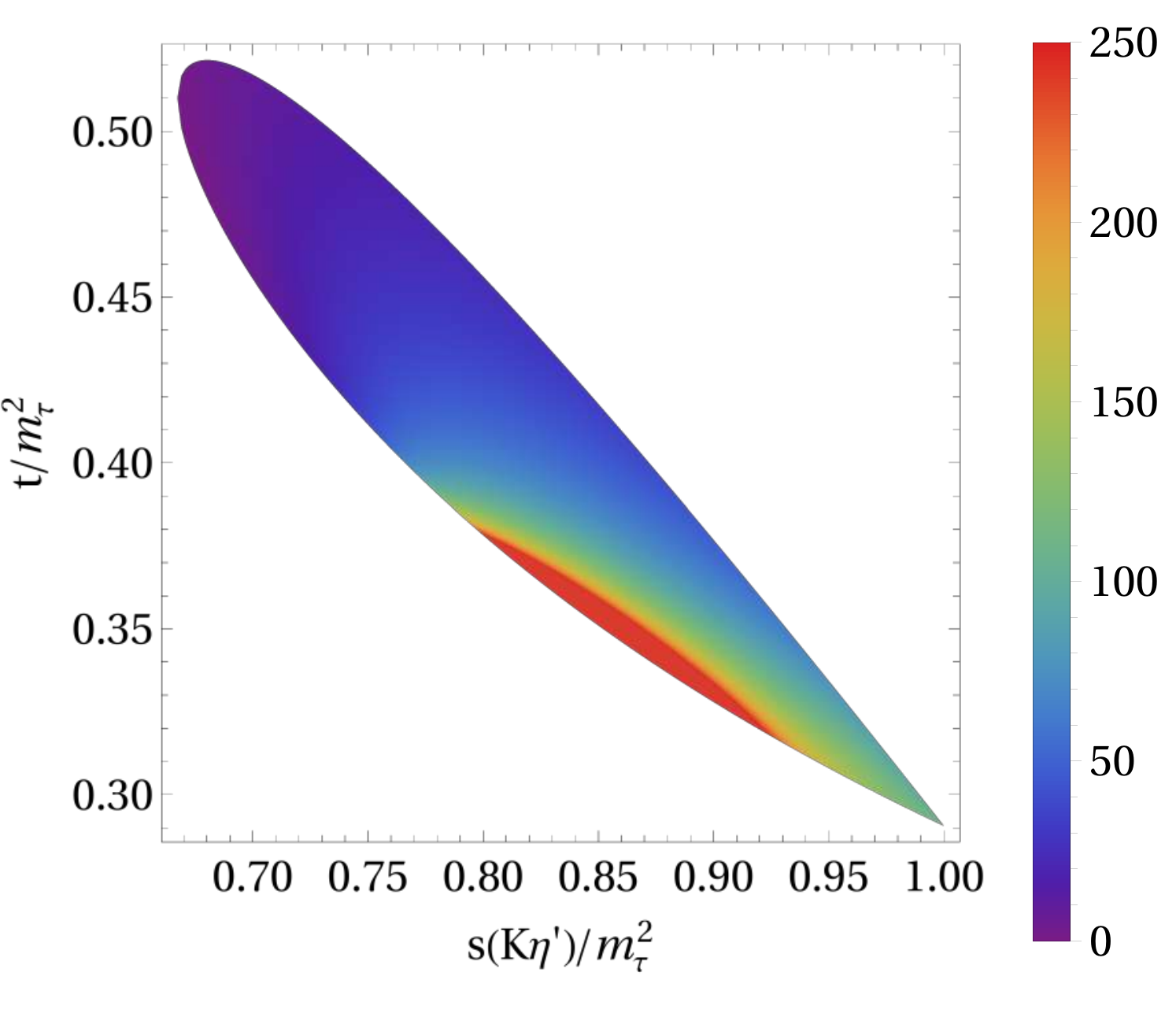}		
\includegraphics[width=7cm]{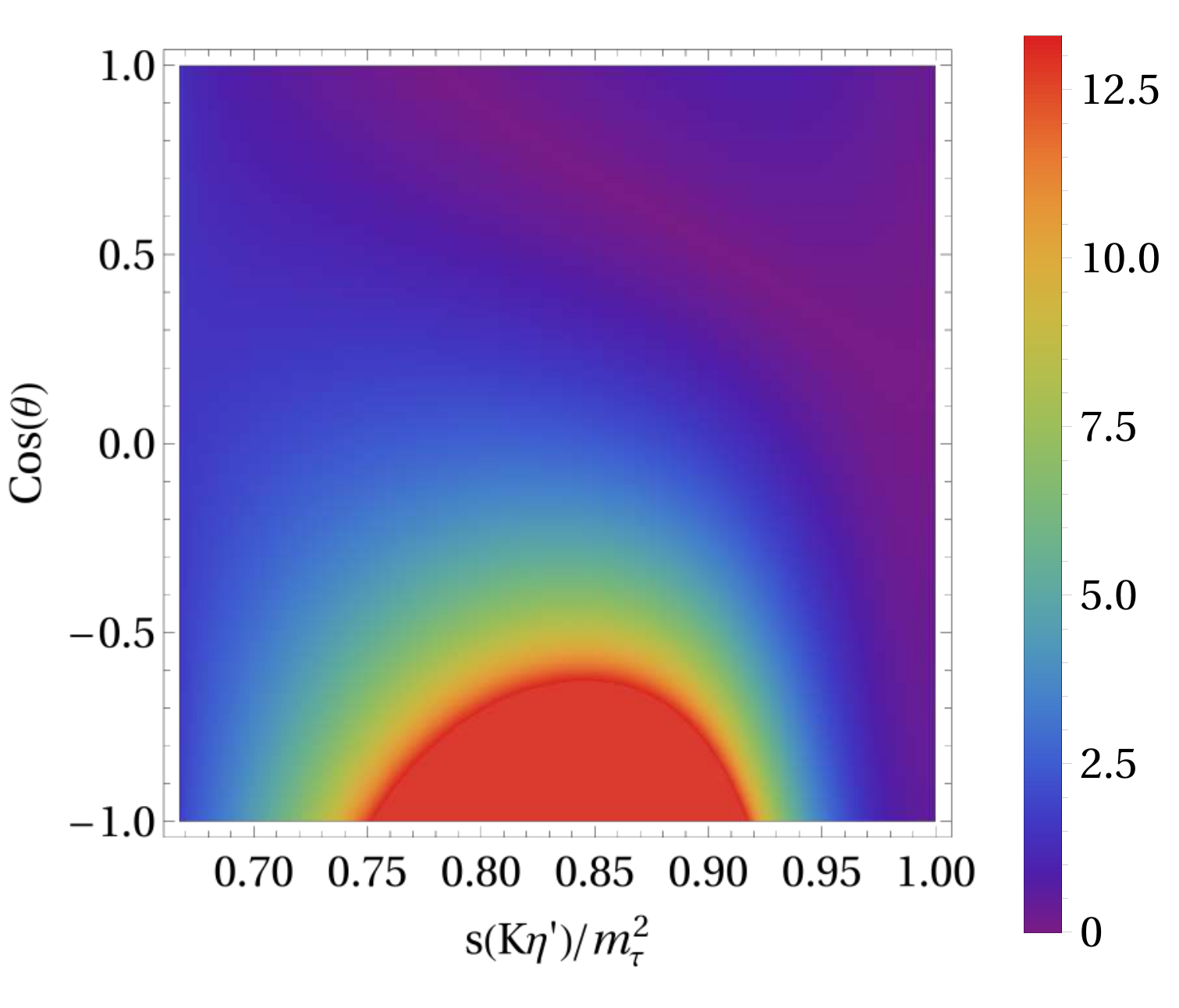}	
\includegraphics[width=7cm]{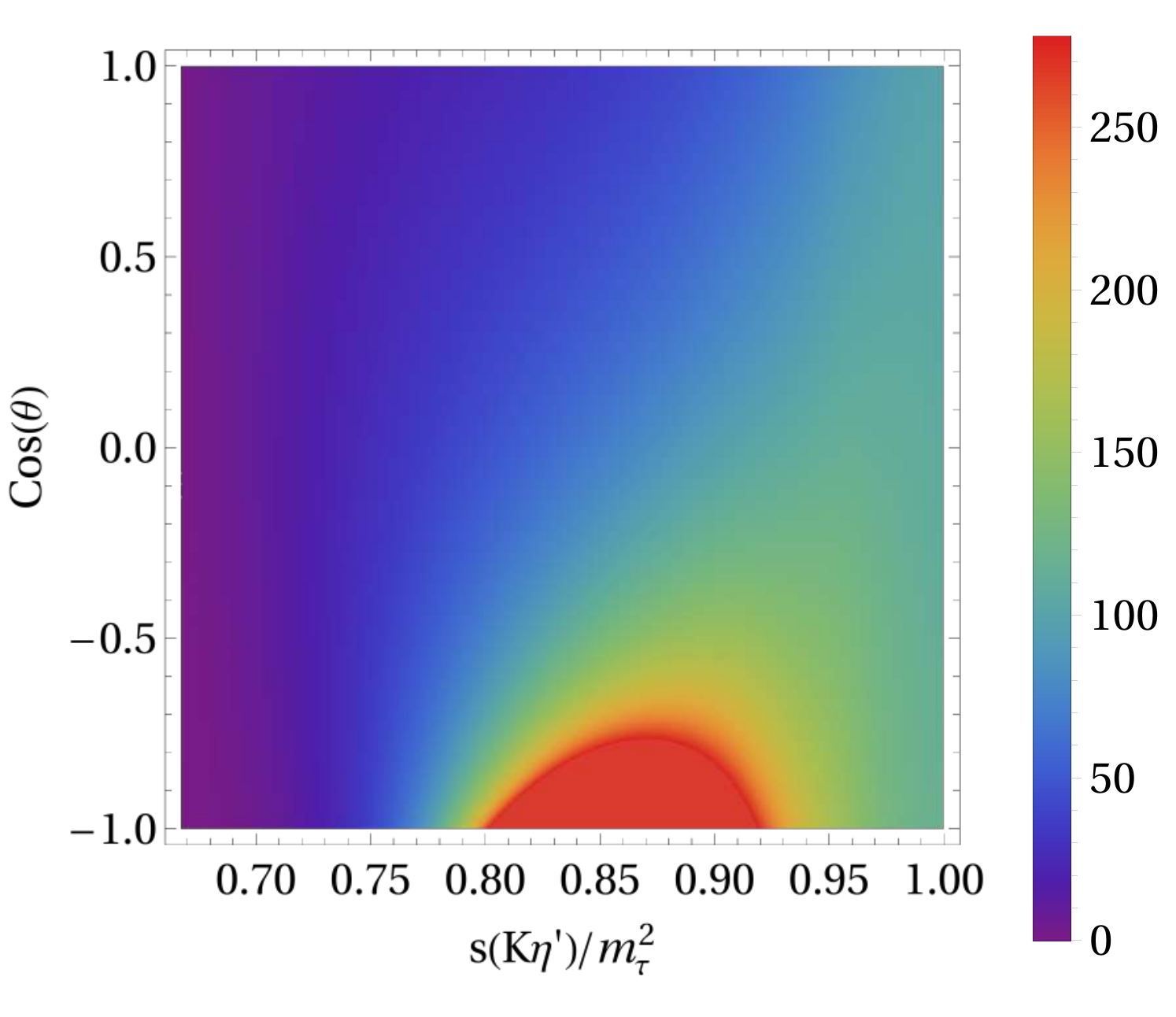}			
\centering			
\caption{Dalitz plot distribution of $\widetilde\Delta\left(\hat{\epsilon}_S,\hat{\epsilon}_T\right)$ in Eq.\,(\ref{DeltaObservable}) for $\tau^{-}\to K^{-}\eta^{\prime}\nu_\tau$ with $(\hat{\epsilon}_S=-0.20,\hat{\epsilon}_T=0)$ (left panels) and $(\hat{\epsilon}_S=0,\hat{\epsilon}_T=14.9)$ (right panels). 
The lower row show the differential decay distribution in the $(s,\cos\theta)$ variables, Eq.\,(\ref{AngularDistribution}). 
The $s$ and $t$ variables are normalized to $m_\tau^2$.}
\label{NPKetaprime}
\end{figure*}

In the case of $\tau^{-}\to K^{-}\eta^{\prime}\nu_{\tau}$, shown in Fig.\,\ref{NPKetaprime}, we use, respectively, $(\hat{\epsilon}_S=-0.20,\hat{\epsilon}_T=0)$ and $(\hat{\epsilon}_S=0,\hat{\epsilon}_T=14.9)$ for the left- and right-plots and the corresponding variations in the Dalitz plot distribution are seen in a reduced and similar region close to $s/m_{\tau}^{2}\sim0.85$ and $t/m_{\tau}^{2}\sim0.35$.
Again, compared to the SM (see upper-right plot in Fig.\,\ref{SMKeta}), these effects appear in a zone of small probability density and will be therefore difficult to be measured.

\begin{figure*}
\includegraphics[width=7cm]{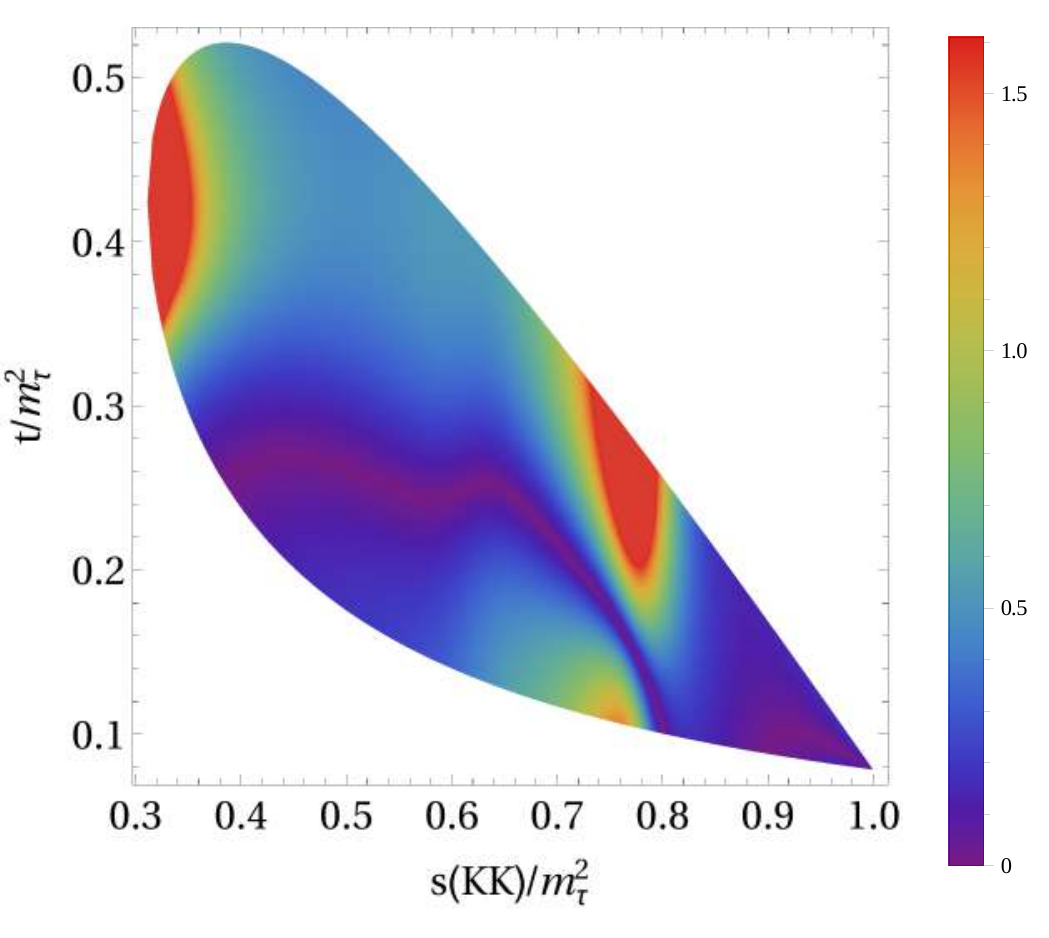}
\includegraphics[width=7cm]{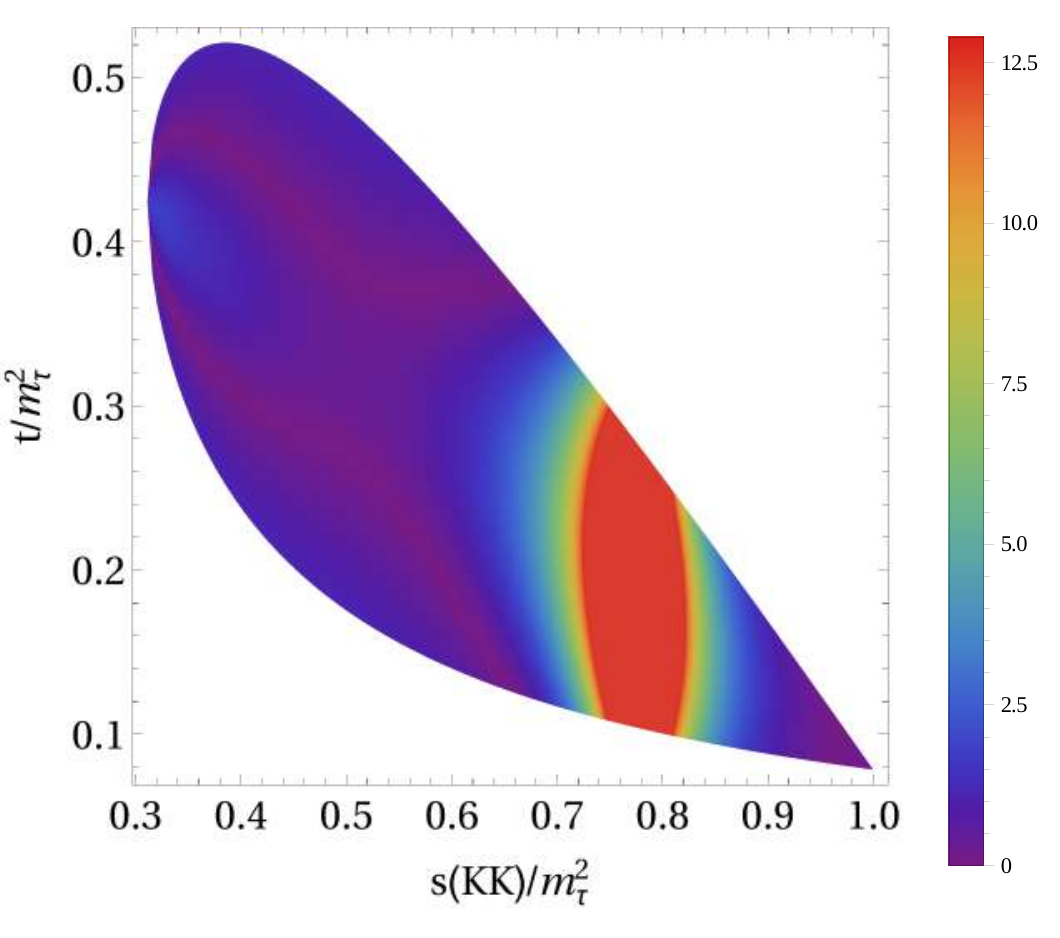}		
\includegraphics[width=7cm]{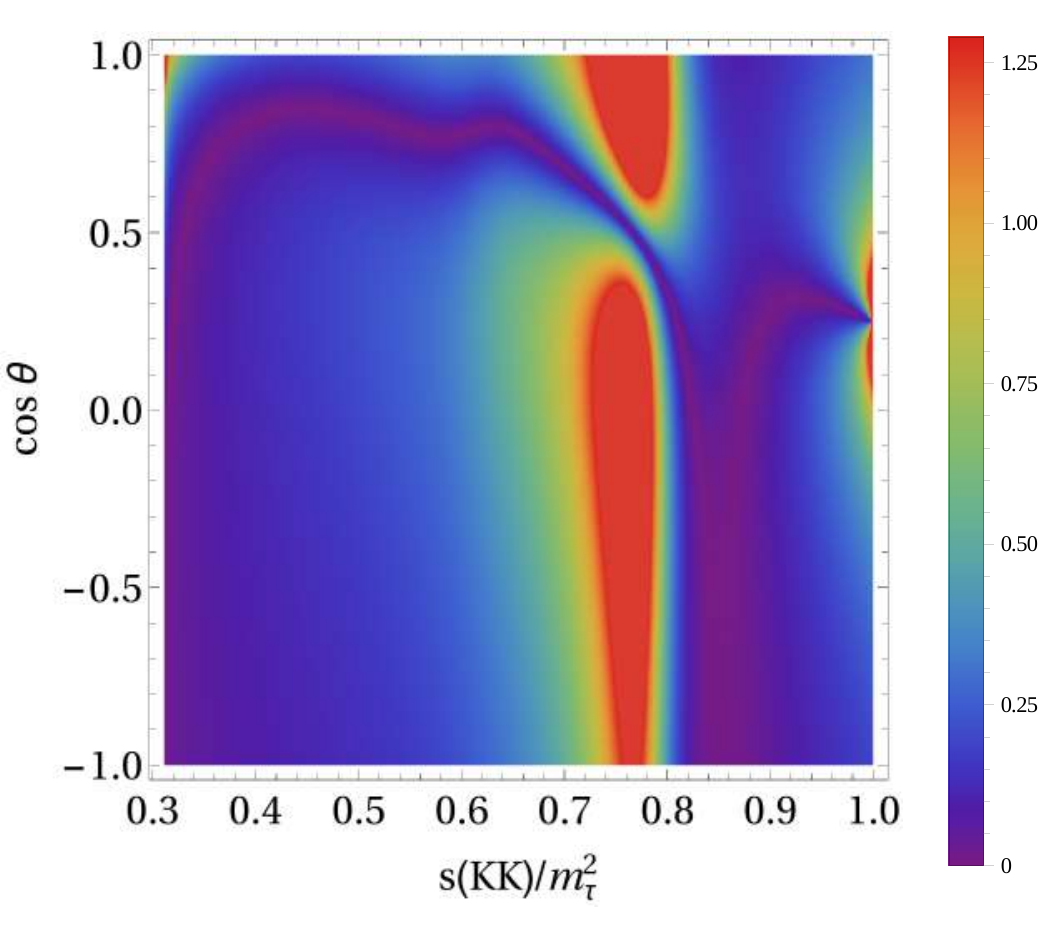}
\includegraphics[width=7cm]{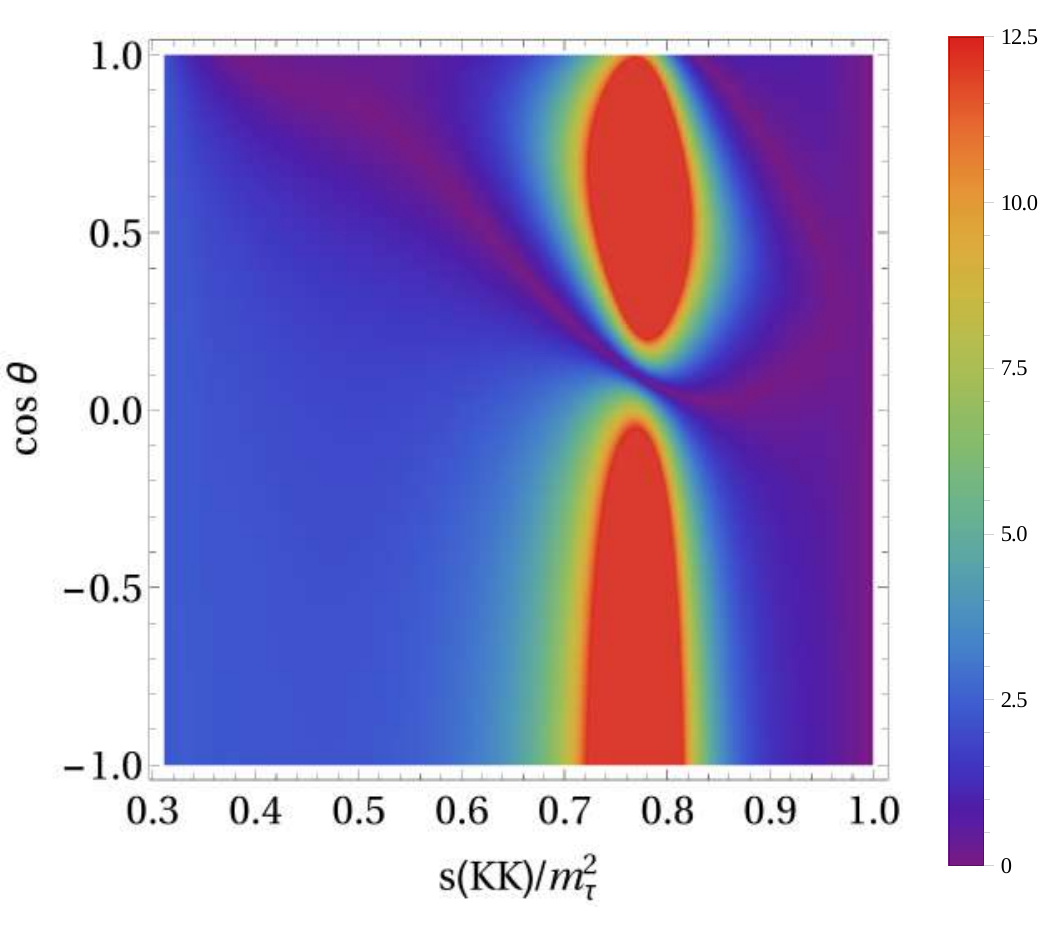}		
\centering			
\caption{Dalitz plot distribution of $\widetilde\Delta\left(\hat{\epsilon}_S,\hat{\epsilon}_T\right)$ in Eq.\,(\ref{DeltaObservable}) for $\tau^{-}\to K^{-}K^{0}\nu_{\tau}$ with $(\hat{\epsilon}_S=0.10,\hat{\epsilon}_T=0)$ (left panels) and $(\hat{\epsilon}_S=0,\hat{\epsilon}_T=0.9)$ (right panels). 
The lower row show the differential decay distribution in the $(s,\cos\theta)$ variables, Eq.\,(\ref{AngularDistribution}). 
The $s$ and $t$ variables are normalized to $m_\tau^2$.}
\label{NPKK}
\end{figure*}

Finally, for $\tau^{-}\to K^{-}K^{0}\nu_{\tau}$, we use $(\hat{\epsilon}_S=0.10,\hat{\epsilon}_T=0)$ and $(\hat{\epsilon}_S=0,\hat{\epsilon}_T=0.9)$ for the left-and right-plots of Fig.\,\ref{NPKK}, respectively.
The effects due to the inclusion of new contributions (see red shaded areas in Fig.\,\ref{NPKK}) appear in the region of the SM Dalitz plot less densely populated (see left plot in Fig.\,\ref{SMKK}) and will again thus be difficult to distinguish.

Had we used other set of values of effective couplings e.g.\,\cite{Gonzalez-Alonso:2016etj}, we would have obtained qualitatively similar results.

\subsection{Angular distribution}

The hadronic mass and angular distributions are also modified by the inclusion of the NP interactions that we are studying.
It is convenient to work in the rest frame of the hadronic $K\eta^{(\prime)}$ system defined by $\vec{p}_{K}+\vec{p}_{\eta^{(\prime)}}=\vec{p}_{\tau}-\vec{p}_{\nu_{\tau}}=0$.
In this frame, the tau lepton and kaon energies are given by $E_{\tau}=(s+m_{\tau}^{2})/2\sqrt{s}$ and $E_{K}=(s+m_{K}^{2}-m_{\eta^{(\prime)}}^{2})/2\sqrt{s}$, and the measurable angle $\theta$ between these two particles can be obtained from the invariant $t$ variable through $t=m_{\tau}^{2}+m_{K}^{2}-2E_{\tau}E_{K}+2|\vec{p}_{K}||\vec{p}_{\tau}|\cos\theta$, where $|\vec{p}_{K}|=\sqrt{E_{K}^{2}-m_{K}^{2}}$ and $|\vec{p}_{\tau}|=\sqrt{E_{\tau}^{2}-m_{\tau}^{2}}$.

The decay distribution in the $(s,\theta)$ variables in the framework of the most general effective interactions is given by:
\begin{widetext}
\begin{eqnarray}
\frac{d^2\Gamma}{d\sqrt{s}d\cos\theta}&=&\frac{G_F^2|V_{us}|^2S_{EW}}{128\pi^3 m_\tau}(1+\epsilon_L+\epsilon_R)^2\left(\frac{m_\tau^2}{s}-1\right)^2|\vec{p}_{K}|\biggl\lbrace(C^S_{K\eta^{(\prime)}})^2(\Delta_{K\pi})^2|F_0^{K\eta^{(\prime)}}(s)|^2\nonumber\\[1ex]
&\times&\left( 1+\frac{s\hat{\epsilon}_S}{m_\tau (m_s-m_u)}\right)^2+16|\vec{p}_{K}|^2 s^2\Bigg|\frac{C^V_{K\eta^{(\prime)}}}{2m_\tau}F_+^{K\eta^{(\prime)}}(s)-\hat{\epsilon}_T F_T^{K\eta^{(\prime)}}(s)\Bigg|^2\nonumber\\[1ex]
&+&4|\vec{p}_{K}|^2s\left(1-\frac{s}{m_\tau^2}\right)\cos^2\theta\left[(C^V_{K\eta^{(\prime)}})^2|F_+^{K\eta^{(\prime)}}(s)|^2-4s\hat{\epsilon}_T^2|F_T^{K\eta^{(\prime)}}(s)|^2\right]+4C^S_{K\eta^{(\prime)}}\Delta_{K\pi}|\vec{p}_{K}|\sqrt{s}\cos\theta \left( 1+\frac{s\hat{\epsilon}_S}{m_\tau (m_s-m_u)}\right)\nonumber\\[1ex]
&\times&\left[C^V_{K\eta^{(\prime)}} \mathrm{Re}[F_0^{K\eta^{(\prime)}}(s)F_+^{*K\eta^{(\prime)}}(s)]-\frac{2s \hat{\epsilon}_T}{m_\tau}\mathrm{Re}[F_T^{K\eta^{(\prime)}}(s)F_0^{*K\eta^{(\prime)}}(s)]\right]\biggr\rbrace\,,
\label{AngularDistribution}
\end{eqnarray}
\end{widetext}
which coincides with the SM result \cite{Escribano:2013bca} when the effective couplings of new interactions are set to zero.

The SM Dalitz plot distribution in the $(s,\cos\theta)$ variables is shown, for the same set of effective couplings discussed previously, in the second row of Fig.\,\ref{SMKeta} for the $K^{-}\eta$ (left) and $K^{-}\eta^{\prime}$ (right) decay modes, and on the plot of the right of Fig.\,\ref{SMKK} for the $K^{-}K^{0}$ channel.

The effects of non-SM interactions on the angular distributions is displayed in the second row of Figs.\,\ref{NPKeta},\ref{NPKetaprime} and \ref{NPKK} for the $K^{-}\eta,K^{-}\eta^{\prime}$ and $K^{-}K^{0}$ decay modes, respectively. 
For the $K^{-}\eta$ channel, the enhanced region near the $K\eta$ threshold in the $(s,t)$ upper-left diagram (the one close to $s$ minimum) is slightly enhanced in a limited region $(\cos\theta>0)$ as it can be seen on the lower-left plot of Fig.\,\ref{NPKeta}, while NP tensor contributions show that the enhanced area for large $t$ translates to nearly minimum values of $\cos\theta$ as it can be observed on the plots of the right. 
For the $K^{-}\eta^{\prime}$ system, both NP scalar and tensor contributions have similar effects in the $(s,\cos\theta)$ plot.
These are given in Fig.\,\ref{NPKetaprime} by the red sunshine area centered at $s/m_{\tau}^{2}\sim0.85$.
Finally, for $\tau^{-}\to K^{-}K^{0}\nu_{\tau}$, in Fig.\,\ref{NPKK} we notice that the enhanced region close to the $s$ minimum in the $(s,t)$ variables of the upper-left plot is washed away in the $(s,\cos\theta)$ variables of the lower-left plot.
In all, we conclude that possible deviations from the SM patterns in near future data will be hard to disentangle in $(s,\cos\theta)$ Dalitz plot analyses.

\subsection{Decay rate}

\begin{figure*}
\includegraphics[width=7cm]{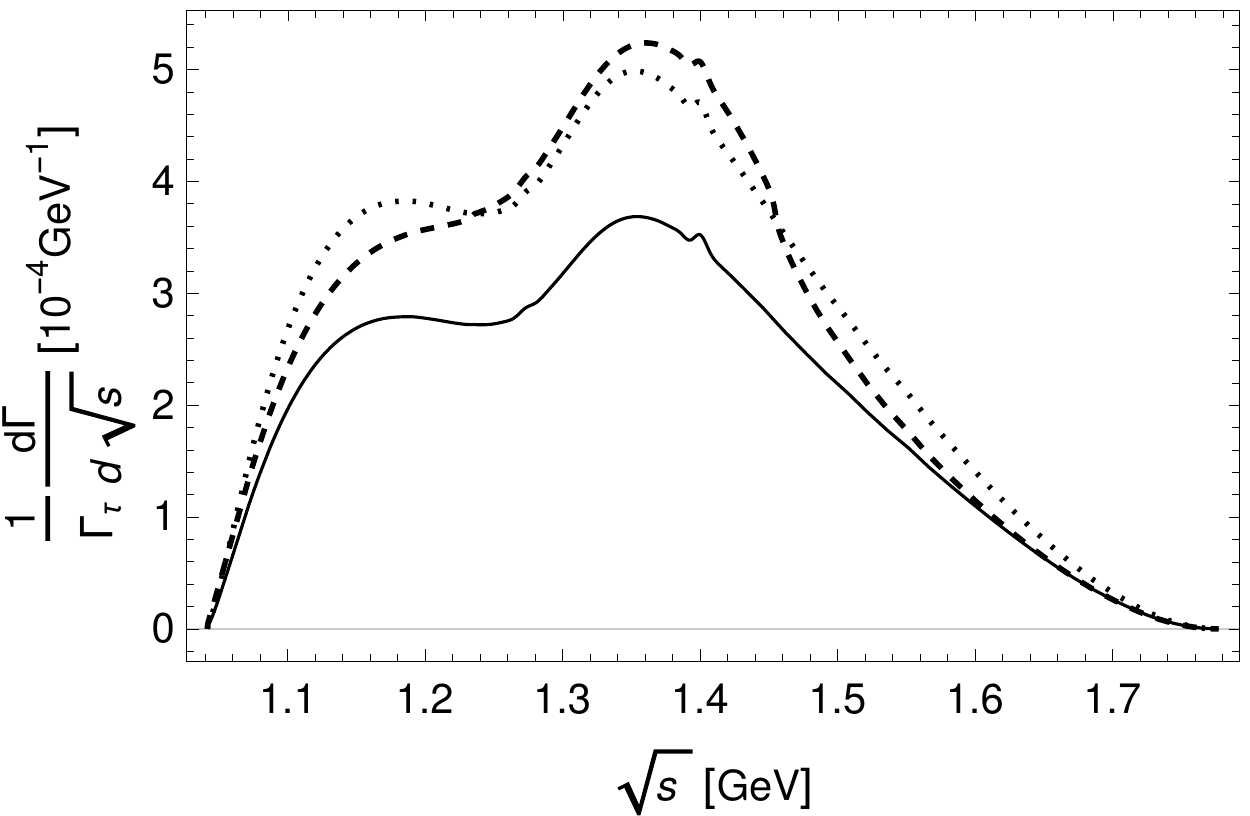}		
\includegraphics[width=7cm]{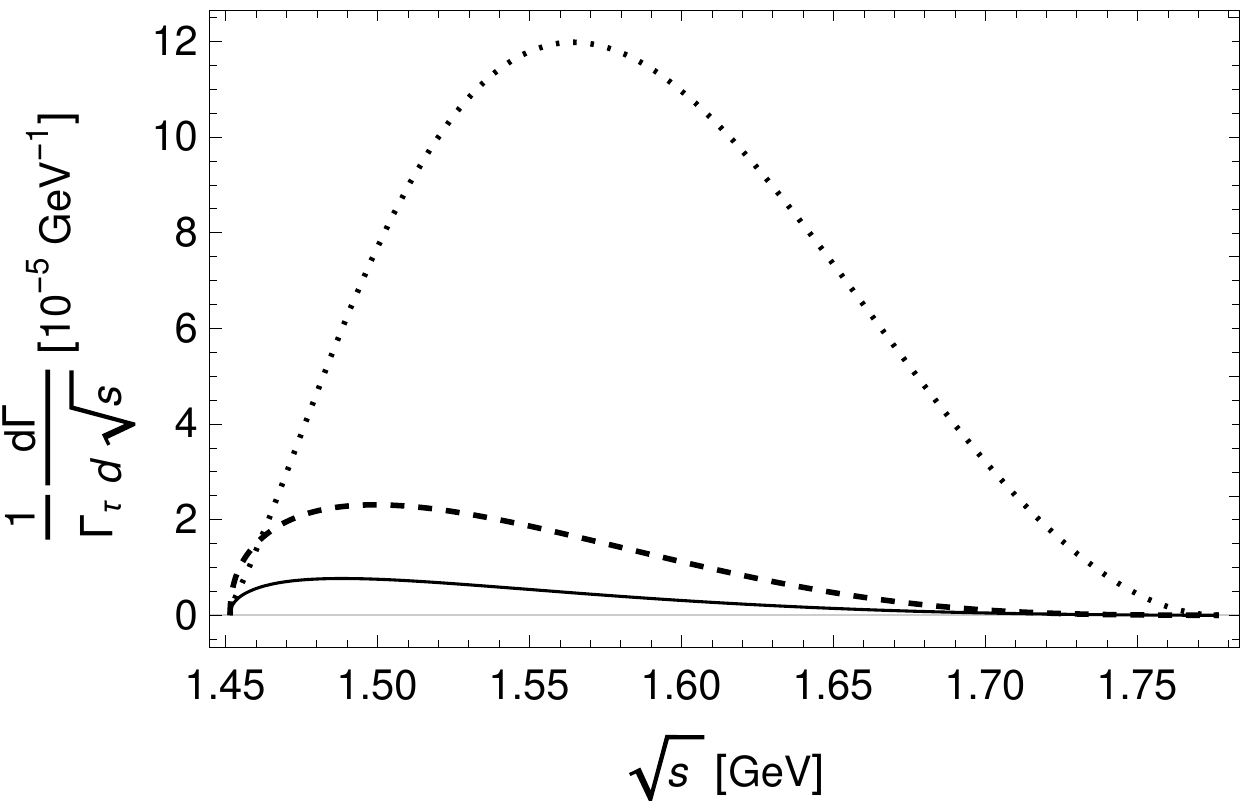}	
\centering			
\caption{
Left: $K^-\eta$ invariant mass distribution in the SM (solid line), and for $\hat{\epsilon}_S=-0.38,\,\hat{\epsilon}_T=0$ (dashed line) and $\hat{\epsilon}_S=0,\,\hat{\epsilon}_T=0.085$ (dotted line). 
Right: $K^- \eta^\prime$ invariant mass distribution in the SM (solid line), and for $\hat{\epsilon}_S=-0.20,\,\hat{\epsilon}_T=0$ (dashed line) and $\hat{\epsilon}_S=0,\,\hat{\epsilon}_T=14.9$ (dotted line). 
Units in axes units are given in $\mathrm{GeV}$ powers and the decay distributions are normalized to the tau decay width.}
\label{DecayRateKetaKetaprime}
\end{figure*}

Integrating Eq.\,(\ref{DoublyDifferentialDecayWidth}) upon the $t$ variable we obtain the $K\eta^{(\prime)}$ invariant mass distribution 
\begin{widetext}
\begin{eqnarray}
\frac{d\Gamma}{d \sqrt{s}}&=&\frac{G_F^2\vert V_{us}\,F_{+}^{K\eta^{(\prime)}}(0)\vert^2 m_\tau^3 S_{EW}}{192\pi^3 \sqrt{s}}(1+\epsilon_L+\epsilon_R)^2\left(1-\frac{s}{m_\tau^2}\right)^2\lambda^{1/2}(s,m_{\eta^{(\prime)}}^2,m_{K}^2)\nonumber\\[1ex]
&\times&\left[X_{VA}+\hat{\epsilon}_SX_S+\hat{\epsilon}_TX_T+\hat{\epsilon}_S^2X_{S^2}+\hat{\epsilon}_T^2X_{T^2}\right],
\label{DecayWidth}
\end{eqnarray}
\end{widetext}
where
\begin{eqnarray}
X_{VA}&=&\frac{(C^V_{K\eta^{(\prime)}})^2}{2s^2}\Big[3|\widetilde{F}_0^{K\eta^{(\prime)}}(s)|^2\Delta_{K\eta^{(\prime)}}^2\nonumber\\[1ex]
&+&|\widetilde{F}_+^{K\eta^{(\prime)}}(s)|^2\left(1+\frac{2s}{m_\tau^2}\right)\lambda(s,m_{\eta^{(\prime)}}^2,m_{K}^2)\Big]\,,
\end{eqnarray}
\begin{eqnarray}
X_S&=&\frac{3}{s\, m_\tau}(C^V_{K\eta^{(\prime)}})^2|\widetilde{F}_0^{K\eta^{(\prime)}}(s)|^2\frac{\Delta_{K\eta^{(\prime)}}^2}{m_s-m_u}\,,\\[1ex]
X_T&=&-\frac{6}{s\,m_\tau}C^V_{K\eta^{(\prime)}}\frac{\mathrm{Re}[F_T^{K\eta^{(\prime)}}(s)F_+^{*K\eta^{(\prime)}}(s)]}{| f_+^{K\eta^{(\prime)}}(0)|^2}\lambda(s,m_{\eta^{(\prime)}}^2,m_{K}^2)\,,\nonumber\\
\end{eqnarray}
\begin{eqnarray}
X_{S^2}&=&\frac{3}{2\,m_\tau^2}(C^V_{K\eta^{(\prime)}})^2|\widetilde{F}_0^{K\eta^{(\prime)}}(s)|^2 \frac{\Delta_{K\eta^{(\prime)}}^2}{\left( m_s-m_u\right)^2}\,,\\[1ex]
X_{T^2}&=&\frac{4}{s}\frac{|F_T^{K\eta^{(\prime)}}(s)|^2}{|F_+^{K\eta^{(\prime)}}(0)|^2} \left(1+\frac{s}{2\,m_\tau^2}\right)\lambda(s,m_{\eta^{(\prime)}}^2,m_{K}^2)\,.
\end{eqnarray}
In Eq.\,(\ref{DecayWidth}) we use $|V_{us}F_{+}^{K^{-}\eta}(0)|=|V_{us}F_{+}^{K^{-}\pi^{0}}(0)\cos\theta_{P}|$ and $|V_{us}F_{+}^{K^{-}\eta^{\prime}}(0)|=|V_{us}F_{+}^{K^{-}\pi^{0}}(0)\sin\theta_{P}|$, with $|V_{us}F_{+}^{K^{-}\pi^{0}}(0)|=0.2165(2)$ \cite{PhysRevD.98.030001}. 
Notice that if one takes $\hat{\epsilon}_S=\hat{\epsilon}_T=0$ we recover the SM result from Eq.\,(2.8) of Ref.\,\cite{Escribano:2013bca}.
The decay distribution in terms of the $K\eta$ and $K\eta^{\prime}$ invariant mass is given, respectively, on the left-and right-plots of Fig.\,\ref{DecayRateKetaKetaprime} for the representative values of the effective couplings.
For the decay $\tau^{-}\to K^{-}\eta\nu_{\tau}$, it can be observed that the deviations with respect to the SM result (solid line) are sizable in the entire energy region of the decay spectrum.
For the $\tau^{-}\to K^{-}\eta^{\prime}\nu_{\tau}$ decay spectrum, we predict a SM branching ratio of $BR_{\rm{SM}}\simeq1.03\times 10^{-6}$. 
This value is found to be totally in line with \cite{Escribano:2013bca} and respects the current experimental upper bound $BR_{\rm{exp}}< 2.4\times 10^{-6}$ at $90\,\%$ C.L. \cite{PhysRevD.98.030001}. 
In this respect, a measurement of this decay mode will be very welcome to further constrain the SM hadronic inputs, a requirement for searches of non-SM interactions.
This measurement should be feasible at Belle-II \cite{Kou:2018nap}.

\begin{figure}
\includegraphics[width=7cm]{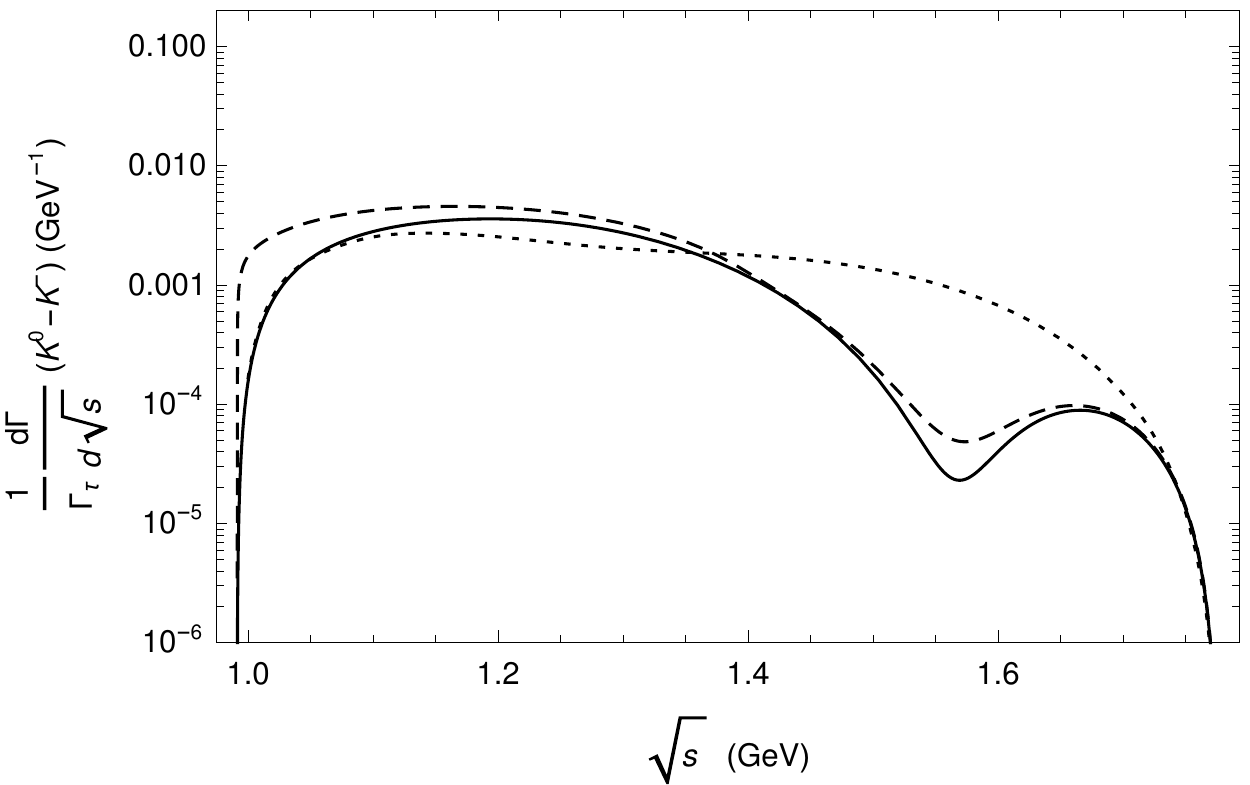}		
\centering			
\caption{Invariant mass distribution for the decay $\tau^{-}\to K^{-}K^{0}\nu_{\tau}$ in the SM (solid line), and for $\hat{\epsilon}_S=0.1,\,\hat{\epsilon}_T=0$ (dashed line) and $\hat{\epsilon}_S=0,\,\hat{\epsilon}_T=0.9$ (dotted line). The decay distribution is normalized to the tau decay width.}
\label{DecayRateKK}
\end{figure}

Regarding the invariant mass distribution of the $\tau^{-}\to K^{-}K^{0}\nu_{\tau}$ transition, it can be obtained after replacing $m_{\eta^{(\prime)}}\to m_{K^{0}}$ and $V_{us}\rightarrow V_{ud}$ in Eq.\,(\ref{DecayWidth}) and using the corresponding Clebsch-Gordan coefficients. 
In Fig.\,\ref{DecayRateKK} we plot the $K^{-}K^{0}$ invariant mass distribution for the SM case (solid line) and for the corresponding effective couplings used for illustration.
In this case, while the (small) effects of non-SM scalar interactions are mostly seen in the first half of the decay spectrum, and in the interference region of the $\rho(1450)$ and $\rho(1700)$ resonances to some extent, the departure from the SM due to tensor interactions is seen on the second half of the spectrum.   

\subsection{Forward-backward asymmetry}

The forward-backward asymmetry for the hadronic $K^{-}\eta^{(\prime)}$ system is defined in analogy to the previous di-meson modes we have studied \cite{Garces:2017jpz,Miranda:2018cpf,Rendon:2019awg}
\begin{equation}
\mathcal{A}_{K\eta^{(\prime)}}(s)=\frac{\int_0^1 d\cos\theta \frac{d^2\Gamma}{ds\,d\cos\theta}-\int_{-1}^0 d\cos\theta \frac{d^2\Gamma}{ds\,d\cos\theta}}{\int_0^1 d\cos\theta \frac{d^2\Gamma}{ds\,d\cos\theta}+\int_{-1}^0 d\cos\theta \frac{d^2\Gamma}{ds\,d\cos\theta}}\,.
\end{equation}

\begin{figure*}
\includegraphics[width=7cm]{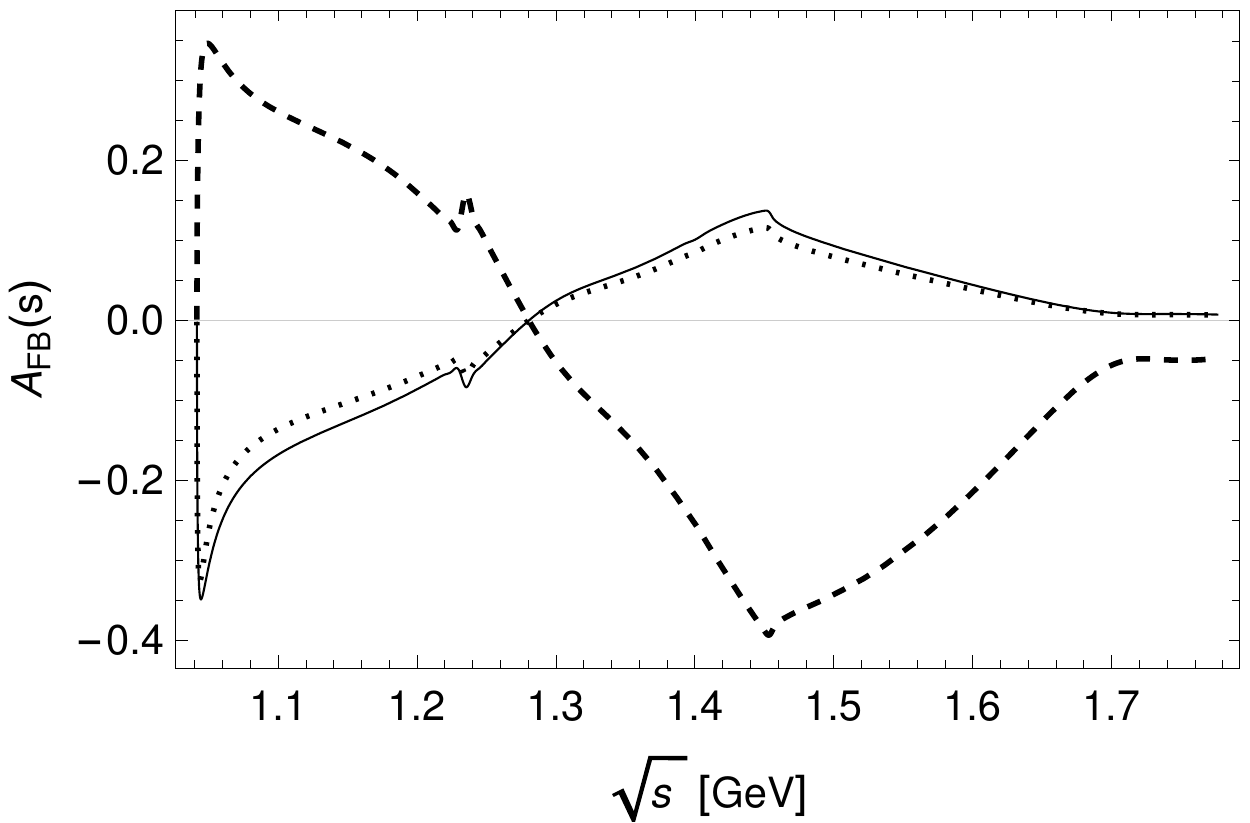}
\includegraphics[width=7cm]{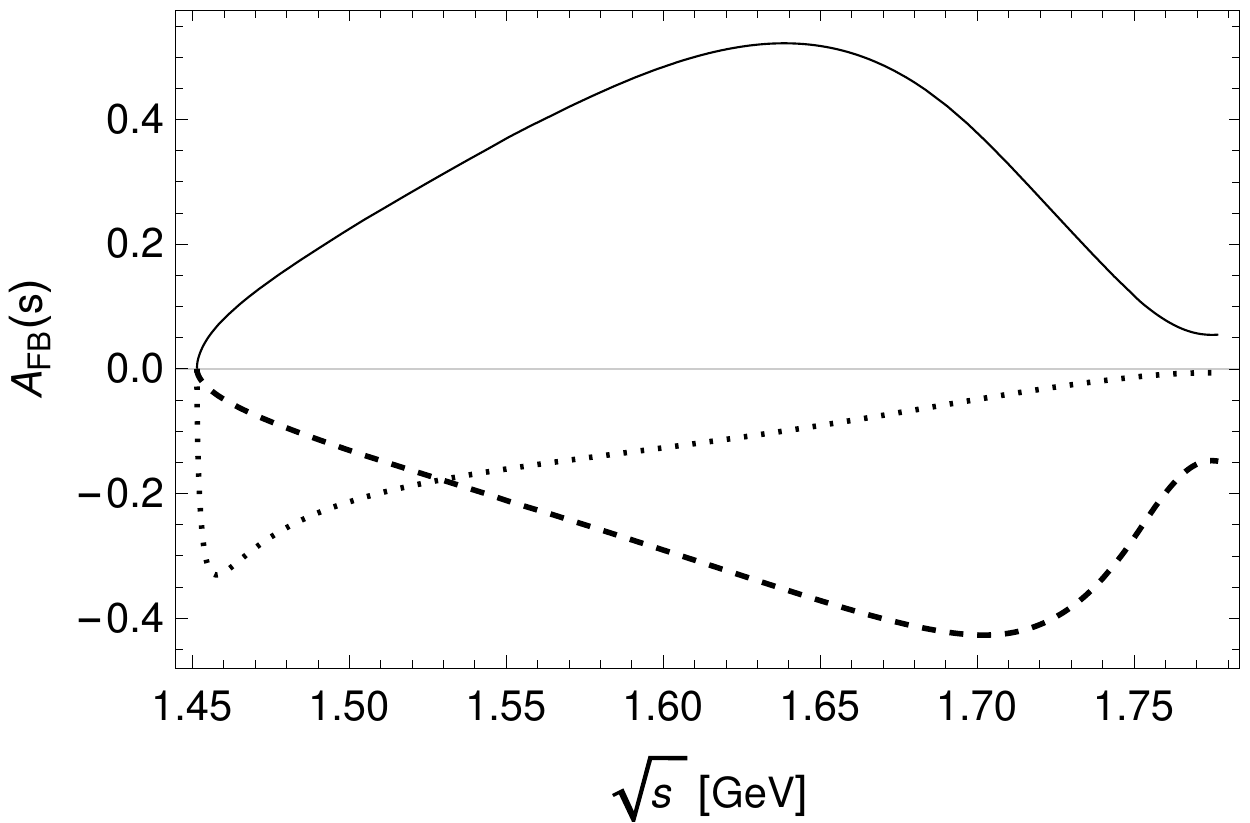}	
\centering			
\caption{Left: Forward-backward asymmetry for the decay $\tau^-\to K^-\eta\nu_\tau$ in the SM (solid line), and for $\hat{\epsilon}_S=-0.38,\,\hat{\epsilon}_T=0$ (dashed line), and $\hat{\epsilon}_T=0.085,\,\hat{\epsilon}_S=0$ (dotted line). 
Right: Forward-backward asymmetry for the decay $\tau^-\to K^-\eta^{\prime}\nu_\tau$ in the SM (solid line), and for $\hat{\epsilon}_S=-0.20,\,\hat{\epsilon}_T=0$ (dashed line), and $\hat{\epsilon}_T=14.9,\,\hat{\epsilon}_S=0$ (dotted line).}\label{AFBKetaKetaprime} 
\end{figure*}

\begin{figure}
\includegraphics[width=7.7cm]{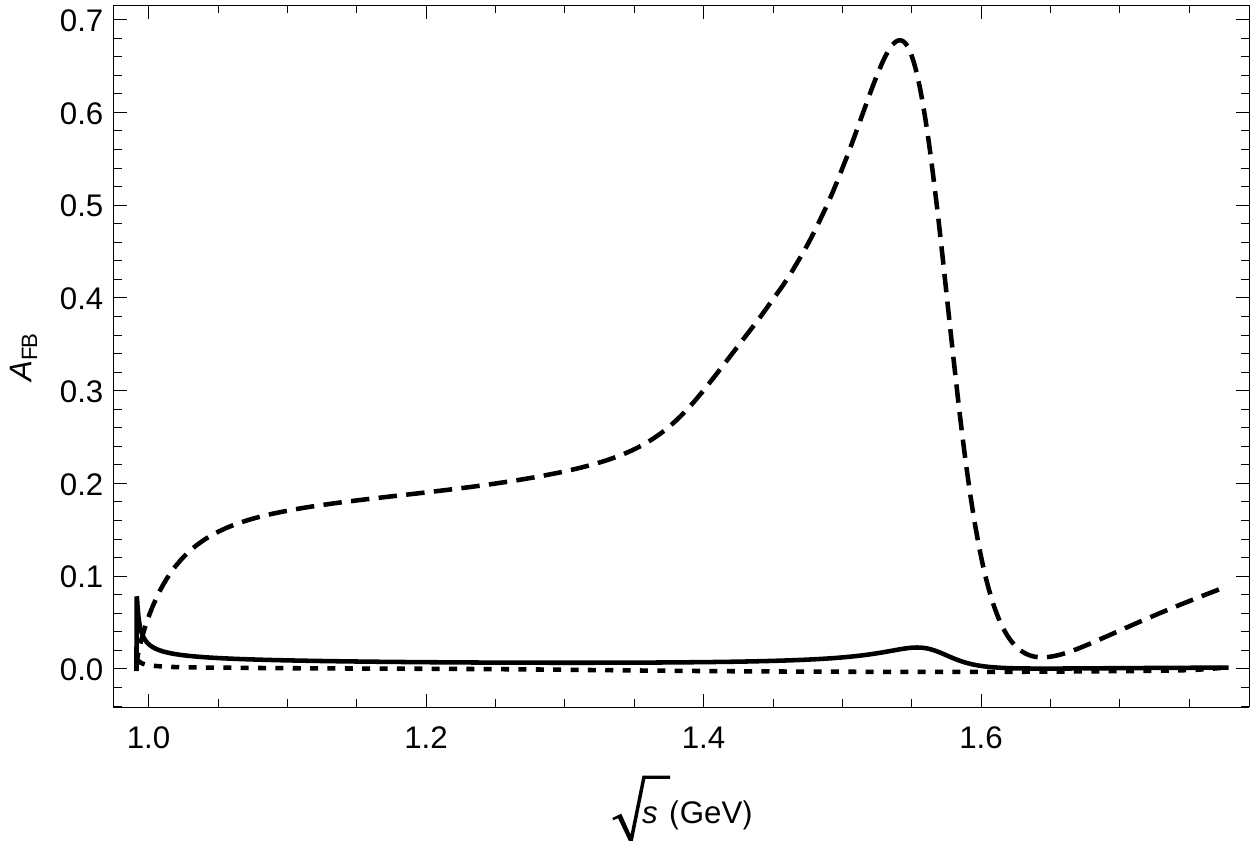}
\centering			
\caption{Forward-backward asymmetry for the decay $\tau^-\to K^-K^{0}\nu_\tau$ in the SM (solid line), and for $\hat{\epsilon}_S=0.1,\,\hat{\epsilon}_T=0$ (dashed line), and $\hat{\epsilon}_T=0.9,\,\hat{\epsilon}_S=0$ (dotted line).}
\label{AFBKK} 
\end{figure}

Inserting Eq.\,(\ref{AngularDistribution}) into the previous expression and integrating upon the $\cos\theta$ variable we obtain its analytical expression
\begin{eqnarray}
\mathcal{A}_{K\eta^{(\prime)}}(s)&=&\frac{3C^S_{K\eta^{(\prime)}}\Delta_{K\pi}\sqrt{\lambda(s,m_{\eta^{(\prime)}}^2,m_{K}^2)}}{2s^2|F_+^{K\eta^{(\prime)}}(0)|^2[X_{VA}+\hat{\epsilon}_SX_S+\hat{\epsilon}_TX_T+\hat{\epsilon}_S^2X_{S^2}+\hat{\epsilon}_T^2X_{T^2}]}\nonumber\\[1ex]
&\times&\left(1+\frac{s\hat{\epsilon}_S}{m_\tau (m_s-m_u)}\right)\Big\{C^V_{K\eta^{(\prime)}}\mathrm{Re}[F_0^{K\eta^{(\prime)}}(s)F_+^{*K\eta^{(\prime)}}(s)]\nonumber\\[1ex]
&-&\frac{2s\,\hat{\epsilon}_T}{m_\tau}\mathrm{Re}[F_T^{K\eta^{(\prime)}}(s)F_0^{*K\eta^{(\prime)}}(s)]\Big\}\,.
\label{AFB}
\end{eqnarray}
Again, replacing $m_{\eta^{(\prime)}}\to m_{K^{0}},m_{s}\to m_{d}$ and $\Delta_{K\pi}\rightarrow \Delta_{KK}$ in Eq.\,(\ref{AFB}) we find the corresponding result for the decay $\tau^{-}\rightarrow K^{-}K^{0}\nu_{\tau}$, $\mathcal{A}_{KK}(s)$ .
The forward-backward asymmetry in the SM case i.e. $\hat{\epsilon}_{S,T}=0$, corresponds to the solid line in Fig.\,\ref{AFBKetaKetaprime} for the decays $K^{-}\eta$ (left plot) and $K^{-}\eta^{\prime}$ (right plot), and in Fig.\,\ref{AFBKK} for the $K^{-}K^{0}$ transition.
For the $K^{-}\eta$ mode, it should not be difficult to measure a non-zero (negative) value near the $K^{-}\eta$ threshold. 
$\mathcal{A}_{K\eta}$ increases with $s$, crosses zero at around $1.28$ GeV and reaches its maximum near $1.45$ GeV, when it decreases up to the upper kinematical limit.
For the $K\eta^{\prime}$ case, the forward-backward asymmetry is a positive increasing observable from the $K\eta^{\prime}$ threshold until around $1.64$ GeV where it has a plateau and decreases afterwards.
Finally, for the $K^{-}K^{0}$ decay channel, the SM $\mathcal{A}_{KK}$ is in general small with a signature right before 1 GeV and small bump at around $1.55$ GeV. 

In these figures, we also display the results arising from considering non-SM scalar and tensor interactions.
For the $K^{-}\eta$ channel, one observes that the tensor case (dotted line) overlaps with the SM prediction thus being difficult unveil its possible effects from the SM contribution.
On the contrary, for non-SM scalar interactions (dashed line), $\mathcal{A}_{K\eta}$ flips sign with respect to the SM slightly before $1.3$ GeV and it gets larger in magnitude as $s$ increases.
If it is possible to measure this observable eventually, this would ease the identification of NP contributions in $\mathcal{A}_{K\eta}$.
The non-standard scalar contribution to the forward-backward asymmetry of the $K\eta^{\prime}$ decay mode is negative and has, to great extent, the same size than the SM ones but with opposite sign.  
The NP tensor contribution, also negative, has a clear non-zero value near threshold and then becomes a decreasing function until the kinematical upper limit of $\sqrt{s}$.
It is clear then that noticeable differences with respect to the SM contribution will be appreciated for quite large values of the new effective couplings.
Similarly, for the $K^{-}K^{0}$ decay, clear non-zero values for the NP coupling of the scalar contributions will unambiguously dominate over the tensor ones.
Therefore, the $\mathcal{A}_{KK}$ would be a good observable for searching non-standard scalar interactions: despite its numerator in Eq.\,(\ref{AFB}) is suppressed by the small value of $\Delta_{K^{-}K^{0}}$; its denominator is further suppressed by the dependence of the $X_{S^{2}}$ on $\Delta_{K^{-}K^{0}}$.

\subsection{Limits on $\hat{\epsilon}_S$ and $\hat{\epsilon}_T$}\label{Limits}

\begin{figure*}
\includegraphics[width=7cm]{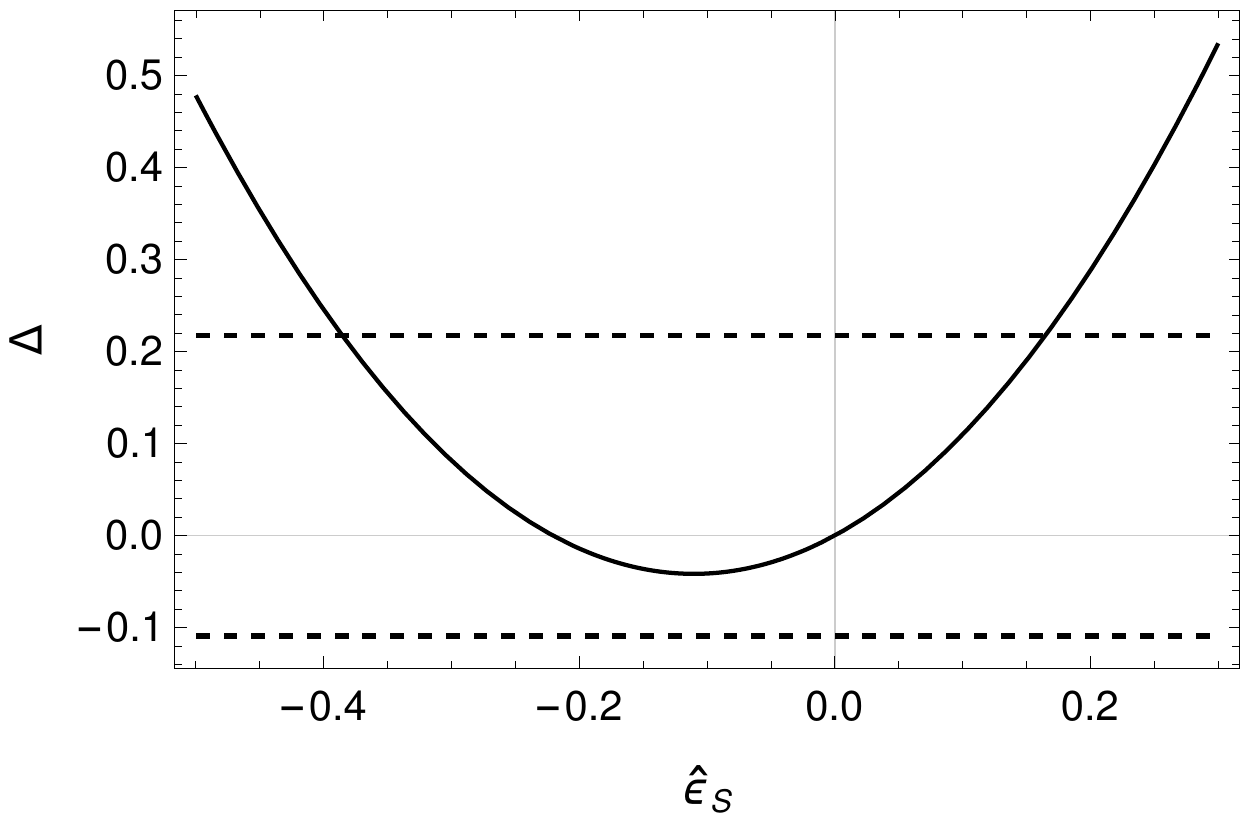}
\includegraphics[width=7cm]{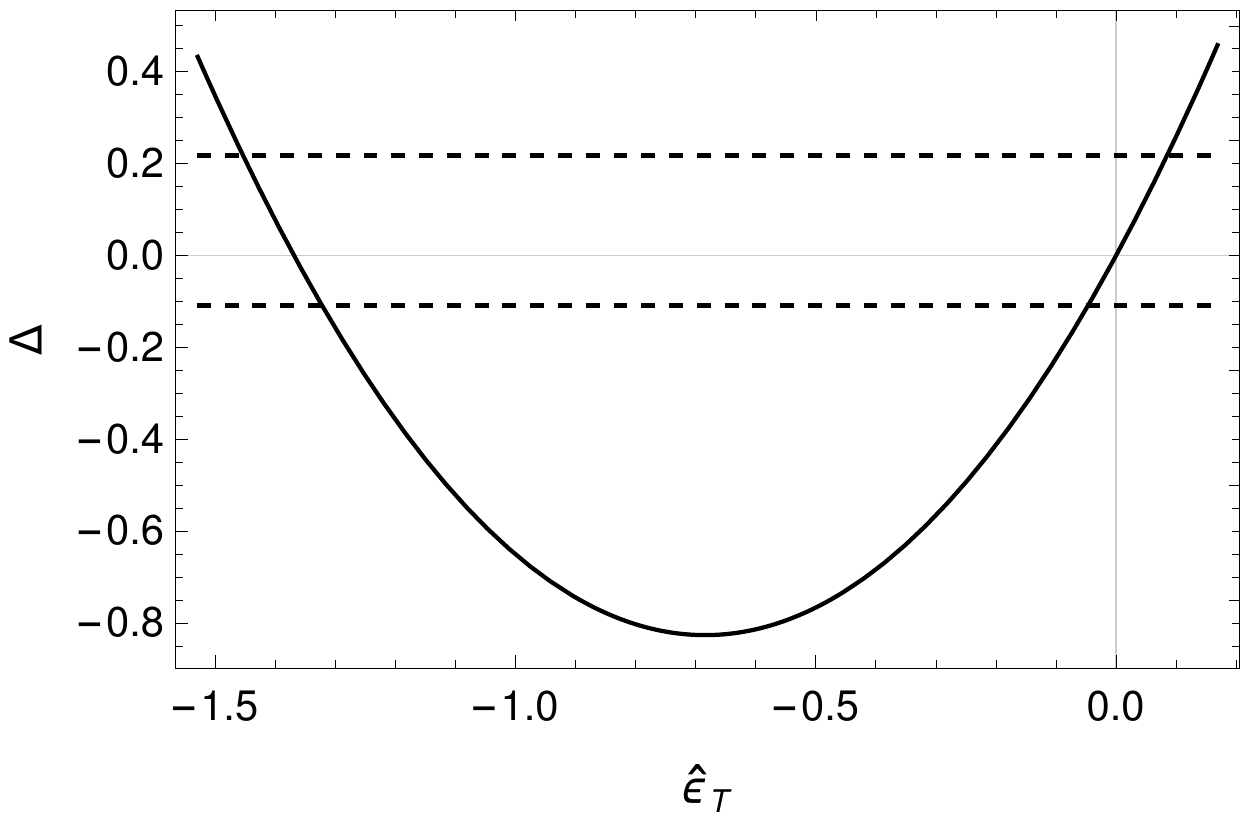}		
\centering			
\caption{$\Delta$ as a function of $\hat{\epsilon}_S$ for $\hat{\epsilon}_T=0$ (left-hand) and $\hat{\epsilon}_T$ for $\hat{\epsilon}_S=0$ (right-hand) for the decay $\tau^-\to K^-\eta\nu_\tau$. 
Horizontal lines represent the values of $\Delta$ according to the current measurement and theory errors (at three standard deviations) of the branching ratio (dashed line).}
\label{DeltaKeta} 
\end{figure*}

\begin{figure*}
\includegraphics[width=7cm]{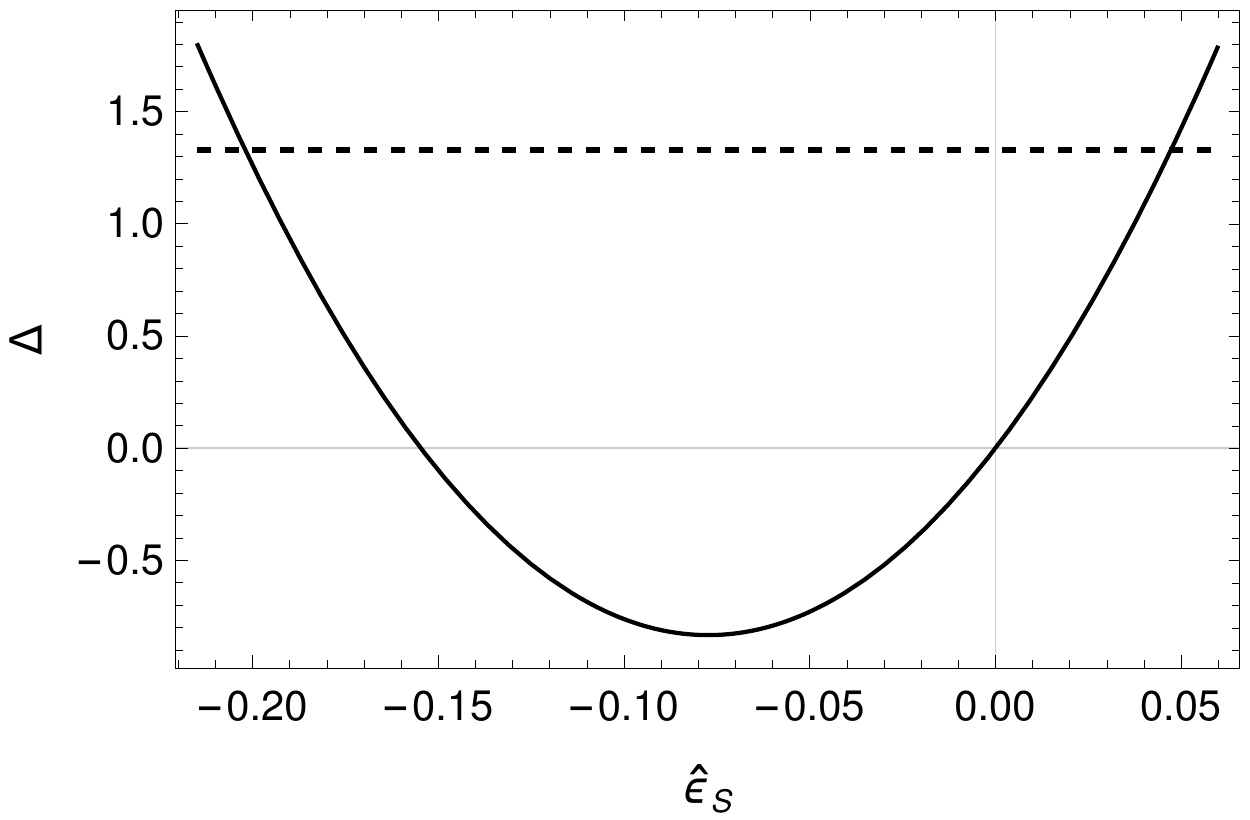}
\includegraphics[width=6.8cm]{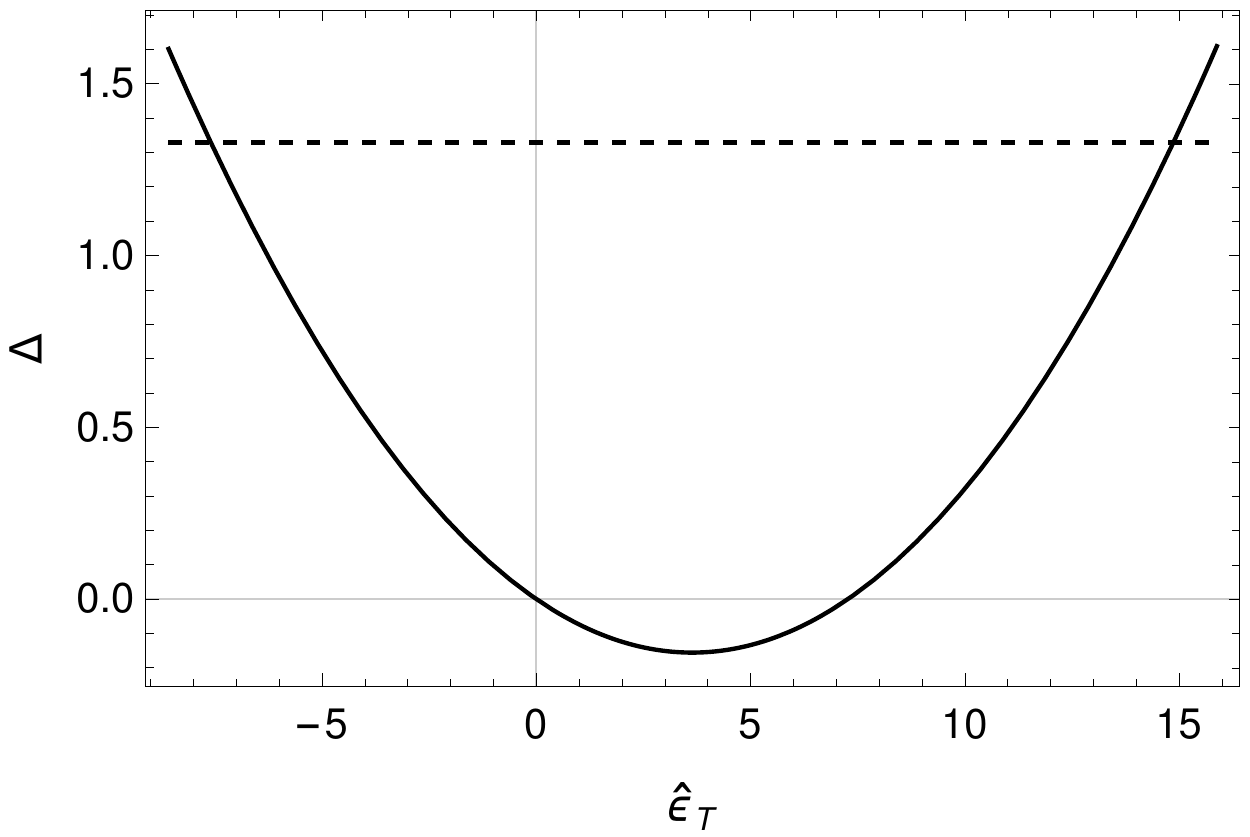}		
\centering			
\caption{$\Delta$ as a function of $\hat{\epsilon}_S$ for $\hat{\epsilon}_T=0$ (left plot) and $\hat{\epsilon}_T$ for $\hat{\epsilon}_S=0$ (right plot) for the decay $\tau^-\to K^-\eta^{\prime}\nu_\tau$. 
Horizontal lines represent the values of $\Delta$ according to the current measurement and theory errors (at three standard deviations) of the branching ratio (dashed line).}
\label{DeltaKetaprime} 
\end{figure*}

\begin{figure*}
\includegraphics[width=6.9cm]{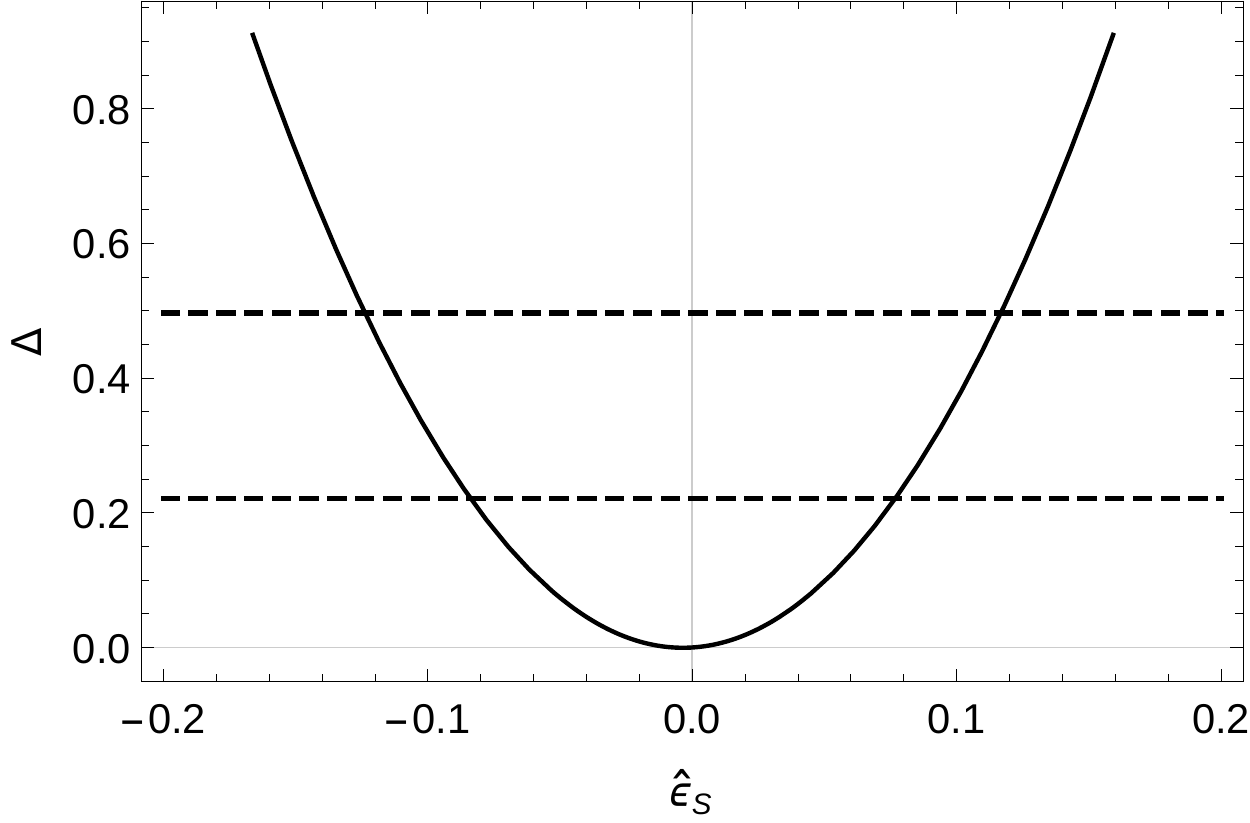}
\includegraphics[width=7cm]{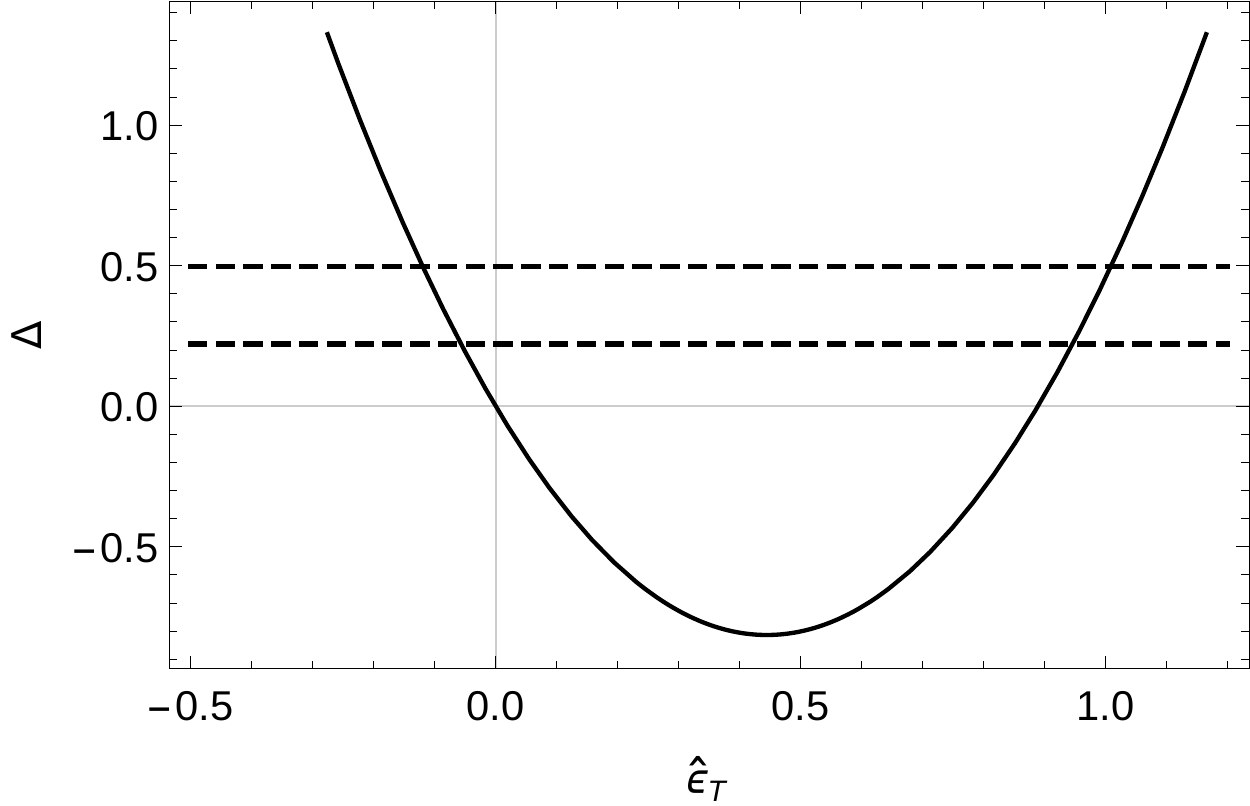}		
\centering			
\caption{$\Delta$ as a function of $\hat{\epsilon}_S$ for $\hat{\epsilon}_T=0$ (left plot) and $\hat{\epsilon}_T$ for $\hat{\epsilon}_S=0$ (right plot) for the decay $\tau^-\to K^{-}K^{0}\nu_\tau$. 
Horizontal lines represent the values of $\Delta$ according to the current measurement and theory errors (at three standard deviations) of the branching ratio (dashed line).}
\label{DeltaKK} 
\end{figure*}

Integrating the invariant mass distribution Eq.\,(\ref{DecayWidth}) upon the $s$ variable one obtains the total decay width which, in turn, depends on the NP effective couplings $\hat{\epsilon}_{S,T}$. 
One can therefore use the experimental branching ratio to set bounds on $\hat{\epsilon}_{S,T}$.
For this purpose, we compare the decay width as obtained by including non-SM interactions, and that we denote by $\Gamma$, with respect to the SM width, $\Gamma^{0}$, obtained by neglecting NP interactions i.e. setting $\hat{\epsilon}_{S,T}=0$.
The relative shift produced by the NP contributions is better accounted for through the following observable:
\begin{equation}
\Delta\equiv \frac{\Gamma-\Gamma^0}{\Gamma^0}=\alpha \hat{\epsilon}_S+\beta \hat{\epsilon}_T+\gamma \hat{\epsilon}_S^2+\delta \hat{\epsilon}_T^2\,.
\label{GammaObservable}
\end{equation}
The numerical values of the coefficients $\alpha,\beta,\gamma$ and $\delta$ for the processes under consideration are found to be: $\alpha=0.85_{-0.09}^{+0.05}$, $\beta=3.7^{+1.2}_{-1.3}$, $\gamma=4.3_{-0.9}^{+0.6}$ and $\delta=3.9^{+3.0}_{-2.2}$ for the $K^{-}\eta$ decay channel; $\alpha=24.2_{-2.7}^{+1.5}$, $\beta=-0.26^{+0.17}_{-0.24}$, $\gamma=175.9^{+23.3}_{-36.6}$ and $\delta=0.10^{+0.28}_{-0.09}$ for the $K^{-}\eta^{\prime}$ mode; and $\alpha=0.24\pm0.01$, $\beta=-3.66^{+0.16}_{-1.74}$, $\gamma=34.4_{-1.4}^{+1.3}$ and $\delta=9.2_{-5.2}^{+1.0}$ for the $K^{-}K^{0}$ transition.
The errors carried by the previous coefficients come from the uncertainty associated to the corresponding form factors (see section \ref{section3}). 
Eq.\,(\ref{GammaObservable}) is a quadratic function of the effective scalar and tensor couplings that can be used to explore the sensitivity of the corresponding decays to the effects of non-SM interactions.
As in Refs.\,\cite{Garces:2017jpz,Miranda:2018cpf,Rendon:2019awg}, we will do this in two different ways.
Firstly, we set one of the couplings to zero and obtain bounds for the other, and viceversa.
The result is shown in Figs.\,\ref{DeltaKeta}, \ref{DeltaKetaprime} and \ref{DeltaKK} for the three decays concerning us, respectively.
In these figures, the horizontal lines represent the current experimental limits on $\Delta$ (at three standard deviations), and 
the resulting bounds for the effective couplings are found to be 
\begin{eqnarray}
&&-0.38\leq\hat{\epsilon}_S\leq0.16\,,\quad\hat{\epsilon}_{T}=0\,,\\[1ex]
&&\hat{\epsilon}_{S}=0\,,\quad\hat{\epsilon}_{T}=[-1.4,-0.7]\cup [-0.047,0.085]\,,
\label{boundsKeta}
\end{eqnarray}
from the decay $\tau^{-}\to K^{-}\eta\nu_{\tau}$ ($BR_{\rm{exp}}=1.55(8)\times10^{-4}$ \cite{PhysRevD.98.030001}),
\begin{eqnarray}
&&-0.20\leq\hat{\epsilon}_S\leq0.05\,,\quad\hat{\epsilon}_{T}=0\,,\\[1ex]
&&\hat{\epsilon}_{S}=0\,,\quad-7.6\leq\hat{\epsilon}_T\leq14.9\,,
\label{boundsKetaprime}
\end{eqnarray}
from the transition $\tau^{-}\to K^{-}\eta^{\prime}\nu_{\tau}$ ($BR_{\rm{exp}}< 2.4\times 10^{-6}$ at $90\,\%$ C.L. \cite{PhysRevD.98.030001}), and 
\begin{eqnarray}
\hat{\epsilon}_{S}&=&[-0.12,-0.08]\cup [0.08,0.12]\,,\quad\hat{\epsilon}_{T}=0\,,\\[1ex]
\hat{\epsilon}_{S}&=&0\,,\quad\hat{\epsilon}_{T}=[-0.12,-0.06]\cup [0.92,0.99]\,,
\label{boundsKK}
\end{eqnarray}
from $\tau^{-}\to K^{-}K^{0}\nu_{\tau}$ ($BR_{\rm{exp}}=1.486(34)\times10^{-3}$ \cite{PhysRevD.98.030001}).
Had we used the BaBar measurement of $\tau^{-}\to K^{-}K_{S}\nu_{\tau}$ ($BR_{\rm{exp}}=0.739(11)(20)\times10^{-3}$ \cite{BaBar:2018qry}), we would have obtained instead
\begin{eqnarray}
\hat{\epsilon}_{S}&=&[-0.12,-0.09]\cup [0.08,0.11]\,,\quad\hat{\epsilon}_{T}=0\,,\\[1ex]
\hat{\epsilon}_{S}&=&0\,,\quad\hat{\epsilon}_{T}=[-0.12,-0.06]\cup [0.93,0.99]\,.
\end{eqnarray}

\begin{figure*}
\includegraphics[width=5.75cm]{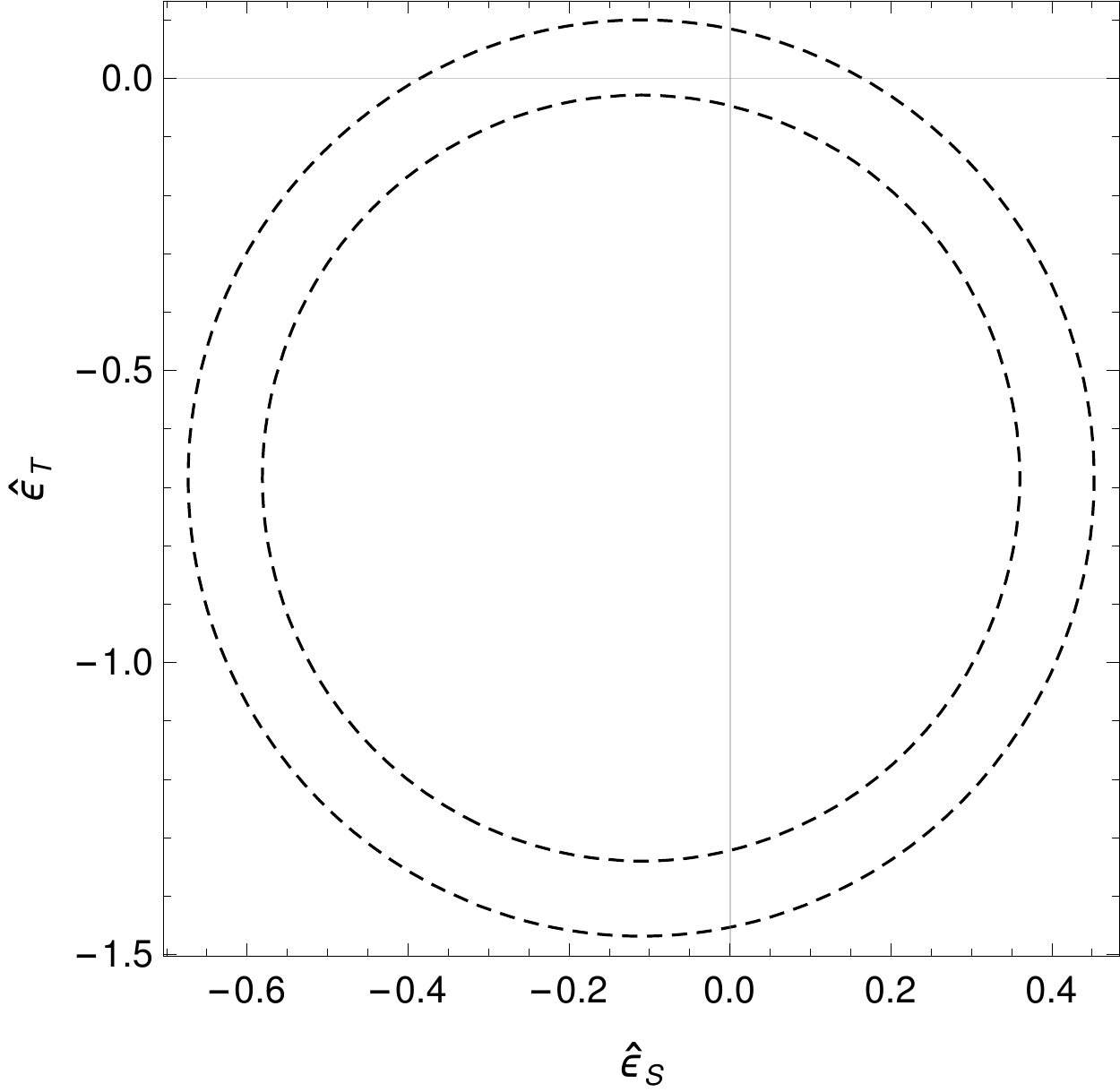}
\hspace{-0.05cm}\includegraphics[width=5.6cm]{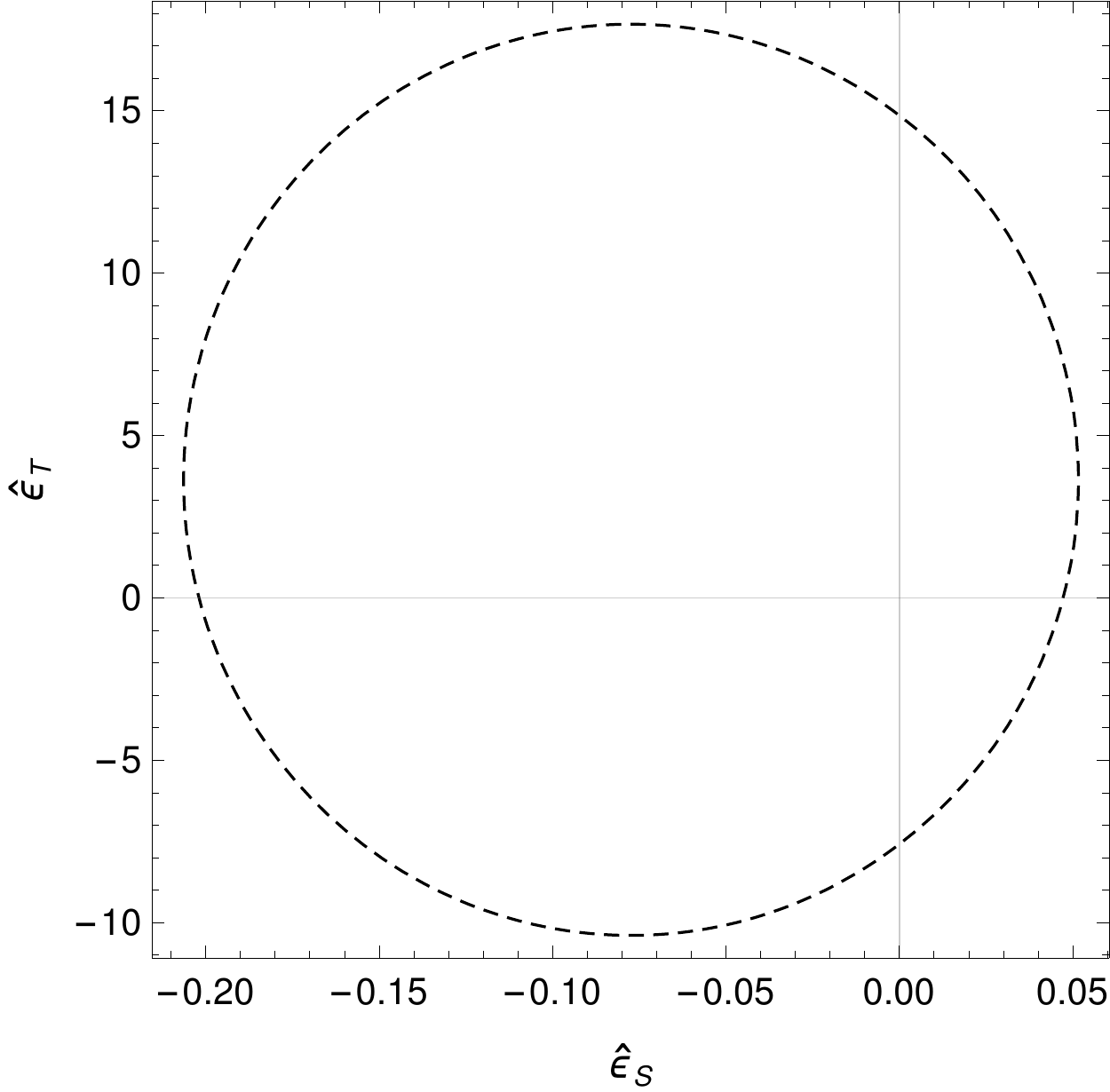}	
\includegraphics[width=5.75cm]{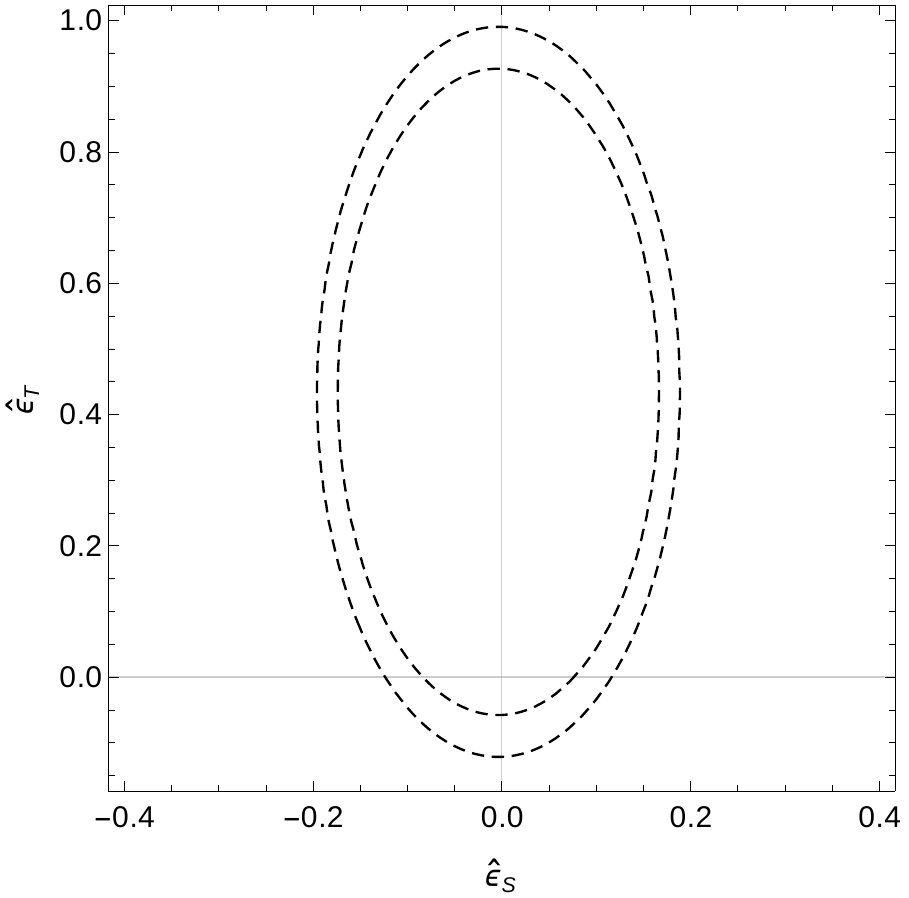}	 
\centering			
\caption{Constraints on the scalar and tensor couplings obtained from $\Delta(\tau^{-}\to K^{-}\eta\nu_\tau)$ (left plot), $\Delta (\tau^-\to K^-\eta^{\prime}\nu_\tau)$ (central plot) and $\Delta (\tau^-\to K^-K^{0}\nu_\tau)$ (right plot) using, respectively, the measured branching ratio (at three standard deviations) and the upper limits of the branching ratio at $90\,\%$ C.L.}
\label{ContourKetaKetaprime}
\end{figure*}

Secondly, we have also set constraints on these couplings from the general case where both are non-vanishing using Eq.\,(\ref{GammaObservable}) as before.
These results are graphically represented by ellipses in the $\hat{\epsilon}_{S}$-$\hat{\epsilon}_{T}$ plane in Fig.\,\ref{ContourKetaKetaprime} for the three decay channels under consideration.

In all, our results for the bounds in the scalar and tensor effective couplings $\hat{\epsilon}_{S}$ and $\hat{\epsilon}_{T}$ that can be obtained at three standard deviations from the current experimental measurement are gathered in Table \ref{ResultsEffectiveCouplings}.
The constraints on the scalar coupling obtained from the $K^{-}\eta$ decay channel is quite symmetric while the tensor coupling has a mild preference for $\hat{\epsilon}_{T}<0$.
The allowed region has the same size for both.
Limits on the scalar coupling from the $K^{-}\eta^{\prime}$ mode favor slightly $\hat{\epsilon}_{S}<0$ while the constraints on the tensor one are much weaker in this case.
Finally, from the $K^{-}K^{0}$ decay, the allowed region for $\hat{\epsilon}_{S}$ is symmetric and shows a small preference over tensor interactions whose coupling prefers to sit on the positive side, $\hat{\epsilon}_{T}>0$.
\begin{table*}
\begin{center}
\begin{tabular}{|l|c|c|c|c|}
\hline
Decay channel& $\hat{\epsilon}_S\,(\hat{\epsilon}_T=0)$ & $\hat{\epsilon}_T\,(\hat{\epsilon}_S=0)$ & $\hat{\epsilon}_S$ & $\hat{\epsilon}_T$ \\
\hline
$\tau^{-}\to K^{-}\eta\nu_{\tau}$ & $\left[-0.38,0.16\right]$ &$\left[-1.4,-0.7\right]\cup\left[-4.7,8.5\right]\cdot 10^{-2}$ & $\left[-0.7,0.5\right]$ & $\left[-1.5,0.1\right]$\\ 
$\tau^{-}\to K^{-}\eta^{\prime}\nu_{\tau}$ & $\left[-0.20,0.05\right]$ & $\left[-7.6,14.9\right]$ & $\left[-0.21,0.05\right]$ & $\left[-10.4,17.7\right]$\\
$\tau^{-}\to K^{-}K^{0}\nu_{\tau}$ & $\left[-0.12,-0.08\right]\cup\left[0.08,0.12\right]$   & $\left[-0.12,-0.06\right]\cup\left[0.92,0.99\right]$ & $\left[-0.2,0.2\right]$ & $\left[-0.12,0.98\right]$\\
\hline
$\tau^{-}\to\pi^{-}\pi^{0}\nu_{\tau}$ \cite{Miranda:2018cpf}& $\left[-1.33,1.31\right]$ &$\left[-0.79,-0.57\right]\cup\left[-1.4,1.3\right]\cdot 10^{-2}$ & $\left[-5.2,5.2\right]$ & $\left[-0.79,0.013\right]$\\ 
$\tau^{-}\to (K\pi)^{-}\nu_{\tau}$ \cite{Rendon:2019awg}& $\left[-0.57,0.27\right]$ & $\left[-0.059,0.052\right]\cup\left[0.60,0.72\right]$ & $\left[-0.89,0.58\right]$ & $\left[-0.07,0.72\right]$\\
$\tau^{-}\to\pi^{-}\eta\nu_{\tau}$ \cite{Garces:2017jpz}& $\left[-8.3,3.9\right]\cdot10^{-3}$   &  $\left[-0.43,0.39\right]$ &  $\left[-0.83,0.37\right]\cdot10^{-2}$ & $\left[-0.55,0.50\right]$\\
$\tau^{-}\to\pi^{-}\eta^{\prime}\nu_{\tau}$ \cite{Garces:2017jpz}& $\left[-1.13,0.68\right]\cdot10^{-2}$   &  $|\hat{\epsilon}_T|<11.4$ &  $\left[-1.13,0.67\right]\cdot10^{-2}$ & $\left[-11.9,11.9\right]$\\
\hline
\end{tabular}
\caption{Constraints on the scalar and tensor couplings obtained (at three standard deviations) through the limits on the current branching ratio measurements. Theory errors are included.}
\label{ResultsEffectiveCouplings}
\end{center}
\end{table*}

In this table, we also compare the results of this work with the constraints we have obtained in previous analyses from the $\pi^{-}\pi^{0}$ \cite{Miranda:2018cpf}, $(K\pi)^{-}$ \cite{Rendon:2019awg} and $\pi^{-}\eta^{(\prime)}$ \cite{Garces:2017jpz} decay channels.
The constraints for the scalar couplings are found to be more precise than those obtained from the di-pion mode, competitive with the limits set from the $(K\pi)^{-}$ decays, and weaker than the bounds coming from the decays $\pi^{-}\eta^{(\prime)}$.
For the tensor couplings, we notice that the $K\eta^{\prime}$ channel gives a much looser limits than the decays $K\eta$ and $K^{-}K^{0}$.
The allowed region of the last two, in turn, is similar than that obtained in our previous analyses but for $\pi^{-}\eta^{\prime}$, which is not competitive restricting tensor interactions.

As a final exercise, we have also determined the effective couplings from a $\chi^{2}$ function in the following way:
\begin{equation}
\chi^{2}=\left(\frac{BR^{\rm{th}}_{K^{-}\eta}-BR^{\rm{exp}}_{K^{-}\eta}}{\sigma_{BR^{\rm{exp}}_{K^{-}\eta}}}\right)^{2}+\left(\frac{BR^{\rm{th}}_{K^{-}K^{0}}-BR^{\rm{exp}}_{K^{-}K^{0}}}{\sigma_{BR^{\rm{exp}}_{K^{-}K^{0}}}}\right)^{2}\,,
\end{equation}
where $BR^{\rm{exp}}_{K^{-}\eta}$ and $\sigma_{BR^{\rm{exp}}_{K^{-}\eta}}$, and $BR^{\rm{exp}}_{K^{-}K^{0}}$ and $\sigma_{BR^{\rm{exp}}_{K^{-}K^{0}}}$, are the measured branching ratio and the corresponding uncertainties of the $K^{-}\eta$ and $K^{-}K^{0}$ decay modes, respectively, and $BR^{\rm{th}}_{K^{-}\eta}$ and $BR^{\rm{th}}_{K^{-}K^{0}}$ are the analogue theoretical expressions obtained upon integrating Eq.\,(\ref{DecayWidth}).
The $\chi^{2}$ function defined above depends solely on $\hat{\epsilon}_{S}$ and $\hat{\epsilon}_{T}$.
Using the experimental values given below Eqs.\,(\ref{boundsKeta}) and (\ref{boundsKK}) we obtain the constraints:
\begin{equation}
\hat{\epsilon}_{S}=0.088^{+0.035}_{-0.056}\,,\quad \hat{\epsilon}_{T}=0.015^{+0.056}_{-0.066}\,,
\end{equation}
where variations up to $3\sigma$ of the measured branching ratios have been taken into account.

Comparing our results with bounds obtained from other low-energy probes, our previous limits are not competitive with semileptonic kaon decays, $\hat{\epsilon}_{S}=(-3.9\pm4.9)\times10^{-4}$ and $\hat{\epsilon}_{T}=(0.5\pm5.2)\times10^{-3}$  \cite{Gonzalez-Alonso:2016etj}, while they are similar than those obtained from hyperon decays \cite{Chang:2014iba}, where $|\hat{\epsilon}_{S}|<4\times10^{-2}$ and $|\hat{\epsilon}_{T}|<5\times10^{-2}$ are found at a $90\%$ C.L.\footnote{For the comparison, we need to assume lepton universality because our study involves the tau lepton, while theirs electrons and muons.
Given the smallness of possible lepton universality violations, this is enough for current precision.
We have also assumed that the corresponding CKM matrix elements do not change under NP interactions, which is the case if $\epsilon(lud)=\epsilon(lus)$ \cite{Descotes-Genon:2018foz}.}.
With respect to the results of Ref.\,\cite{Cirigliano:2018dyk}, obtained also from hadronic tau decays (strangeness-conserving transitions only), our corresponding limits are less precise.
However, the use of all available data of all possible di-meson tau decays (see Table \ref{ResultsEffectiveCouplings}) could allow us improve the knowledge in this respect.
Such analysis is our next step plan. 

\section{Conclusions}\label{conclusions}

Hadronic tau lepton decays remain to be an advantageous tool for the investigation of the hadronization of QCD currents in the non-perturbative regime of the strong interaction.
In this paper, we have studied the decays $\tau^-\to K^-(\eta^{(\prime)},K^0) \nu_\tau$ in the presence of non-Standard Model scalar and tensor interactions.
We have focused our analysis on setting bounds on the corresponding New Physics couplings from the current experimental measurements of these decays.
This has been possible due to the satisfactory knowledge we have on the necessary Standard Model hadronic input, the form factors.
For the description of the participating vector and scalar form factors, we have employed previous results based on constraints from Chiral Perturbation Theory supplemented by dispersion relations and experimental data.
On the contrary, there are no experimental data to help us constructing the required tensor form factor and, therefore, it has been described under theoretical arguments solely.
Within this framework, we have set limits (see Table \ref{ResultsEffectiveCouplings}) on the non-standard scalar and tensor couplings, $\hat{\epsilon}_{S}$ and $\hat{\epsilon}_{T}$, respectively, using the measured branching ratios, and have studied their effects on different phenomenological observables including Dalitz plot and angular distributions, the decay rate and the forward-backward asymmetry.
The present analysis completes our series of dedicated studies of two-meson tau decays \cite{Garces:2017jpz,Miranda:2018cpf,Rendon:2019awg} that have shown the complementary role that tau decays can play in restricting non-standard interactions.
Despite our bounds on the NP couplings are not as precise as those placed, for example, from semileptonic kaon decays \cite{Gonzalez-Alonso:2016etj}, and the corresponding effects are very challenging to identify, we hope our works can serve as a motivation for the experimental tau physics groups at Belle-II to measure the different observables we have discussed.

\begin{acknowledgements}
\,
The work of S.GS has been supported in part by the National Science Foundation (PHY-1714253) and by the U.S. Department of Energy under Grants No. DE-FG02-87ER40365. 
The work of J. A. Miranda and J. J. Rend\'on has been granted by their Conacyt scholarships. 
P. R. thanks Conacyt funding through projects 250628 (Ciencia Básica) and Fondo SEP-Cinvestav 2018 (No. 142). 
We are indebted to Zhi-Hui Guo, Matthias Jamin and Jos\'e Antonio Oller for providing us with scalar form factors' tables and to Rafel Escribano for fruitful discussion on the topic.

\end{acknowledgements}

\end{document}